\documentclass[a4paper,12pt,preprintnumbers]{article}

\usepackage{jheppub}
\usepackage{mathrsfs}
\usepackage{verbatim}
\usepackage{booktabs}

\usepackage[latin1]{inputenc} 
\usepackage{latexsym} 
\usepackage{graphicx} 
\usepackage{amsfonts}
\usepackage{amsmath} 
\usepackage{amssymb} 
\usepackage{parskip}
\usepackage{caption}
\usepackage{subcaption}
\usepackage{array}
\usepackage{color}
\usepackage[toc,page]{appendix}
\usepackage{rotating}
\usepackage{afterpage}

\usepackage[normalem]{ulem}
\usepackage{bbm}
\usepackage{slashed}
\usepackage{epsfig,graphicx,bm,color} % Include figure files
\usepackage{dcolumn}% Align table columns on decimal point
\usepackage{url}
\usepackage{bm}% bold math
\usepackage{hyperref}% add hypertext capabilities
\usepackage{breakurl}
\usepackage{multirow}
\usepackage{soul} 
\setstcolor{red}
\usepackage{hyperref}

\newcommand{\be}{\begin{equation}}
\newcommand{\ee}{\end{equation}}
\newcommand{\bea}{\begin{eqnarray}}
\newcommand{\eea}{\end{eqnarray}}
\newcommand{\met}{\ensuremath{\slashed{E}_T}}

\newcommand{\lsim}{\mathrel{\mathop{\kern 0pt \rlap
  {\raise.2ex\hbox{$<$}}}
  \lower.9ex\hbox{\kern-.190em $\sim$}}}
\newcommand{\gsim}{\mathrel{\mathop{\kern 0pt \rlap
  {\raise.2ex\hbox{$>$}}}
  \lower.9ex\hbox{\kern-.190em $\sim$}}}

\definecolor{orange}{RGB}{255,127,0}

\begin{document}

% title to be defined
\title{\center{The BSM-AI project: \\ SUSY-AI -- generalizing LHC limits on supersymmetry with machine learning}}

\author[a,b]{Sascha~Caron,}
\author[c]{Jong~Soo~Kim,}
\author[c,d]{Krzysztof~Rolbiecki,}
\author[e]{Roberto~Ruiz~de~Austri,}
\author[a]{Bob Stienen}

\preprint{IFT-UAM/CSIC-16-042}

\affiliation[a]{Institute for Mathematics, Astro- and Particle Physics IMAPP, Radboud Universiteit, Nijmegen, The Netherlands}
\affiliation[b]{Nikhef, Amsterdam, The Netherlands}
\affiliation[c]{Instituto de F\'{\i}sica Te\'{o}rica UAM/CSIC, Madrid, Spain}
\affiliation[d]{Faculty of Physics, University of Warsaw, Warsaw, Poland}
\affiliation[e]{Instituto de F\'isica Corpuscular, IFIC-UV/CSIC, Valencia, Spain}

%%%%%%%%%%%%%%%%%%%%%%%%%%%%%%%%%%%%%%%%%%%%%%%%%%%%%%%%

\abstract{A key research question at the Large Hadron Collider is the test of models of new physics. Testing if a particular parameter set of such a model is excluded by LHC data is a challenge: it requires time consuming generation of scattering events, simulation of the detector response, event reconstruction, cross section calculations and analysis code to test against several hundred signal regions defined by the ATLAS and CMS experiments. In the BSM-AI project we approach this challenge with a new idea. A machine learning tool is devised to predict within a fraction of a millisecond if a model is excluded or not directly from the model parameters. A first example is SUSY-AI, trained on  the phenomenological supersymmetric standard model (pMSSM). About $300\,000$ pMSSM model sets -- each tested against 200 signal regions by ATLAS -- have been used to train and validate SUSY-AI. The code is currently able to reproduce the ATLAS exclusion regions in 19 dimensions with an accuracy of at least $93\%$. It has been validated further within the constrained MSSM and the minimal natural supersymmetric model, again showing high accuracy. SUSY-AI and its future BSM derivatives will help to solve the problem of recasting LHC results for any model of new physics.

SUSY-AI can be downloaded from \url{http://susyai.hepforge.org/}. An on-line interface to the program for quick testing purposes can be found at \url{http://www.susy-ai.org/}.}

%%%%%%%%%%%%%%%%%%%%%%%%%%%%%%%%%%%%%%%%%%%%%%%%%%%%%%%%

\maketitle

\section{Introduction}
The ATLAS and CMS experiments at the Large Hadron Collider (LHC) have  analyzed the full Run~1 and a small fraction of the Run~2 data set and no evidence of new physics has been found. In particular, there is no trace of supersymmetry (SUSY) in conventional searches. 

Both collaborations have explored intensively the impact of the null results in the context of simplified models \cite{Aad:2015iea,Aad:2015eda,Chatrchyan:2013sza} as well as in complete models like the constrained minimal supersymmetric standard model (cMSSM) \cite{Aad:2015iea}. In addition, both experiments have started to focus on {\it natural} SUSY scenarios addressing the emerging little hierarchy problem \cite{Aad:2015pfx,Khachatryan:2015pwa}. Finally, the ATLAS and CMS data sets were successfully interpreted with sampling of the 19-dimensional phenomenological minimal supersymmetric standard model (pMSSM) with specific priors, astrophysical constraints and particle physics constraints from the Higgs physics, electroweak precision observables and direct LEP2 limits~\cite{Aad:2015baa,Khachatryan:2016nvf}. 

The estimation of the number of expected signal events for a fixed point 
in the SUSY parameter space can take from several minutes to hours 
in CPU time when full detector simulations with GEANT are performed. In addition, it requires to model all LHC SUSY searches. Therefore 
any attempt to use LHC data on models like the pMSSM \cite{Djouadi:1998di} with 19 dimensions 
is cumbersome. 

In order to overcome this problem, several tools have been developed 
in the past few years to recast the LHC results: for simplified models, \texttt{Fastlim}~\cite{Papucci:2014rja} and \texttt{SModelS}~\cite{Kraml:2013mwa} have been introduced. 
Both tools recast LHC searches on new physics scenarios without relying on the slow Monte Carlo (MC) event generation and detector simulation. 
However, realistic models or scenarios with high-dimensional parameter space do not fulfill the assumptions of simplified models and the full MC
 event generation is usually required. For this reason, tools like \texttt{NLL-fast}~\cite{Beenakker:1996ch,Kulesza:2008jb,Kulesza:2009kq} as a fast cross section calculator, the recasting projects 
\texttt{CheckMATE}~\cite{Drees:2013wra,Kim:2015wza} and \texttt{MadAnalysis}~\cite{Conte:2012fm} based on \texttt{Delphes}~\cite{deFavereau:2013fsa}, which is a fast detector simulator, were developed. 
However, \texttt{CheckMATE} as well as \texttt{MadAnalysis} still require MC event generation and thus testing model 
points still takes a few tens of minutes.

Machine learning (ML) is becoming a powerful tool for the analysis of complex and large data sets, 
successfully assisting scientists in numerous fields of science and technology. An example of this is
the use of boosted decision trees \cite{rok2008} in the analyses that led to the Higgs discovery at the LHC in 2012 \cite{Aad:2012tfa, Chatrchyan:2012xdj}. 
Moreover, recently there have been applications to SUSY phenomenology in coverage studies \cite{Bridges:2010de}, 
in the study of the  cMSSM~\cite{Buckley:2011kc} and in the reconstruction of the cMSSM parameters 
\cite{Bornhauser:2013aya}.

In this work we propose the use of ML methods to explore in depth LHC constraints on the rich SUSY phenomenology.
In particular, we investigate the use of classifiers to predict whether 
a point in the pMSSM parameter space is excluded or not in light of the results of the full set of ATLAS Run~1 data, avoiding time consuming MC 
simulations.  We show that decision tree classifiers like the {Random Forest (RF) algorithm} perform very well 
in the pMSSM. Similar results have been obtained for other MSSM realizations {such} as the natural SUSY model and the cMSSM.
The method discussed here allows for a quick analysis of large data sets and can  be coupled with recasting tools to resolve the remaining ambiguities
by generating more training data. 
It could also be used in projects aiming at fits
of the multidimensional parameter space of the pMSSM or derived models, like e.g.~\cite{deVries:2015hva,Strege:2014ija}.

The paper is structured as follows. In Section~\ref{sec:atlasandpmssm} we recap the search for SUSY by ATLAS in the context of the pMSSM. In Section~\ref{sec:mlandclassification} we briefly review the machine learning techniques used in our analysis.  In Section \ref{sec:training} we present a procedure of generating the ML classifier. The validation and performance of the classifier are described in Section~\ref{sec:5}: in the pMSSM framework in Section~\ref{sec:pmssm}, for the natural SUSY model in Section~\ref{sec:nsusy} and for the constrained MSSM in Section~\ref{sec:cmssm}. Section~\ref{sec:limitations} discusses limitations of the code. Finally, we summarize our findings in Section~\ref{sec:conclusions}. In Appendix~\ref{app:splitting} we discuss a comparison of different estimation methods and in Appendix~\ref{app:projections} we provide additional validation information.

%%%%%%%%%%%%%%%%%%%%%%%%%%%%%%%%%%%%%%%%%%%%%%%%%%%%%%%%

\section{The pMSSM and ATLAS SUSY searches \label{sec:atlasandpmssm}}
The MSSM with $R$-parity conservation is uniquely described by its particle spectrum and the superpotential \cite{Drees:2004jm},
\begin{equation}
W=\epsilon_{ij}\left[(h_L)_{mn}H_1^iL_m^j \overline{E}_n+(h_D)_{mn}H_1^i Q_m^j \overline{D}_n-(h_U)_{mn}H_2^i Q_m^j \overline{U}_n-\mu H_1^iH_2^j\right],
 \end{equation}
where $\epsilon_{ij}$ is the antisymmetric $SU(2)$ tensor with $\epsilon_{12}=+1$. $h_L$, $h_D$, $h_U$ and $\mu$ denote the lepton-, down-type and up-type Yukawa couplings and 
the Higgs superpotential mass parameter, respectively. Generation indices are denoted by $m$ and $n$. The chiral superfields have the following gauge quantum numbers under the Standard Model (SM) group $G=SU(3)_C\times SU(2)_L\times U(1)_Y$: 
\begin{eqnarray}
L:(1,2,-1/2),\quad \overline{E}:(1,1,1),\quad Q:(3,2,1/6),\quad \overline{U}:(\bar{3},1,-2/3), \nonumber\\
\overline{D}:(3,1,1/3),\quad H_1:(1,2,-1/2),\quad H_2:(1,2,1/2),
 \end{eqnarray}
while the vector multiplets have the following charges under $G$:
\begin{eqnarray}
g:(8,1,0),\quad W:(1,3,0),\quad B:(1,1,0).
\end{eqnarray}
All kinetic terms and gauge interactions must be consistent with supersymmetry and be invariant under $G$.
Since the origin of supersymmetry breaking is unknown, one approach to addressing this issue is avoiding explicit assumptions about a SUSY-breaking mechanism. It is then common to write down the most general supersymmetry breaking terms consistent with the gauge symmetry and the $R$-parity conservation~\cite{Girardello:1981wz},
\begin{eqnarray}\label{eq:susybreaking}
V_{\rm{soft}}&=&m_1^2|H_1|^2+m_2^2|H_2|^2-m_{12}^2(\epsilon_{ij} H_1^i H_2^j+\rm{h.c.})\nonumber\\
&+&(M_{\tilde Q}^2)_{mn}\tilde Q_m^{i*}\tilde Q_n^i+(M_{\tilde U}^2)_{mn}\tilde U_m^{i*}\tilde U_n^i+(M_{\tilde D}^2)_{mn}\tilde D_m^{i*}\tilde D_n^i\nonumber\\
&+&(M_{\tilde L}^2)_{mn}\tilde L_m^{i*}\tilde L_n^i+(M_{\tilde E}^2)_{mn}\tilde E_m^{i*}\tilde E_n^i+\nonumber\\
&+&\epsilon_{ij}[(h_L A_L)_{mn}\tilde H_1^i\tilde L_m^j\tilde E_n+(h_D A_D)_{mn}\tilde H_1^i\tilde Q_m^j\tilde D_n+(h_U A_U)_{mn}\tilde H_2^i\tilde Q_m^j\tilde U_n+\rm{h.c.}]\nonumber\\
&+&\frac{1}{2}[M_3\tilde g\tilde g+M_2\tilde W^a\tilde W^a+M_1 \tilde B \tilde B+\rm{h.c.}].
\end{eqnarray}
Here, $M_{\tilde Q}^2$, $M_{\tilde U}^2$, $M_{\tilde D}^2$, $M_{\tilde L}^2$ and $M_{\tilde E}^2$ are $3\times3$ Hermitian matrices in generation space, $(h_L A_L)$, $(h_D A_D)$ and $(h_U A_U)$ are complex $3\times3$ trilinear scalar couplings and $m_1^2$, $m_2^2$ as well as $m_{12}^2$ correspond to the SUSY-breaking Higgs masses. $M_1$, $M_2$ and $M_3$ denote the $U(1)_Y$, $SU(2)_L$ and $SU(3)_C$ gaugino masses, respectively. The fields with a tilde are the supersymmetric partners of the corresponding SM field in the respective supermultiplet.
Most new parameters of the MSSM are introduced by Eq.~\eqref{eq:susybreaking} and a final count yields 105 genuine new parameters \cite{Haber:1997if}. One can reduce the 105 MSSM parameters to $19$ by imposing phenomenological constraints, which define the so-called phenomenological MSSM (pMSSM)~\cite{Djouadi:1998di,Berger:2008cq}. In this scheme, one assumes the following: (i) all the soft SUSY-breaking parameters are real, therefore the only source of CP-violation is the CKM matrix; (ii) the matrices of the sfermion masses and the trilinear couplings are diagonal, in order to avoid FCNCs at the tree-level; (iii) first and second sfermion generation universality to avoid severe constraints, for instance, from  $K^0$--$\bar{K}^0$ mixing.

The sfermion mass sector is described by the first and second generation universal squark masses $M_{\tilde{Q}_1}\equiv (M_{\tilde Q})_{nn}$, $M_{\tilde{U}_1}\equiv (M_{\tilde U})_{nn}$ and $M_{\tilde{ D}_1}\equiv (M_{\tilde D})_{nn}$ for $n=1,2$, 
the third generation squark masses $M_{\tilde{Q}_3}\equiv (M_{\tilde Q})_{33}$, $M_{\tilde{U}_{33}}\equiv (M_{\tilde U})_{33}$ and $M_{\tilde{D}_3}\equiv (M_{\tilde D})_{33}$, the first and second generation 
slepton mass $M_{\tilde{L}_1}\equiv (M_{\tilde L})_{nn}$, $M_{\tilde{E}_1}\equiv (M_{\tilde E})_{nn}$ for $n=1,2$ and the third generation slepton masses $M_{\tilde{L}_3}\equiv (M_{\tilde L})_{33}$ and $M_{\tilde{E}_3}\equiv (M_{\tilde E})_{33}$. The trilinear couplings of the sfermions enter in the off-diagonal parts of the sfermion mass matrices. Since these entries are proportional to the Yukawa couplings of the respective fermions, we can approximate the trilinear couplings associated with the first and second generation fermions to be zero. Instead, the third generation trilinear couplings are described by the  parameters $A_t \equiv (A_U)_{33}$,  $A_b\equiv (A_D)_{33}$ and $A_\tau\equiv (A_L)_{33}$. 

After the application of the electroweak symmetry breaking conditions, the Higgs sector can be fully described by the ratio of the Higgs vacuum expectation values, $\tan \beta$, and the soft SUSY-breaking Higgs mass parameters $m_{i}^2$. Instead of the Higgs masses, we choose to use the higgsino mass parameter $\mu$ and the mass of the pseudoscalar Higgs, $m_A$, as input parameters, as they are more directly related to the phenomenology of the model. 

\begin{table}
\begin{center}
\begin{tabular}{lll}
\hline
Reference & Final State & Category\\
\hline
\cite{Aad:2014wea} & $0$ lepton + $2$--$6$ jets + \met &Inclusive \\
\cite{Aad:2013wta} &  $0$ lepton + $7$--$10$ jets + \met &          \\
\cite{Aad:2015mia} &  $1$ lepton + jets + \met &         \\
\cite{Aad:2014mra} &  $\tau(\tau/\ell)$ + jets + \met    &      \\
\cite{Aad:2014pda} &   SS/$3$ lepton + jets + \met     &    \\
\cite{Aad:2014lra} &  $b$-jets + $0/1$ lepton + \met &          \\
\cite{Aad:2015zva} & monojet &           \\ \hline
\cite{Aad:2014bva} & $0$ lepton stop search&Third generation \\
\cite{Aad:2014kra} &  $1$ lepton stop search &  squarks       \\
 \cite{Aad:2014qaa} &  $2$ lepton stop search &        \\
\cite{Aad:2014nra} & monojet search &          \\
\cite{Aad:2014mha} &   stop search with $Z$ in final state &        \\
\cite{Aad:2013ija} &  2$b$-jets sbottom search  &       \\
\cite{Aad:2015pfx}       &  asymmetric stop search  & \\ \hline
\cite{Aad:2015jqa} & $1$ lepton plus Higgs final state & Electroweak \\
\cite{Aad:2014vma} &  dilepton final state   &       \\
\cite{Aad:2014yka} &   $2\tau$ final state&         \\
\cite{Aad:2014nua} &  trilepton final state     &     \\
\cite{Aad:2014iza} &    four-lepton final state &        \\
\cite{Aad:2013yna} &   disappearing track  &      \\
\hline
\cite{Aad:2012pra,ATLAS:2014fka} &  Long-lived particle search & Other \\
\cite{Aad:2014vgg} & $H/A\rightarrow\tau\tau$ search &           \\ \hline
\end{tabular}
\end{center}
\caption{The experimental analyses used in the ATLAS study~\cite{Aad:2015baa}. The middle column denotes the final
  state for which the analysis is optimized, and the third column
  shows the target scenario of this analysis.}
\label{tab:lhc_searches}
\end{table}

The final ingredients of our model are the three gaugino masses: the bino mass $M_1$, the wino mass $M_2$, and the gluino mass $M_3$. The above parameters describe the 19-dimensional realization of the pMSSM, which encapsulates all phenomenologically relevant features of the full model that are of interest for dark matter and collider experiments. 

The ATLAS study~\cite{Aad:2015baa} considered 22 separate ATLAS analyses of the Run 1 summarized in Table~\ref{tab:lhc_searches}. These studies cover a large number of different final-state topologies, disappearing tracks, long-lived charged particles as well as the search for heavy MSSM Higgs bosons. Reference~\cite{Aad:2015baa} combines all searches and the corresponding signal regions in order to derive strict constraints on the pMSSM. For this purpose, $5\times10^8$ model points were sampled within the ranges shown in Table~\ref{tab:softbreaking_pmssm}. 
The model points had to satisfy preselection cuts following closely the procedure described in Ref.~\cite{CahillRowley:2012cb}. All selected points had to pass the precision electroweak and flavor constraints summarized in Table~\ref{tab:constraints}. These include the electroweak parameter $\Delta \rho$~\cite{Baak:2012kk}, the branching ratios for rare $B$ decays~\cite{Amhis:2012bh,DeBruyn:2012wk,CMS:2014xfa,Mahmoudi:2008tp}, the SUSY contribution to the muon anomalous magnetic moment $\Delta(g-2)_{\mu}$~\cite{Aoyama:2012wk,Bennett:2006fi}, and the $Z$ boson width~\cite{ALEPH:2005ab} {and LEP limits on the production of SUSY particles~\cite{LEPconst}}. {Furthermore, thermally produced dark matter relic density is required to be at or below the Planck measured value~\cite{Ade:2015xua}. Finally, the constraint on the Higgs boson mass~\cite{Aad:2015zhl,Hahn:2013ria} was applied. After this preselection $310\,327$ model points remained, for which the production cross sections for all final states were computed.}

\begin{table*}[t!]
\begin{center}
\begin{tabular}{lll}
\hline
Parameter & Description & Scanned range \\
\hline
$m_{\tilde L_1}$& \small{1$^{\rm st}$/2$^{\rm nd}$ gen. $SU(2)$ doublet soft breaking slepton mass} & \small{$[90 \ {\rm GeV},\, 4 \ {\rm TeV}]$}\\
$m_{\tilde E_1}$ & \small{1$^{\rm st}$/2$^{\rm nd}$ gen. $SU(2)$ singlet soft breaking slepton mass} & \small{$[90 \ {\rm GeV},\, 4 \ {\rm TeV}]$}\\
$m_{\tilde L_3}$& \small{3$^{\rm rd}$ gen. $SU(2)$ doublet soft breaking slepton mass} & \small{$[90 \ {\rm GeV},\, 4 \ {\rm TeV}]$}\\
$m_{\tilde E_3}$ & \small{3$^{\rm rd}$ gen. $SU(2)$ singlet soft breaking slepton mass} & \small{$[90 \ {\rm GeV},\, 4 \ {\rm TeV}]$}\\
$m_{\tilde Q_1}$& \small{1$^{\rm st}$/2$^{\rm nd}$ gen. $SU(2)$ doublet soft breaking squark mass} & \small{$[200 \ {\rm GeV},\, 4 \ {\rm TeV}]$}\\
$m_{\tilde U_1}$ & \small{1$^{\rm st}$/2$^{\rm nd}$ gen. $SU(2)$ singlet soft breaking squark mass} & \small{$[200 \ {\rm GeV},\, 4 \ {\rm TeV}]$}\\
$m_{\tilde D_1}$ & \small{1$^{\rm st}$/2$^{\rm nd}$ gen. $SU(2)$ singlet soft breaking squark mass} & \small{$[200 \ {\rm GeV},\, 4 \ {\rm TeV}]$}\\
$m_{\tilde Q_3}$& \small{3$^{\rm rd}$ gen. $SU(2)$ doublet soft breaking squark mass} & \small{$[100 \ {\rm GeV},\, 4 \ {\rm TeV}]$}\\
$m_{\tilde U_3}$ & \small{3$^{\rm rd}$ gen. $SU(2)$ singlet soft breaking squark mass} & \small{$[100 \ {\rm GeV},\, 4 \ {\rm TeV}]$}\\
$m_{\tilde D_3}$ & \small{3$^{\rm rd}$ gen. $SU(2)$ singlet soft breaking squark mass} & \small{$[100 \ {\rm GeV},\, 4 \ {\rm TeV}]$}\\
\hline
$A_{t}$ & \small{Stop trilinear coupling} & \small{$[-8 \ {\rm TeV},\, 8 \ {\rm TeV}]$}\\
$A_{b}$ & \small{Sbottom trilinear coupling} & \small{$[-4 \ {\rm TeV},\, 4 \ {\rm TeV}]$}\\
$A_{\tau}$ & \small{Stau trilinear coupling} & \small{$[-4 \ {\rm TeV},\, 4 \ {\rm TeV}]$}\\
\hline
$|\mu|$ & \small{Higgsino mass parameter} & \small{$[80 \ {\rm GeV},\, 4\ {\rm TeV}]$}\\
$|M_1|$ & \small{Bino mass parameter} & \small{$[0 \ {\rm TeV},\, 4 \ {\rm TeV}]$}\\
$|M_2|$ & \small{Wino mass parameter} & \small{$[70 \ {\rm GeV},\, 4 \ {\rm TeV}]$}\\
$M_3$ & \small{Gluino mass parameter} & \small{$[200 \ {\rm GeV},\, 4 \ {\rm TeV}]$}\\
\hline
$M_A$ & \small{Pseudoscalar Higgs mass} & \small{$[100 \ {\rm GeV},\, 4 \ {\rm TeV}]$}\\
$\tan\beta$ & \small{Ratio of vacuum expectation values} & \small{$[1, \, 60]$} \\
\hline
\end{tabular}
\end{center}
\caption{Variable input parameters of the ATLAS pMSSM scan %defined at the scale $Q=\sqrt{m_{\tilde{t}_1 } m_{\tilde{t}_2 } }$, 
 and the range over which these parameters are scanned.
 \label{tab:softbreaking_pmssm}}
\end{table*}

All models with production cross sections larger than a threshold were further processed. Matched truth level MC event samples with up to one additional parton in the matrix element were generated and efficiency factors\footnote{{The efficiency factor tells the number of Monte Carlo events passing the experimental selections relative to the full MC sample size for each parameter point. Since the number of simulated MC events typically exceeds the nominal number by a factor of a few (to reduce statistical fluctuations) the final number of expected events in each signal region is obtained by multiplying the efficiency, integrated luminosity and cross section.}} were determined for each signal region and the final yield was determined. For points that could not be classified with at least 95\% certainty using this method, a fast detector simulation based on GEANT4 was performed \cite{Aad:2015baa}. The exclusion of the model points was then determined using 22 different analyses, taking into account almost 200 signal regions covering a large spectrum of final-state signatures. The exclusion of a model point is decided by the analysis with the best expected sensitivity. We follow this approach here.

\begin{table*}[t!]
\begin{center}
\begin{tabular}{lll}
\hline
Parameter & Minimum Value & Maximum Value \\
\hline
$\Delta\rho$ & $-0.0005$ & $0.0017$\\
$\Delta(g-2)_{\mu}$ & $-17.7\times10^{-10}$ & $43.8\times10^{-10}$\\
BR($b\rightarrow s\gamma)$ & $2.69\times10^{-4}$ & $3.87\times10^{-4}$\\
BR($B_s\rightarrow \mu^+\mu^-)$ & $1.6\times10^{-9}$ & $4.2\times10^{-9}$\\
BR($B^+\rightarrow \tau^+\nu_\tau)$ & $66\times10^{-6}$ & $161\times10^{-6}$\\
$\Omega_{\tilde\chi_1^0}h^2$ & $-$ & $0.1208$\\
$\Gamma_{\rm invisible}(Z)$ & $-$ & $2$ MeV\\
Masses of charged sparticles & 100 GeV & $-$\\
$m_{\tilde\chi_1^\pm}$ & $103$ GeV & $-$\\
%$m_{\tilde U_{1}}$, $m_{\tilde D_{1}}$ %$m_{\tilde c_{1,2}}$, $m_{\tilde s_{1,2}}$  & $200$ GeV & $-$\\
$m_h$ & $124$ GeV & $128$ GeV\\
\hline
\end{tabular}
\end{center}
\caption{Preselection cuts for the pMSSM benchmark points~\cite{Aad:2015baa}.
 \label{tab:constraints}}
\end{table*}

%%%%%%%%%%%%%%%%%%%%%%%%%%%%%%%%%%%%%%%%%%%%%%%%%%%%%%%%

\section{Machine learning and classification \label{sec:mlandclassification}}
In terms of ML the problem considered in this paper is a classification problem and there are several methods for addressing it. We will focus on decision tree classifiers and in particular on the random forest classifier~\cite{Breiman:RF}, which was found to give the best results in the present case, compared to other ML methods like AdaBoost~\cite{Freund:1997}, $k$-Nearest Neighbors~\cite{Cover:1967} and Support Vector Machines~\cite{Boser:1992}, among others.

In the following we present an introduction to decision trees and the random forest classifier, aiming at providing a basic understanding of the algorithms. For more complete texts on the subject, the reader is referred to  Refs.~\cite{Witten:2005} and \cite{Bishop:2006}. A more detailed and technical description of the Random Forest algorithm can be found in Ref.~\cite{Breiman:RF}.

\subsection{Decision trees and random forest}
In a classification problem the goal is to classify a parameter set (attribute set), $\vec{y} = \{y_1,...,y_N\}$, by assigning it a class label $C$, corresponding to the class it belongs to. The procedure starts with \textit{training} a classifier by presenting parameter sets and the corresponding class labels, in order to learn patterns that the input data follow. Though this basic principle is the same for all classification algorithms, the specific implementation differs depending on the particular problem.

\begin{figure}
\begin{center}
\includegraphics[width=0.9\textwidth]{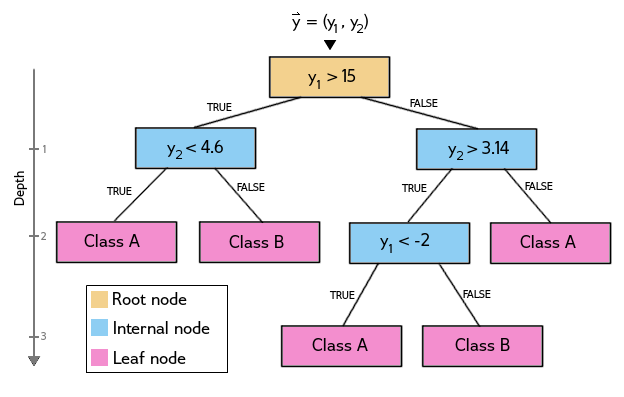}
\caption{Graphical representation of a sample decision tree.}
\label{fig:dtree}
\end{center}
\end{figure}

Decision trees are often used as a method to approach a classification problem. An example of a decision tree is shown in Figure~\ref{fig:dtree}. In this example, the tree classifies a 2-dimensional attribute set $\vec{y} = (y_1, y_2)$ as either class \textit{A} or class \textit{B}.

A decision tree consists of multiple nodes. Every node specifies a test performed on the attribute set arriving at that node. The result of this test determines to which node the attribute set is sent next. In this way, the attribute set moves down the tree. This process is repeated until the final leaf node is reached, i.e.\ the node with no further nodes connected to it. At the final node no test is performed, but a class label is assigned to the set, specifying its class according to the classifier. The depth of the tree is the maximum number of nodes, as shown in Figure~\ref{fig:dtree}.  

Because the tree works on the entire parameter space, every test performed in each node can also be interpreted as a cut in this space. By creating a tree with multiple nodes, the parameter space is split into disjunct regions, each having borders defined by the cuts in the root and internal nodes, and a classification defined by a leaf node.

One of the drawbacks of decision trees is that they are prone to \textit{overtraining}: they have a tendency to learn every single data point as an expression of a true feature of the underlying pattern, yielding decision boundaries with more detailed features than actually present in the underlying pattern. Although this overtraining may cause a better prediction when classifying the training data, such classifiers generally perform poorly on new data sets.

A simple, yet crude method to fix this problem is known as \textit{pruning}: training the entire tree, but cutting away all nodes beyond a certain maximum depth. This effectively reduces the amount of details the tree can distinguish in the learned data pattern, since fewer cuts are made in the parameter space. With a maximum depth set to a certain value, classification will not be perfect and mistakes will be made in predictions for the training. Lone data points in sparse regions or individual data points with a classification different from a classification of the data points around it will therefore be learned less efficiently, thus reducing their influence on the trained classifier.

In the Random Forest algorithm, multiple decision trees are combined into a single classifier, creating an ensemble classifier. Classification of an attribute set is then decided by a majority vote: the ratio of trees that predict class $C$ and the total number of trees is taken as a measure of probability the attribute set belongs to class $C$. The class with the highest score is assigned to the attribute set. This method averages out fluctuations that cause overtraining in individual trees.

Another method to overcome overtraining is to implement a random attribute selection, meaning that at each node the cuts are applied only to a subset of attributes. In a single decision tree this would introduce a massive error in predictions, but since the random forest is an ensemble classifier, its predictions actually improve.

In addition to these two methods, random forests also use \textit{bagging} to reduce overtraining even more. In bagging, each decision tree is trained on a random selection of $n$ model points out of $N$ available in the training set. The sampling is done with replacement, meaning a single model point can be selected multiple times. In bagging, $n$ is conventionally chosen to be equal to the total amount of model points available, which means each tree is trained on approximately $63\%$ of points for large $N$~\cite{breimano}. By using this procedure, the contribution of a single data point to the learned pattern is reduced, making the classifier more focused on the collective patterns. 

Overtraining of a classifier is difficult to express quantitatively, but it can be tested qualitatively using an independent test set, with which one can estimate the fraction of incorrectly predicted data points for general datasets (i.e.\ datasets other than the training set). This fraction is called the \textit{generalization error}. A high generalization error is a possible indicator of overtraining. Normally the check on this error is performed by splitting the available data into training and testing subsets. The training set is used to train the classifier, while the testing set is used to test the predictions of the algorithm. This splitting of the dataset is, however, not the procedure we followed here. Since random forests employ \textit{bagging}, a single data point is used only in training of a part of the trees. Testing of the algorithm can thus be done by letting all data points be predicted by those trees that did not use a particular data point in their training. As with a test set, one can now obtain a fraction incorrectly predicted data points, and thus estimate the generalization error. This method is called \textit{out-of-bag estimation} (OOB). The obvious advantage of this procedure is that all the data can be used in training without the need to split the sample into the training and testing sets, hence improving the general prediction quality of the algorithm. It was shown in Ref.~\cite{breimano} that this method provides an error estimate as good as train-test split method; see Appendix~\ref{app:splitting} for a direct comparison in our case.

Though random forest is in general not very susceptible to overtraining due to the bagging procedure, its performance depends on the number of trees it contains: the more trees, the less overtraining will occur, because predictions are averaged over many individual trees thus reducing undesired fluctuations. The number of decision trees inside the forest is a configuration parameter that has to be set before starting a training, as are the maximum depth of the decision trees and the number of features used in the random attribute selection at each node for example.

In this work we used the RF implementation in the \texttt{scikit-learn} Python package~\cite{scikit-learn} (version 0.17.1).

%%%%%%%%%%%%%%%%%%%%%%%%%%%%%%%%%
\subsection{Performance of a classifier}
%%%%%%%%%%%%%%%%%%%%%%%%%%%%%%%%

Given a classifier and a testing set, there are four possible outcomes in case of binary classification (i.e.\ ``positive'' and ``negative''). If the true classification is positive and the prediction by the classifier is positive, then the attribute set is counted as a true positive (TP). If the classifier classifies the set as negative, it is counted as false negative (FN).  If, on the contrary, the attribute set is truly negative and it is classified as negative, it is counted as a true negative (TN) and if it is classified as positive, it is counted as a false positive (FP).

With this, one can define the true positive rate (TPR) as the ratio of the positives correctly classified and the actual positive data points. The false positive rate (FPR) is the ratio of negatives incorrectly classified and the total truly negative data points.

A receiver operating characteristic (ROC) graph is a two-dimensional plot in which the TPR is plotted on the vertical axis and the FPR is plotted on the horizontal axis.\footnote{TPR and FPR may also be called sensitivity and specificity in the literature.} The ROC graph shows a relative trade-off between benefits (true positives) and costs (false positives). Every discrete classifier produces an (FPR, TPR) pair for a specified cut on the classifier output corresponding to a single point in the ROC space. The lower left point $(0,0)$ represents a strategy of never getting a positive classification; such a classifier commits no false positive errors but also does not predict true positives. The opposite strategy, of unconditionally assigning positive classifications, is represented by the upper right point $(1,1)$. The point $(0, 1)$ represents the perfect classification. 

The ROC curve is a two-dimensional representation of a classifier performance. A common method used to compare classifiers is the  area under the ROC curve (AUC)~\cite{fawcett:2006roc}. Its value is always between 0 and 1. Because a random classification produces a diagonal line between $(0,0)$ and $(1,1)$, which corresponds to $\mathrm{AUC}=0.5$, no realistic classifier should have AUC less than 0.5. A perfect classifier has a AUC equal to 1. Each point on the curve corresponds to a different choice of the classifier output value that separates data points classified as allowed or excluded.

%%%%%%%%%%%%%%%%%%%%%%%%%%%%%%%%%%%%%%%%%%%%%%%%%%%%%%%%

\section{Training of SUSY-AI \label{sec:training}}
The classifier was trained using the data points generated by ATLAS as discussed in Section~\ref{sec:atlasandpmssm}. 
The set of parameters used in this classification task is shown in Table~\ref{tab:pmssmvars}. We follow here the SLHA-1 standard~\cite{Skands:2003cj} and provide the respective block names and parameter numbers. All input variables are defined at the SUSY scale $Q=\sqrt{m_{\tilde{t_1} } m_{\tilde{t_2} } }$.

\begin{table}
  \centering
  \begin{tabular}{p{2cm}p{2cm}p{1cm}p{1cm}p{2cm}p{2cm}p{1cm}}
  \cline{1-3}
  \cline{5-7}
    Parameter   & Block & No. & & Parameter   & Block & No.\\
   \cline{1-3}
   \cline{5-7}
    $M_1$ & MSOFT & 1 & &$m_{\widetilde{U}_3}$ & MSOFT & 46\\  
    $M_2$ & MSOFT & 2 & &$m_{\widetilde{D}_1}$ & MSOFT & 47\\  
    $M_3$ & MSOFT & 3 & &$m_{\widetilde{D}_3}$ & MSOFT & 49\\  
    $m_{\widetilde{L}_1}$ & MSOFT & 31 & &$A_t$ & AU & 3,3\\  
    $m_{\widetilde{L}_3}$ & MSOFT & 33 & &$A_b$ & AD & 3,3\\  
    $m_{\widetilde{E}_1}$ & MSOFT & 34 & &$A_\tau$ & AE & 3,3\\  
    $m_{\widetilde{E}_3}$ & MSOFT & 36 & &$\mu$ & HMIX & 1\\  
    $m_{\widetilde{Q}_1}$ & MSOFT & 41 & &$M_A^2$ & HMIX & 4\\  
    $m_{\widetilde{Q}_3}$ & MSOFT & 43 & &$\tan\beta$ & HMIX & 3\\ \cline{5-7}  
    $m_{\widetilde{U}_1}$ & MSOFT & 44 & & & & \\  
    \cline{1-3}
  \end{tabular}
  \caption{Variables used for training in the pMSSM. The variables are identified according to the \texttt{SLHA-1} standard~\cite{Skands:2003cj}, given in order used by \texttt{SUSY-AI}.}
  \label{tab:pmssmvars}
\end{table}

The class labels were generated by the exclusion analysis performed in Ref.~\cite{Aad:2015baa}. From the 22 analyses that were used by ATLAS, the exclusion of each point is decided  by means of the signal region with the best expected sensitivity at a given point. To excluded data points we assign a class label~\texttt{0}, to allowed data points a class label~\texttt{1}. {Note that the current version only uses the combined classification, without making a distinction which particular analysis excluded a given parameter point.}

In order to find an optimal configuration of the classifier, we used a grid search as an automated investigation method, {which varies the hyperparameters (number of trees, maximal depth, number of features at each node) to find a configuration with a maximal value for a figure of merit, which is the OOB score in our case. The result of this search yielded the following parameters:} 900 decision trees, a maximal depth of 30 nodes and a maximum number of features considered at each node of 12. The training was performed including the out-of-bag estimation technique for the creation of an estimate for the generalization accuracy. 

\begin{figure}
\centering
  \centering
  \includegraphics[width=0.7\textwidth]{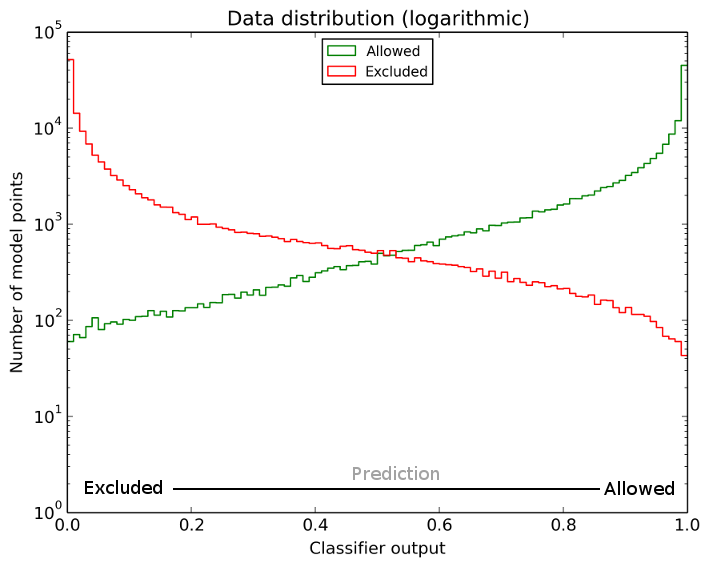}
  \caption{Distribution of the number of true allowed or excluded points by \mbox{ATLAS}~\cite{Aad:2015baa} as a function of the classifier output. 
  \label{fig:classification_histogram}}
\end{figure}

Figure \ref{fig:classification_histogram} shows a histogram of all data points with a classification prediction determined by the classifier. 
The horizontal axis shows the classifier output 
while the vertical axis shows a number of points for a given output with a true label \texttt{1} (green histogram) or \texttt{0} (red histogram). From the figure, one can conclude that although a vast majority of the points is classified correctly (the allowed points pile-up at a classifier output of $1.0$, while the excluded points pile-up at a classifier output of $0.0$), some of the points fall into the categories of false positives or false negatives.
A perfect classification is therefore not possible, and one has to make cuts in this diagram to make the classification binary. Setting the cut at $0.5$ would mean the truly excluded data points with a value for the output of $0.5$ or more would be classified as allowed, while the truly allowed data points with a classifier output of $0.5$ or less would be classified as excluded.

The desired location of the cut depends on the required properties of the classifier. For example, when one would like to avoid false positives the preferred value should be close to $1.0$, while the value close to $0.0$ will result in many true positive points being classified as positive for a price of many true negative points wrongly classified as positives. Typically, the neutral choice would be at a point of intersection of red and green histograms, in our example at $0.535$. It is assumed that newly added points will follow the same distribution as a function of the classifier output.

\begin{figure}
  \centering
  \includegraphics[width=0.7\textwidth]{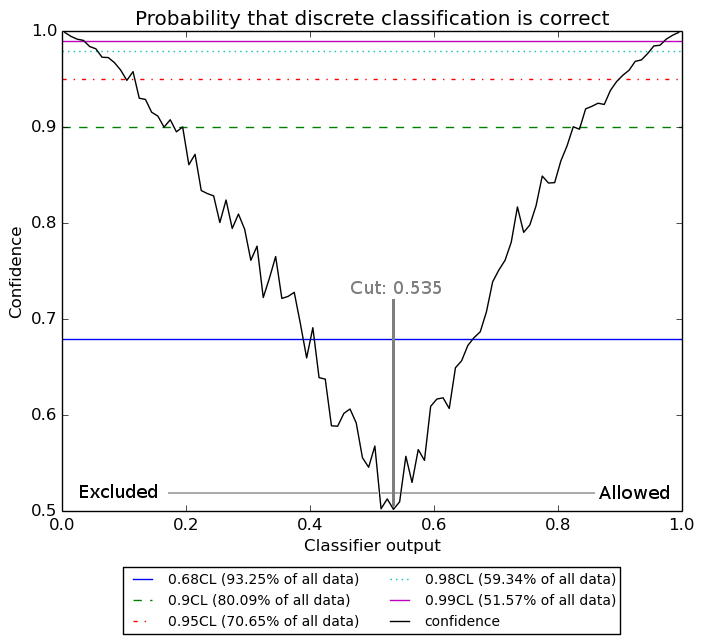}
  \caption{Confidence level that the classification is correct, as defined in the text. The horizontal lines indicate specific confidence levels.}
  \label{fig:cls}
\end{figure}

Another possibility, which we adopt here, is to plot the ratio of the majority class and the total number of points for each bin, as showed in Figure~\ref{fig:cls}. This provides a frequentist confidence level that a point with a given classifier output is truly allowed or excluded. The horizontal lines for typically used confidence levels are also shown in the figure.

\begin{figure}
	\centering
	\begin{subfigure}[b]{0.475\textwidth}
    	\includegraphics[width=\textwidth]{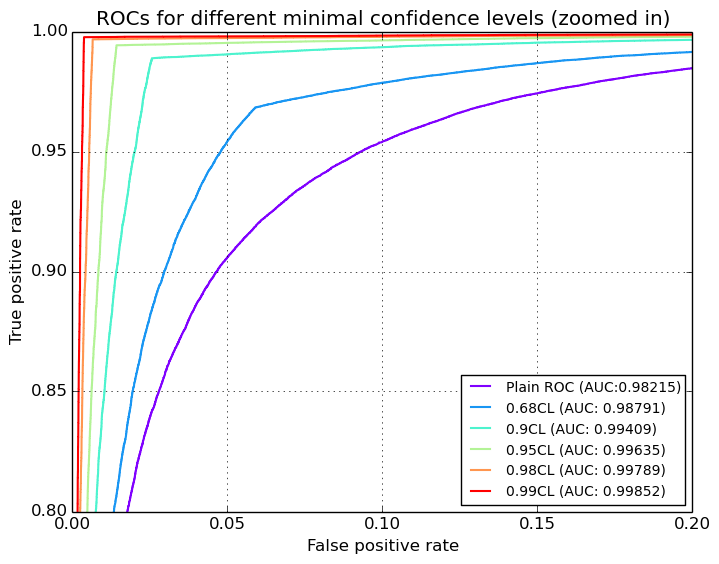}
		\caption{%\BS{ROC curves for different minimum CLs following from analysis of Figure~\ref{fig:cls}. Note that only the upper left corner of the full ROC curve is shown.}
  		\label{fig:rocsforcls}}
  	\end{subfigure}
    \hfill
    \begin{subfigure}[b]{0.515\textwidth}
    	\includegraphics[width=\textwidth]{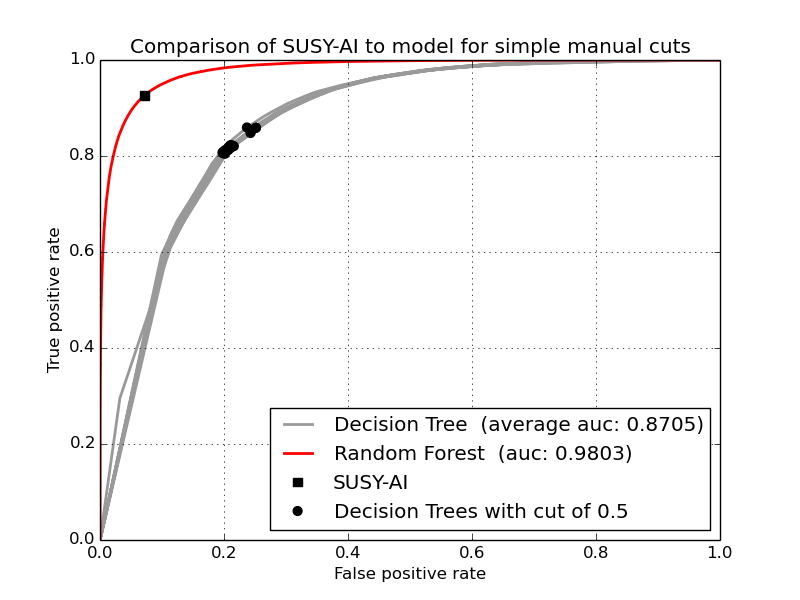}
  		\caption{%\BS{ROCs for SUSY-AI and for a simple decision tree modeling manual cuts on the dataset. The marker indicates SUSY-AI performance.}}
	  	\label{fig:dtcomparisonroc}}
 	\end{subfigure}
  
	\caption{(a) {ROC curves for different minimum CLs following from analysis of Figure~\ref{fig:cls}. Note that only the upper left corner of the full ROC curve is shown.} (b) ROCs for SUSY-AI and 20 decision trees (ROCs overlap here). The decision trees model educated manual cuts on the dataset. The square marker indicates the location of SUSY-AI performance without a cut on CL, while discs denote different decision trees. \label{fig:rocs}}
\end{figure}

Using this confidence level (CL) method, one can use the trained classifier to provide both a classification and a measure of confidence in that classification. By demanding a specific CL (and so the probability of the wrong classification) one determines the classifier output, which can be read from Figure~\ref{fig:cls}. For example, for a confidence level of 99\%, the classifier output for a given point should be below $0.05$ or above $0.95$, while for a confidence level of 95\% the predicted probabilities below $0.133$ or above $0.9$ are sufficient. Trimming the dataset using the limits determined by this method yields better results for the classification, but also implies that further analyses have to be done on the points that were cut away. The improved quality of classification can be seen by the increasing AUC for higher confidence levels in the ROC curve in Figure~\ref{fig:rocsforcls}. This plot was generated by varying the decision cut on the classifier output between $0.0$ and $1.0$ and plotting the result as a function of FPR vs TPR for a given CL cut.

\begin{figure}
  {\centering
  \includegraphics[width=\textwidth]{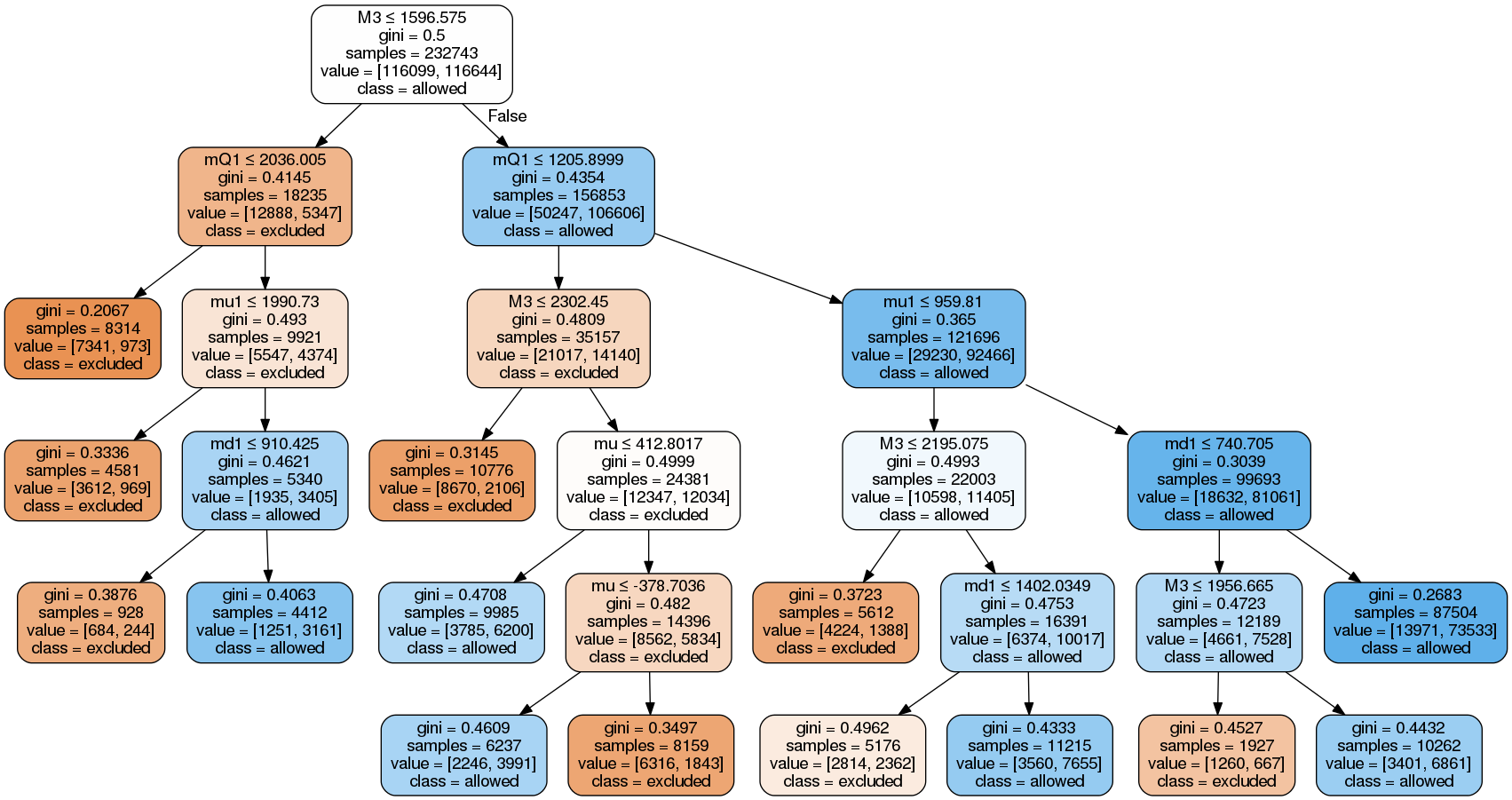}%[width=0.7\textwidth]
  \caption{Example of a decision tree modeling educated manual cuts, trained with a 75:25 ratio of training and testing data. The figure has been created with \texttt{GraphViz}~\cite{Gansner00anopen}. {The information shown in the nodes are, respectively: a cut that is made in the parameter space (line 1), an impurity measure which is minimized in the training (line 2), a number of model points in the node (line 3), the distribution of model points over the classes [excluded, allowed] (line 4) and a label of the majority class (line 5).}
  \label{fig:exampledt}}
  }
\end{figure}

In order to demonstrate how SUSY-AI outperforms a simple decision tree we make a comparison of both methods in Figure~\ref{fig:dtcomparisonroc}, which shows ROC curves for both of them. We study here $\mathcal{O}(20)$ simple decision trees (with a maximum depth of 5) that model a set of simple cuts one would put on the data set manually. The decision trees were trained with a train:test split dataset according to a ratio 75:25, cf.\ Appendix~\ref{app:splitting}. An example of a decision tree trained in this way can be seen in Figure~\ref{fig:exampledt}. For the purpose of this exercise, the cut of $0.5$ was imposed on the classifier output.  The square marker in the figure shows the actual location of SUSY-AI on the RF ROC curve. Clearly, for any choice of FPR SUSY-AI outperforms the simple decision trees. The difference is particularly visible once we take low FPR. Note that the SUSY-AI ROC curve plotted here does not take into account CL cuts discussed in the previous paragraph. Once this taken into account the advantage of the package increases even further.

%%%%%%%%%%%%%%%%%%%%%%%%%%%%%%%%%%%%%%%%%%%%%%%%%%%%%%%%

\section{Performance tests of SUSY-AI\label{sec:5}}
In this section we study the performance of SUSY-AI on a sample that was initially used for its training and on two specific SUSY models. The first one is a natural SUSY model that focuses on only several chosen parameters (with the rest effectively decoupled) of the 19-dimensional pMSSM parameter space, but fulfills almost all the constraints of the original sample. The second one is the constrained MSSM defined by high-scale parameters. It generally contains all particles from the pMSSM spectrum, but with the constraints from dark matter relic abundance and Higgs physics being relaxed. In the last subsection we discuss validation performance on two additional ad-hoc models and provide a general discussion of SUSY-AI validation and applicability for models not specified in this paper, which users could try on their own.

\subsection{Performance in the 19-dimensional pMSSM \label{sec:pmssm}}
To validate the performance of SUSY-AI, all possible two-dimensional projections of the 19-dimensional pMSSM parameter space have been searched for differences between the classification by SUSY-AI and ATLAS. Figures~\ref{fig:kwilts} and \ref{fig:kwilts_mass} show as an example the classification in the $M_1$--$M_2$ and $m_{\tilde{g}}$--$m_{\tilde{\chi}^0_1}$ plane. Various other classification plots are shown in Figures~\ref{fig:appplot01}--\ref{fig:appplot08} in Appendix~\ref{app:projections}.

One could expect that misclassified points are primarily located at a border of allowed and excluded regions of the 19-dimensional parameter space. To test this hypothesis, we bin every possible two-dimensional projections of parameter space and calculate the ratio of allowed points to the total number of points in each bin. We compare the true classification and out-of-bag prediction.\footnote{{A comparison of out-of-bag estimation and the validation via splitting of data in a training and testing set is discussed in Appendix~\ref{app:splitting}}.} The fraction of misclassified points can be plotted in the same manner and provides information on prediction errors in different parts of parameter space; see Figures~\ref{fig:kwilts} and~\ref{fig:kwilts_mass}.

The classification has been studied for different cases: including all points, including only points within the 95\% CL limit and for points within the 99\% CL limit, as shown in the different rows of Figures~\ref{fig:kwilts} and~\ref{fig:kwilts_mass}. As expected, the difference between the original classification and the predicted classification becomes smaller when demanding a higher confidence level. Figures~\ref{fig:appplot01}--\ref{fig:appplot08} in Appendix~\ref{app:projections} further support the hypothesis that the misclassified points indeed gather around decision boundaries in the 19-dimensional parameter space.

\afterpage{
%\begin{figure}[p]
%\afterpage{
    \begin{sidewaysfigure}
	\centering
	\begin{subfigure}[b]{0.03\textwidth}
	    \hspace{0.5cm}
	\end{subfigure}
	\hfill
	\begin{subfigure}[b]{0.22\textwidth}
	    \begin{center}
		Number of model points
	    \end{center}
	    \vspace{0.5cm}
	\end{subfigure}
	\hfill
	\begin{subfigure}[b]{0.22\textwidth}
	    \begin{center}
		True classification
	    \end{center}
	    \vspace{0.5cm}
	\end{subfigure}
	\hfill
	\begin{subfigure}[b]{0.22\textwidth}
	    \begin{center}
		Prediction by classifier
	    \end{center}
	    \vspace{0.5cm}
	\end{subfigure}
	\hfill
	\begin{subfigure}[b]{0.22\textwidth}
	    \begin{center}
		Ratio of misclassified points
	    \end{center}
	    \vspace{0.5cm}
	\end{subfigure}
	\hfill

	\begin{subfigure}[b]{0.03\textwidth}
      \rotatebox{90}{\hspace{2cm}All data}
      \end{subfigure}
	\hfill
	\begin{subfigure}[b]{0.23\textwidth}
	    \includegraphics[width=\textwidth]{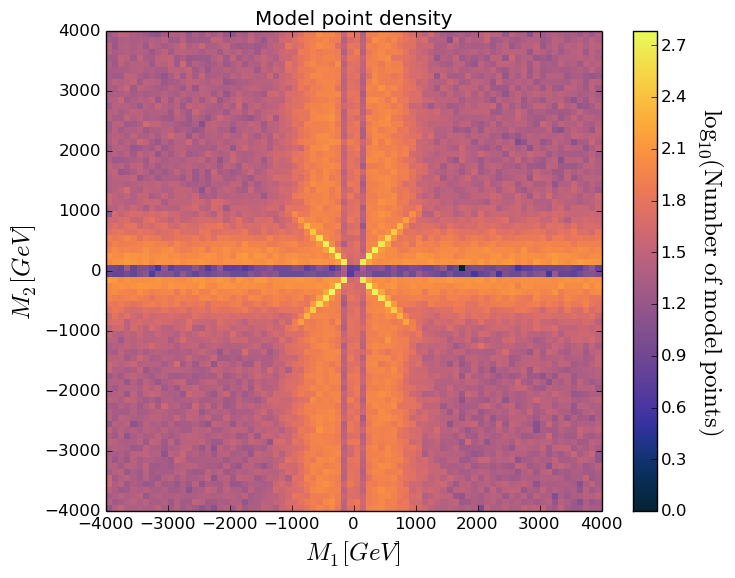}
	\end{subfigure}
	\hfill
	\begin{subfigure}[b]{0.23\textwidth}
	    \includegraphics[width=\textwidth]{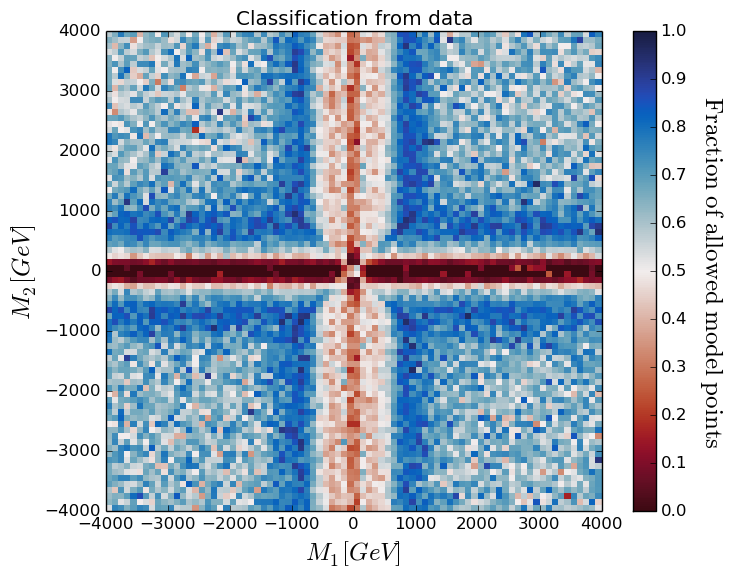}
	\end{subfigure}
	\hfill
	\begin{subfigure}[b]{0.23\textwidth}
	    \includegraphics[width=\textwidth]{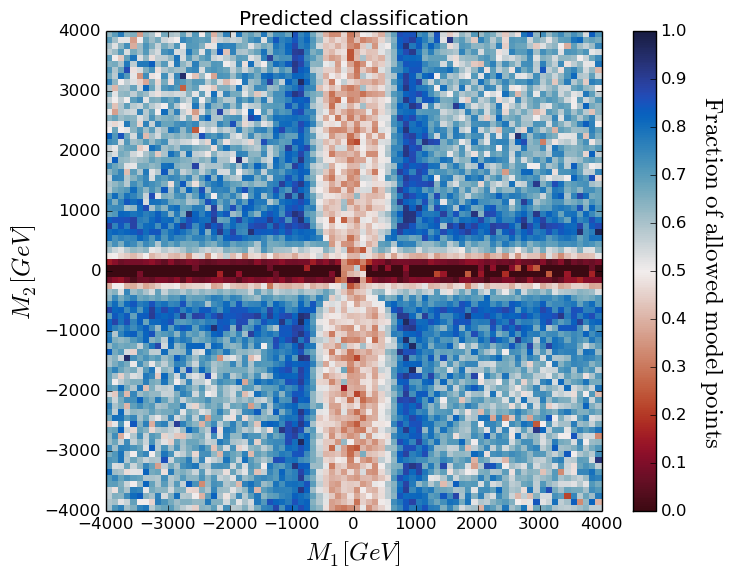}
	\end{subfigure}
	\hfill
	\begin{subfigure}[b]{0.23\textwidth}
	    \includegraphics[width=\textwidth]{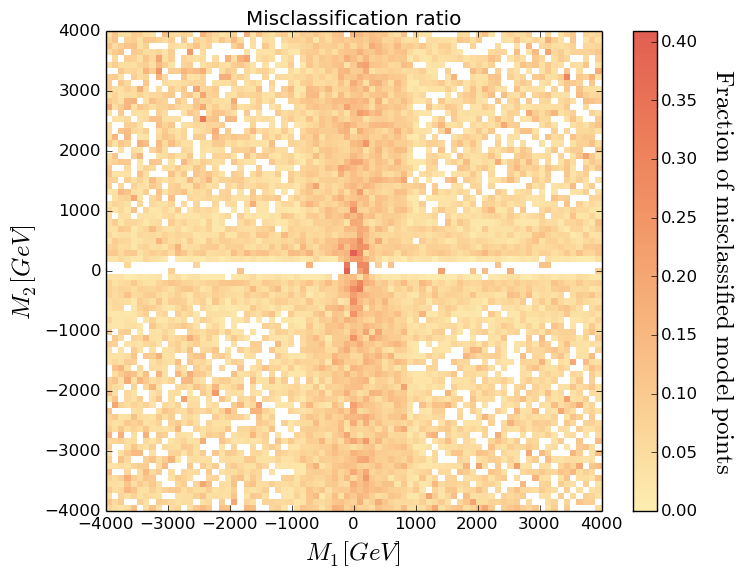}
	\end{subfigure}
	
	\begin{subfigure}[b]{0.03\textwidth}
	    \rotatebox{90}{\hspace{2.15cm}95\-CL}
	\end{subfigure}
	\hfill
	\begin{subfigure}[b]{0.23\textwidth}
	    \includegraphics[width=\textwidth]{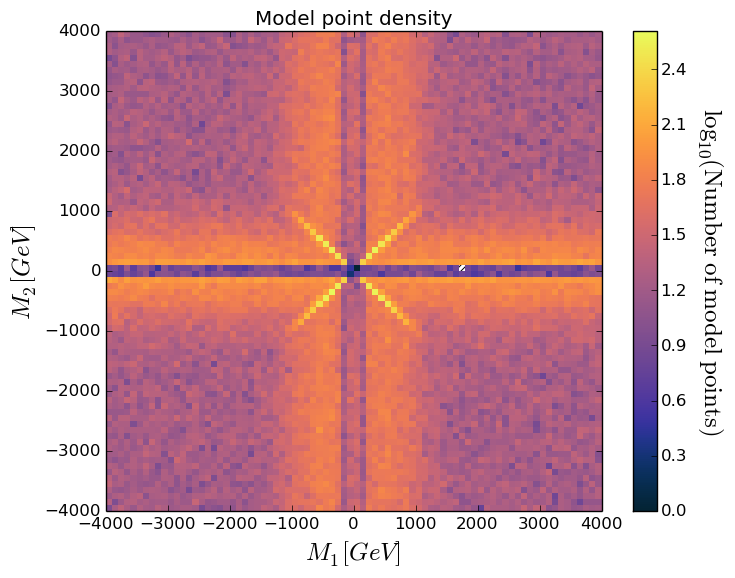}
	\end{subfigure}
	\hfill
	\begin{subfigure}[b]{0.23\textwidth}
	    \includegraphics[width=\textwidth]{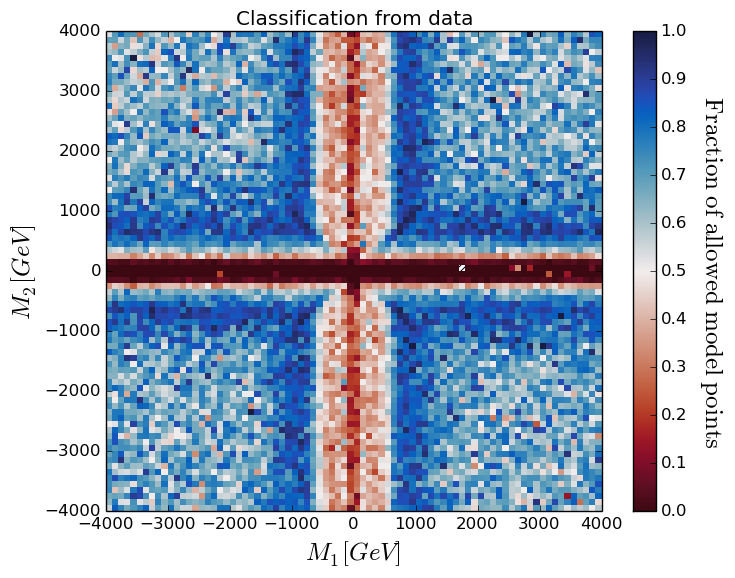}
	\end{subfigure}
	\hfill
	\begin{subfigure}[b]{0.23\textwidth}
	    \includegraphics[width=\textwidth]{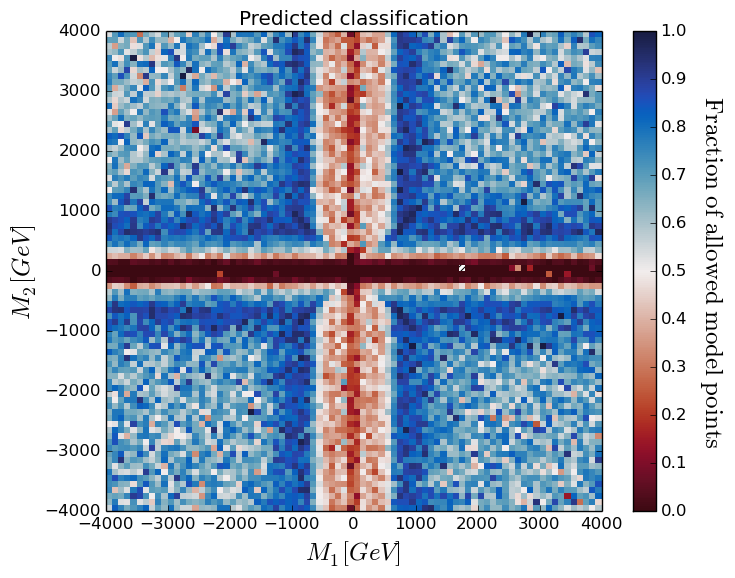}
	\end{subfigure}
	\hfill
	\begin{subfigure}[b]{0.23\textwidth}
	    \includegraphics[width=\textwidth]{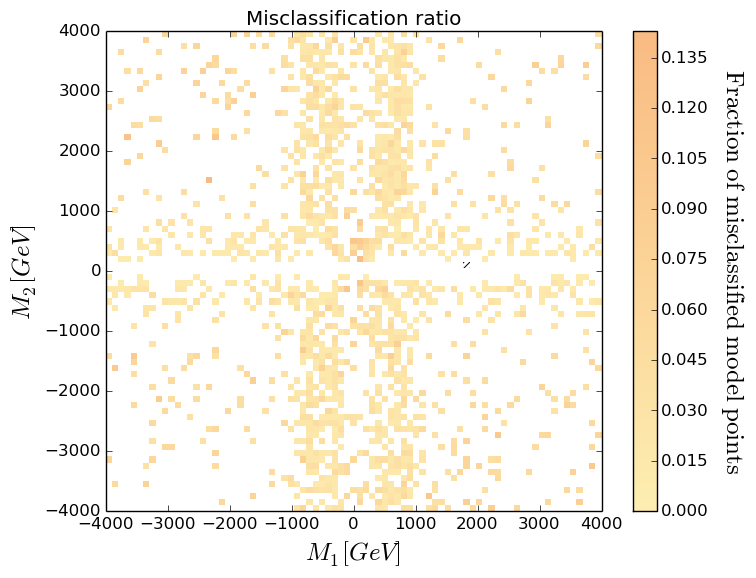}
	\end{subfigure}  
	
	\begin{subfigure}[b]{0.03\textwidth}
	    \rotatebox{90}{\hspace{2.15cm}99\-CL}
	\end{subfigure}
	\hfill
	\begin{subfigure}[b]{0.23\textwidth}
	    \includegraphics[width=\textwidth]{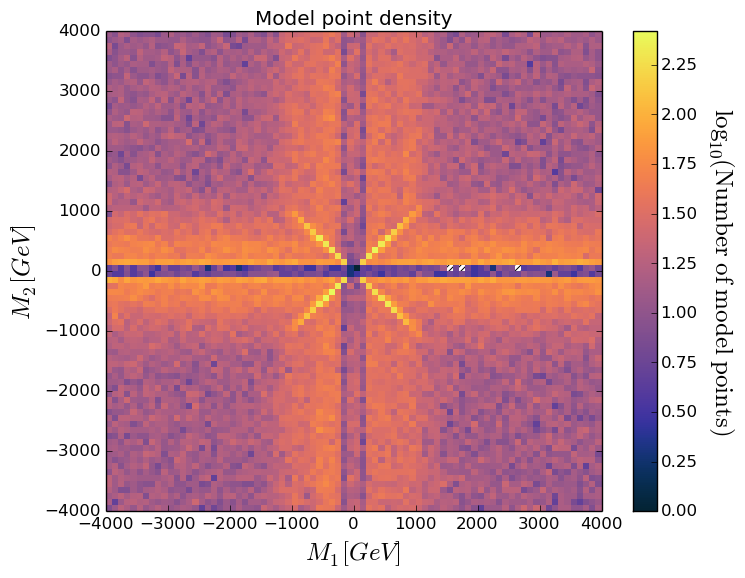}
	\end{subfigure}
	\hfill
	\begin{subfigure}[b]{0.23\textwidth}
	    \includegraphics[width=\textwidth]{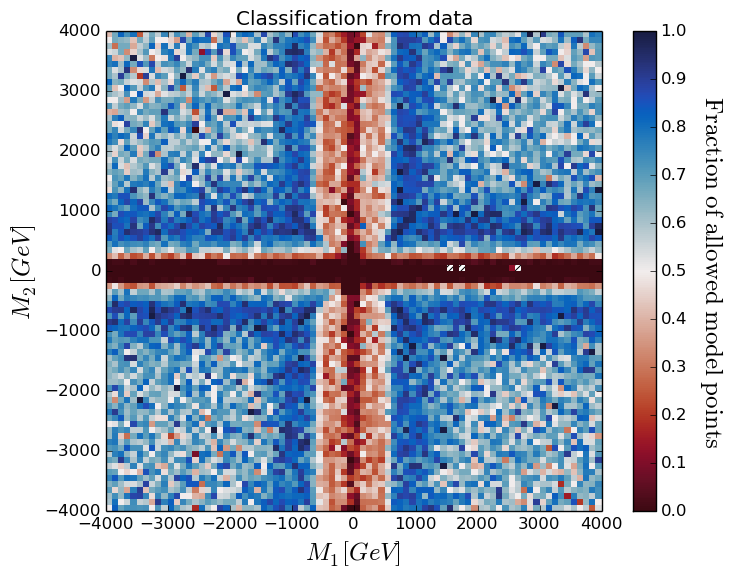}
	\end{subfigure}
	\hfill
	\begin{subfigure}[b]{0.23\textwidth}
	    \includegraphics[width=\textwidth]{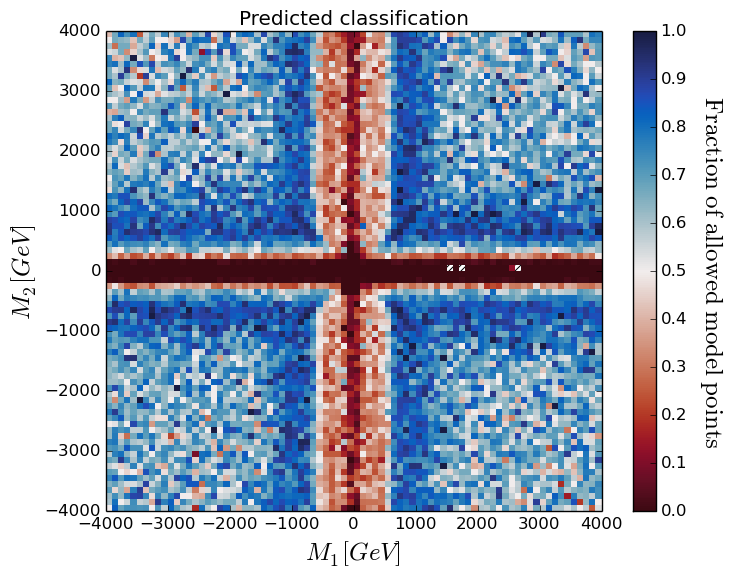}
	\end{subfigure}
	\hfill
	\begin{subfigure}[b]{0.23\textwidth}
	    \includegraphics[width=\textwidth]{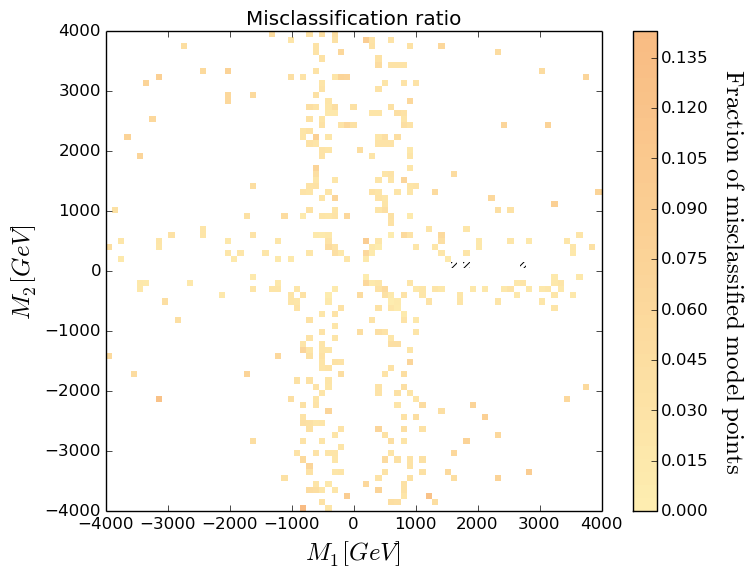}
	\end{subfigure}
	\caption{Color histograms for a projection of the 19-dimensional pMSSM parameter space on the $M_1$--$M_2$ plane. The color in the second and third column indicates the fraction of allowed data points for the true classification and the out-of-bag prediction, respectively. The last column shows the fraction of misclassified model points by the prediction, with white areas denoting no misclassifications.\label{fig:kwilts}}
    \end{sidewaysfigure}
%    \clearpage 
%} 
%\end{figure}
\clearpage}

\afterpage{
%\begin{figure}[p]
%\afterpage{
    \begin{sidewaysfigure}
	\centering
	\begin{subfigure}[b]{0.03\textwidth}
	    \hspace{0.5cm}
	\end{subfigure}
	\hfill
	\begin{subfigure}[b]{0.22\textwidth}
	    \begin{center}
		Number of model points
	    \end{center}
	    \vspace{0.5cm}
	\end{subfigure}
	\hfill
	\begin{subfigure}[b]{0.22\textwidth}
	    \begin{center}
		True classification
	    \end{center}
	    \vspace{0.5cm}
	\end{subfigure}
	\hfill
	\begin{subfigure}[b]{0.22\textwidth}
	    \begin{center}
		Prediction by classifier
	    \end{center}
	    \vspace{0.5cm}
	\end{subfigure}
	\hfill
	\begin{subfigure}[b]{0.22\textwidth}
	    \begin{center}
		Ratio of misclassified points
	    \end{center}
	    \vspace{0.5cm}
	\end{subfigure}
	\hfill

	\begin{subfigure}[b]{0.03\textwidth}
      \rotatebox{90}{\hspace{2cm}All data}
      \end{subfigure}
	\hfill
	\begin{subfigure}[b]{0.23\textwidth}
	    \includegraphics[width=\textwidth]{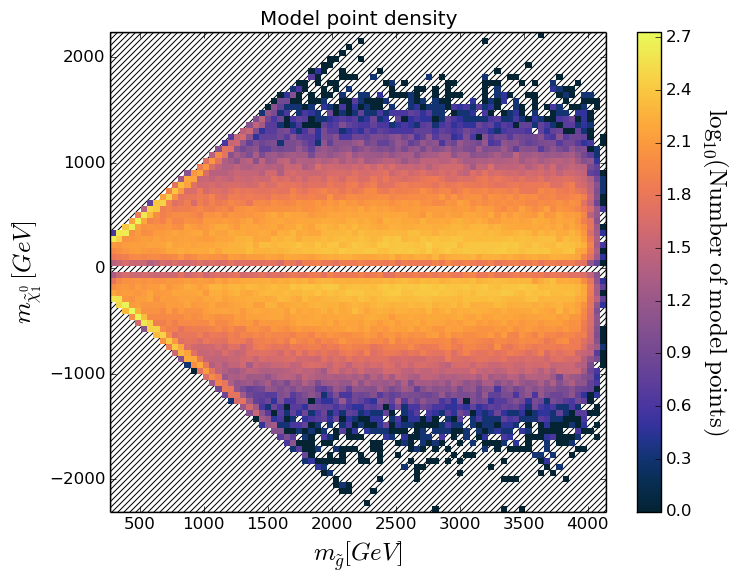}
	\end{subfigure}
	\hfill
	\begin{subfigure}[b]{0.23\textwidth}
	    \includegraphics[width=\textwidth]{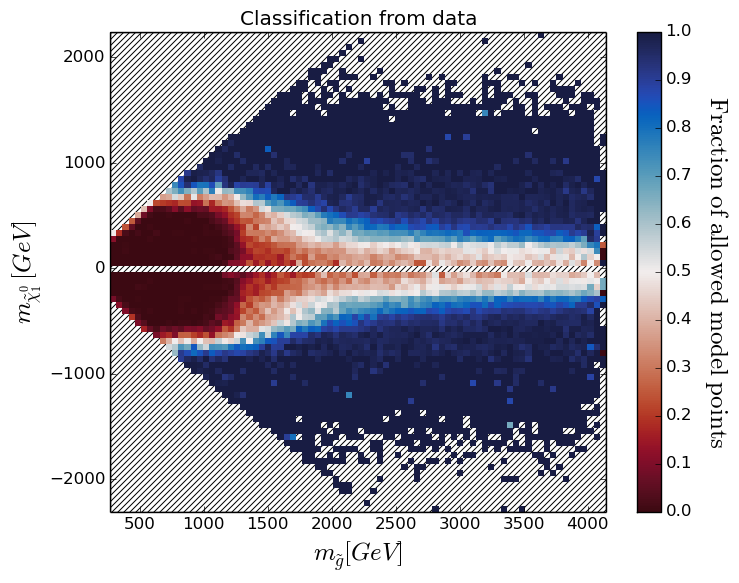}
	\end{subfigure}
	\hfill
	\begin{subfigure}[b]{0.23\textwidth}
	    \includegraphics[width=\textwidth]{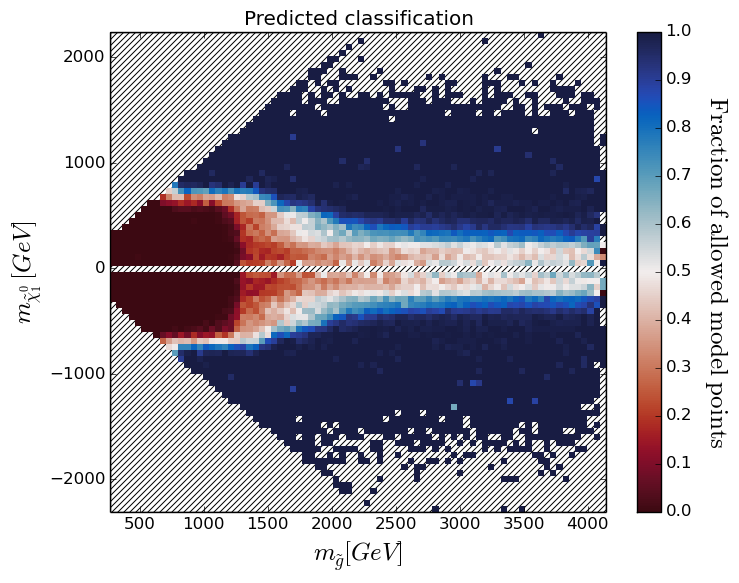}
	\end{subfigure}
	\hfill
	\begin{subfigure}[b]{0.23\textwidth}
	    \includegraphics[width=\textwidth]{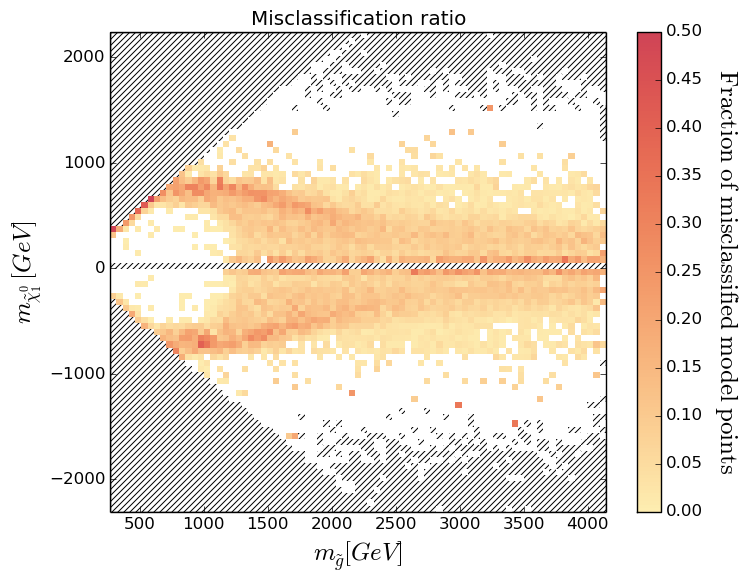}
	\end{subfigure}
	
	\begin{subfigure}[b]{0.03\textwidth}
	    \rotatebox{90}{\hspace{2.15cm}95\-CL}
	\end{subfigure}
	\hfill
	\begin{subfigure}[b]{0.23\textwidth}
	    \includegraphics[width=\textwidth]{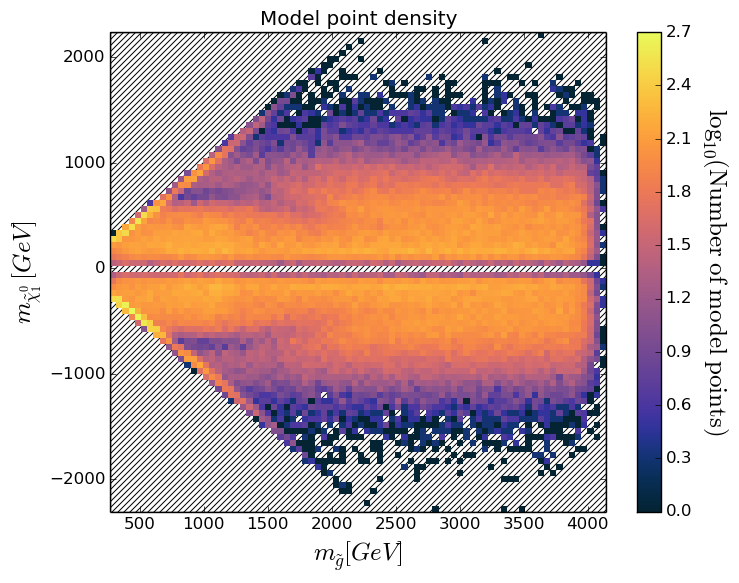}
	\end{subfigure}
	\hfill
	\begin{subfigure}[b]{0.23\textwidth}
	    \includegraphics[width=\textwidth]{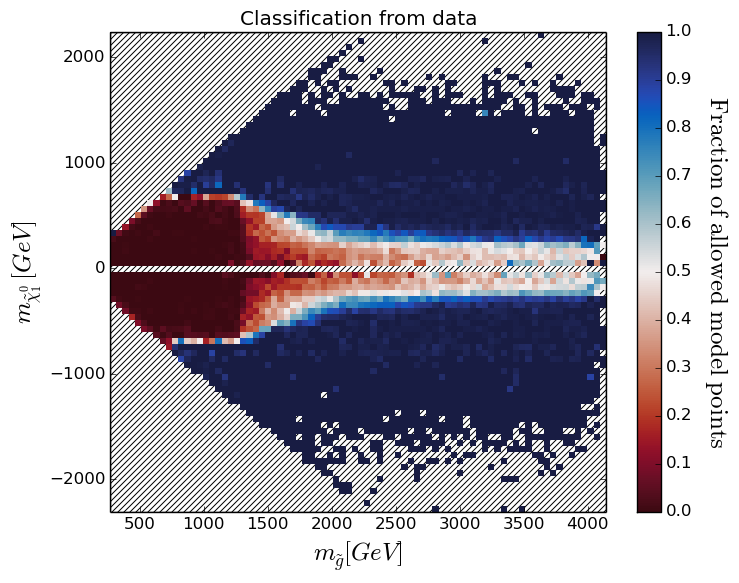}
	\end{subfigure}
	\hfill
	\begin{subfigure}[b]{0.23\textwidth}
	    \includegraphics[width=\textwidth]{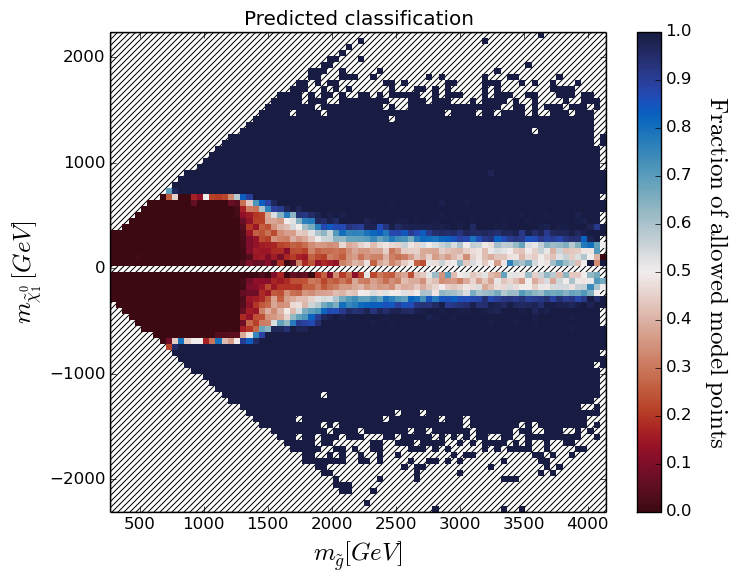}
	\end{subfigure}
	\hfill
	\begin{subfigure}[b]{0.23\textwidth}
	    \includegraphics[width=\textwidth]{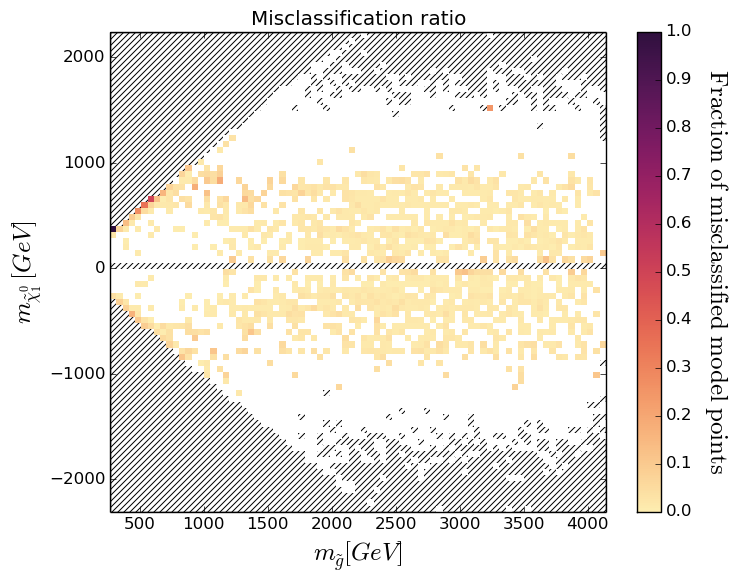}
	\end{subfigure}  
	
	\begin{subfigure}[b]{0.03\textwidth}
	    \rotatebox{90}{\hspace{2.15cm}99\-CL}
	\end{subfigure}
	\hfill
	\begin{subfigure}[b]{0.23\textwidth}
	    \includegraphics[width=\textwidth]{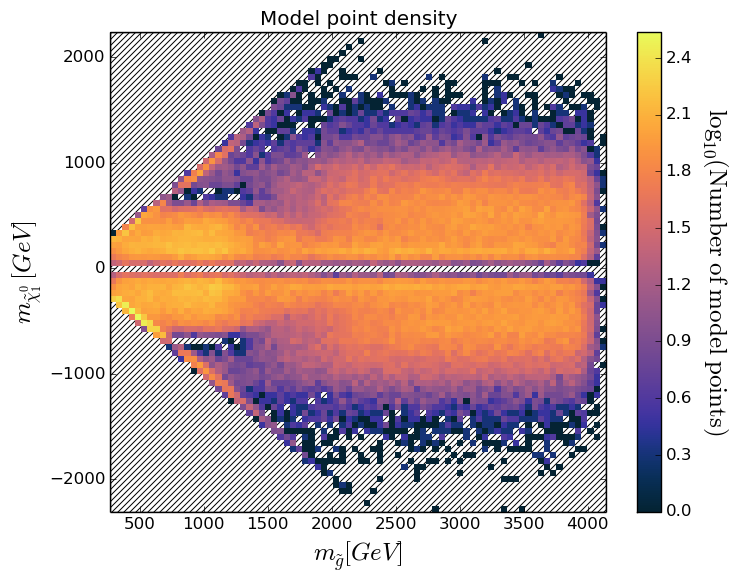}
	\end{subfigure}
	\hfill
	\begin{subfigure}[b]{0.23\textwidth}
	    \includegraphics[width=\textwidth]{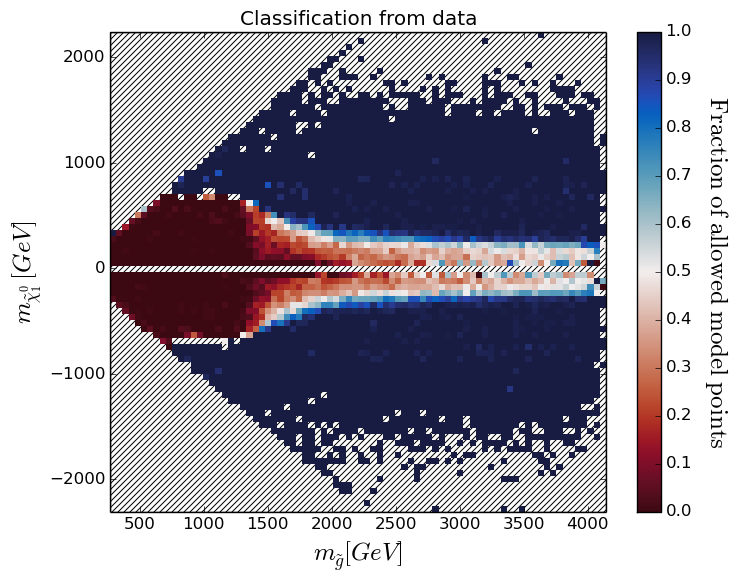}
	\end{subfigure}
	\hfill
	\begin{subfigure}[b]{0.23\textwidth}
	    \includegraphics[width=\textwidth]{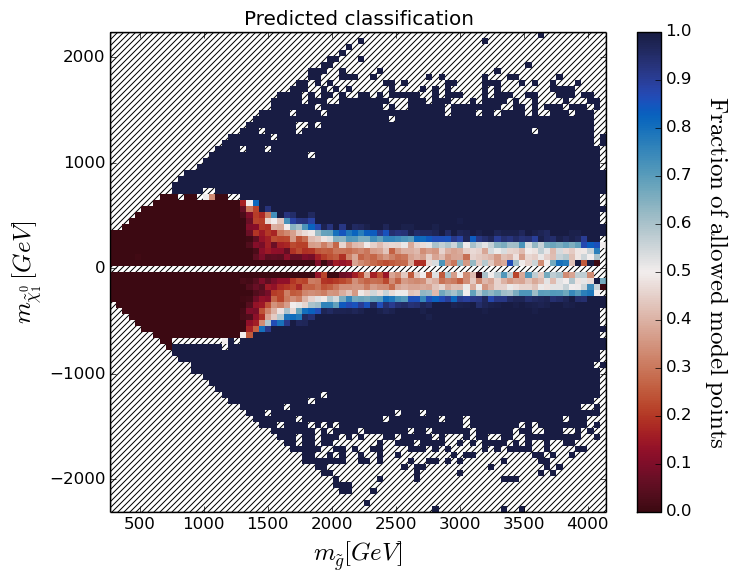}
	\end{subfigure}
	\hfill
	\begin{subfigure}[b]{0.23\textwidth}
	    \includegraphics[width=\textwidth]{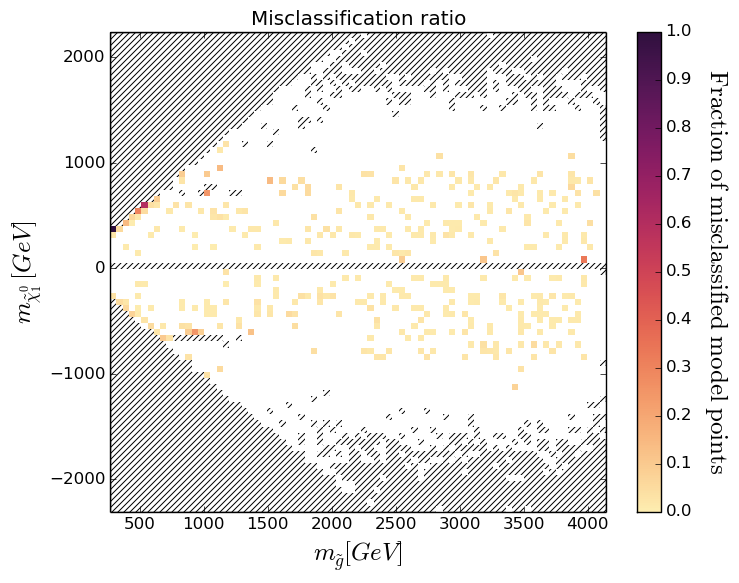}
	\end{subfigure}
	\caption{Color histograms for a projection of the 19-dimensional pMSSM parameter space on the $m_{\tilde{g}}$--$m_{\tilde{\chi}^0_1}$ plane. The color in the second and third column indicates the fraction of allowed data points for the true classification and the out-of-bag prediction, respectively. The last column shows the fraction of misclassified model points, with white areas denoting no misclassifications. The dashed bins contain no data points.
	\label{fig:kwilts_mass}}
    \end{sidewaysfigure}
%    \clearpage
%}
%\end{figure}
\clearpage
}

Without a confidence level cut, SUSY-AI classifies 93.2\% of the data correctly at the working point with the classifier output cut of $0.535$. This can be compared with the performance of the simple decision tree in Figure~\ref{fig:dtcomparisonroc}, which is markedly worse for any value of the false positive rate. Comparing SUSY-AI at the working point, $\mathrm{FPR}_\textrm{SUSY-AI}=0.112$, $\mathrm{TPR}_\textrm{SUSY-AI}=0.960$, with the decision tree at the same FPR we obtain just $\mathrm{TPR}_\mathrm{DT}=0.647$. Alternatively, looking at true and false negatives we have: $\mathrm{FNR}_\textrm{SUSY-AI}=0.089$, $\mathrm{TNR}_\textrm{SUSY-AI}=0.947$, while the decision tree at the same FNR we obtain just $\mathrm{TNR}_\mathrm{DT}=0.670$, and at the same TNR: $\mathrm{TNR}_\mathrm{DT}=0.660$. The last result is particularly worth noting as it means that for a decision tree that correctly excludes 95\% of points the rate of \textit{incorrect} exclusions is at 66\%.
With the confidence level cut of 95\%, corresponding to 70\% of the full dataset, the correct classification by the RF is increased to 99.0\%. Finally the confidence level cut of 99\%, corresponding to 50\% of the full dataset, yields a correct classification of 99.7\%. 

One might think that the volume of the 19-dimensional parameter space is so large that the data points become too sparse to make reliable classification possible. Our results show otherwise. The first reason for this is that the size of the parameter space in the pMSSM is inherently reduced by sampling restrictions. Of the $500\cdot10^6$ model points sampled, only $3\cdot10^5$ survived restrictions on the Higgs mass, non-existence of tachyons and color breaking minima, correct electroweak symmetry breaking etc. This decreases the number of points needed for training since the volume of the valid and relevant parameter space is reduced. Moreover, only the part of parameter space with a complicated decision boundary shape ($< 4$~TeV) has to be sampled. Furthermore, we want to stress that the DM constraint already excludes benchmark points with an LSP different from the lightest neutralino and thus a non-trivial cut into parameter space is performed. Moreover, the majority of the bino-like LSP points are concentrated with masses below 100~GeV and in particular at the $Z$ and Higgs boson pole, have a higgsino or wino NLSP for co-annihilation or have colored scalars (usually stops) or staus as the NLSP candidate. Usually, benchmark points with a wino and higgsino LSP are constrained to masses below 1.5~TeV. 

Secondly, the experimental constraints on the pMSSM significantly reduce the allowed SUSY parameter space. Only less than 1 out of 100 randomly sampled SUSY parameter points were selected after the constraints applied by ATLAS. The $300\,000$ training points, therefore, represent a much larger set of randomly selected parameters. The classifier remains valid nevertheless, since one only needs to sample the part of parameter space where the decision boundary shape will change as a function of a particular feature `X'. This happens in the low-energy range therefore justifies the upper cut of $< 4$~TeV. Another relevant issue here is the coverage of the compressed spectrum region where one might expect poor performance. The ATLAS scan, however, covers fairly well compressed spectra and provides training data also in these regions.

The final reason is that not all 19 dimensions of the pMSSM are phenomenologically relevant. For example, the production of gluinos and squarks, which is the main search channel at the LHC, depends mainly on the squark masses, the gluino mass, and the electroweakino mass parameters $M_1$, $M_2$ and $\mu$, while the trilinear couplings and $\tan \beta$ only have a small impact on the predictions.

This can be exposed by investigating features' importance. Every node in a decision tree is a condition on a single feature and splits the dataset into two parts. The locally optimal condition is chosen based on a measure called \textit{impurity}. In our case, we implement the \textit{Gini impurity} which is given by
\begin{equation}
I=\sum_{i=1}^{C} f_i \cdot (1-f_i) = 1-\sum_{i=1}^{C}f_i^2,
\end{equation}
where $C$ is the total number of classes and $f_i$ the fraction of class $i$ in this node. The smaller the Gini impurity, the purer the dataset at the given node. Minimizing this value
during training guarantees that model points will be split according to their class label. 

After training a tree, it can be computed for that tree how much each feature $j$ decreases the weighted impurity: the impurity change weighted with the fraction of model points it influences, summed over all nodes making a split on feature $j$ in that tree:
\begin{equation}
\sum_{k \in \textrm{nodes splitting }j}^{} \frac{\textrm{model points at node }k}{\textrm{total number of model points}} \cdot \textrm{impurity change}.
\end{equation}
This weighted impurity change for each feature can be averaged for the forest and the features can be ranked according to this measure. The result of this exercise for SUSY-AI is shown in Table~\ref{tab:featureimportances}, where the features' importance are listed.
% http://blog.datadive.net/selecting-good-features-part-iii-random-forests/
 
\begin{table}
\centering
\begin{tabular}{llcll}
\cline{1-2}
\cline{4-5}
{Parameter} & {Importance} & & {Parameter} & {Importance} \\
\cline{1-2}
\cline{4-5}
%\texttt{mL1} & 0.020722329872 & \hspace{0.75cm} & \texttt{M1} & 0.0576284287751 \\
\texttt{mL1} & 0.021 & \hspace{0.75cm} & \texttt{M1} & 0.058 \\
%\texttt{me1} & 0.0188857795865 & & \texttt{M2} & 0.164160591853 \\
\texttt{me1} & 0.019 & & \texttt{M2} & 0.164 \\
%\texttt{mL3} & 0.0138133304874 & & \texttt{mu} & 0.130095649773 \\
\texttt{mL3} & 0.014 & & \texttt{mu} & 0.130 \\
%\texttt{me3} & 0.0143471537963 & & \texttt{M3} & 0.241522918816 \\
\texttt{me3} & 0.014 & & \texttt{M3} & 0.242 \\
\cline{1-2}
\cline{4-5}
%\texttt{mQ1} & 0.0791210937176 & & \texttt{At} & 0.0132556452076 \\
\texttt{mQ1} & 0.079 & & \texttt{At} & 0.013 \\
%\texttt{mu1} & 0.0662304185267 & & \texttt{Ab} & 0.0116453579659 \\
\texttt{mu1} & 0.066 & & \texttt{Ab} & 0.012 \\
%\texttt{md1} & 0.036920220724 & & \texttt{Atau} & 0.0117438282989 \\
\texttt{md1} & 0.037 & & \texttt{Atau} & 0.012 \\
\cline{4-5}
%\texttt{mQ3} & 0.0261421684941 & & \texttt{mA2} & 0.0307053005307 \\
\texttt{mQ3} & 0.026 & & \texttt{mA2} & 0.031 \\
%\texttt{mu3} & 0.0181801062761 & & \texttt{tanbeta} & 0.0191273137473 \\
\texttt{mu3} & 0.018 & & \texttt{tanbeta} & 0.019 \\
\cline{4-5}
%\texttt{md3} & 0.0257523635519 & & & \\
\texttt{md3} & 0.026 & & & \\
\cline{1-2}
\end{tabular}
\caption{Features' importance for the trained RF classifier.
\label{tab:featureimportances} }
\end{table} 
 
One can see that a subset of all features have a significantly higher contribution to the final prediction by SUSY-AI. We investigated a reduction of the number of features taken as an input by SUSY-AI and the reduction yielded classifiers with systematically lower quality.

From the above discussion one can see that the effective dimensionality of the problem is significantly reduced. With this in mind let us make several remarks about the uncertainty of a decision boundary. Firstly, we note that it does not scale with $1/N$, where $N$ is the number of points in each dimension. The error actually scales with $0.5\cdot V/(N+1)$ where $V$ is the allowed volume. Let us assume $V=1$ and four points in the unit box placed at $0.2$, $0.4$, $0.6$, $0.8$ (a grid spacing) and that the first two points are excluded. The algorithm would guess the limit to be at $0.5$ (i.e.\ between points 2 and 3). The uncertainty of this guess is only $0.1$ and not $0.25$ even for a grid spacing of points. In a 9-dimensional space this means $4^9 = 262144$ points. Taking into account that some of the features are relatively unimportant and with the constraints on the parameter space that reduce the effective volume (e.g.\ the physical vacuum, a viable DM candidate, etc.), it becomes plausible that our sample provides sufficient coverage of the parameter space.

In addition, the separation of the excluded and allowed regions close to the decision boundary becomes better defined when
applying confidence level cuts that remove
model points not classified with a high enough certainty. We note that in the future the probability for correct classification by SUSY-AI will be improved with more training data.

\subsection{Performance in a pMSSM submodel: the 6-dimensional natural SUSY model \label{sec:nsusy}}
In order to further test the performance of the trained classifier, a cross-check has been performed on two models: the 6-dimensional natural SUSY~\cite{Drees:2015aeo} and the 5-dimensional CMSSM. The natural SUSY sample is contained within the limits specified in Table~\ref{tab:softbreaking_pmssm}, however, one might worry that this specific part of the parameter space was too sparsely sampled. We show here that nevertheless the prediction of SUSY-AI is reliable. On the other hand, the CMSSM sample relaxes some of the constraints of the training sample, like the Higgs mass or dark matter abundance. Still, we demonstrate the exclusion limits are correctly reproduced.

In Ref.~\cite{Drees:2015aeo} limits were presented on the parameter space of the natural supersymmetry based on Run~1 SUSY searches. The authors considered 22000 model points in a 6-dimensional parameter space listed in Table~\ref{tab:softbreaking}. The mass spectra consist of higgsinos as the lightest supersymmetric particle, as well as light left-handed stops and sbottoms, right-handed stops and gluinos, while assuming a SM-like Higgs boson. All remaining supersymmetric particles and supersymmetric Higgs bosons were decoupled. All benchmark scenarios have to satisfy low-energy limits such as the $\rho$ parameter \cite{Drees:1990dx}, LEP2 constraints~\cite{Abbiendi:2002vz,Abdallah:2003xe,Abbiendi:2003sc} and have to be consistent with the measured dark matter relic density~\cite{Ade:2015xua}, i.e.\ the total cold dark matter energy density is used as an upper limit on the LSP abundance. However, no constraints from $b$-physics experiments have been imposed. In summary, our natural SUSY sample fulfills the ATLAS pMSSM constraints, except the $b$-physics limits.

\begin{table*}[t]
\begin{center}
\begin{tabular}{lll}
\hline
Parameter & Description & Scanned Range \\
\hline
$m_{\tilde Q_3}$& \small{3$^{\rm rd}$ generation $SU(2)$ doublet soft
breaking squark mass} & $[0.1 \ {\rm TeV},\, 1.5 \ {\rm TeV}]$\\
$m_{\tilde U_3}$ & \small{3$^{\rm rd}$ generation $SU(2)$ singlet soft
breaking squark mass} & $[0.1 \ {\rm TeV},\, 1.5 \ {\rm TeV}]$\\
$M_3$ & \small{Gluino mass parameter} & $[0.1 \ {\rm TeV},\, 3.0 \ {\rm TeV}]$\\
$A_{t}$ & \small{Stop trilinear coupling} & $[-3.0 \ {\rm TeV},\, 3.0 \ {\rm TeV}]$\\
$\mu$ & \small{Higgsino mass parameter} & $[0.1 \ {\rm TeV},\, 0.5\ {\rm TeV}]$\\
$\tan\beta$ & \small{Ratio of vacuum expectation values} & $[1, \, 20]$ \\
\hline
\end{tabular}
\end{center}
\caption{Input parameters of the natural SUSY scenario of Ref.~\cite{Drees:2015aeo}, and
  the range over which these parameters were scanned.} 
 \label{tab:softbreaking} 
\end{table*}

The event generation was performed with {\tt Pythia~8.185} \cite{Sjostrand:2014zea} as well as with {\tt Madgraph} \cite{Alwall:2014hca} interfaced with the shower generator {\tt Pythia~6.4} \cite{Sjostrand:2006za} for matched event samples. The truth level MC events were then passed over to {\tt CheckMATE} \cite{Drees:2013wra,Kim:2015wza}. It consists of a simulation of the detector response with a modified {\tt Delphes} \cite{deFavereau:2013fsa} where the settings have been re-tuned to resemble the responses of the ATLAS detector. 

\begin{table}
\begin{center}
\begin{tabular}{llll}\hline
Reference & Final State & $\mathcal{L}$ [fb$^{-1}$] & \#SR\\
\hline
1308.2631 (ATLAS) \cite{Aad:2013ija} & 0$\ell$ + 2$b$-jets + \met & 20.1 & 6 \\
1403.4853 (ATLAS) \cite{Aad:2014qaa} & 2$\ell$ + \met & 20.3 & 12 \\ 
1404.2500 (ATLAS) \cite{Aad:2014pda} & SS 2$\ell$ or 3$\ell$ & 20.3 & 5 \\
1407.0583 (ATLAS) \cite{Aad:2014kra} & 1$\ell$ + ($b$)-jets + \met & 20.0 & 27\\
1407.0608 (ATLAS) \cite{Aad:2014nra} & monojet + \met & 20.3 & 3\\
1303.2985 (CMS) \cite{Chatrchyan:2013mys} & $\alpha_{T}$ + $b$-jets & 11.7& 59\\
ATLAS-CONF-2012-104 \cite{ATLAS-CONF-2012-104} & 1$\ell$ + $\geq$4 jets + \met & 5.8 & 2\\ 
ATLAS-CONF-2013-024 \cite{ATLAS-CONF-2013-024} & 0$\ell$ + 6 (2$b$)-jets + \met & 20.5 & 3\\ 
ATLAS-CONF-2013-047 \cite{ATLAS-CONF-2013-047} & 0$\ell$ + 2--6 jets+\met & 20.3 & 10\\ 
ATLAS-CONF-2013-061 \cite{ATLAS-CONF-2013-061} & 0--1$\ell$ + $\geq$3$b$-jets + \met & 20.1& 9 \\ 
ATLAS-CONF-2013-062 \cite{ATLAS-CONF-2013-062} & 1--2$\ell$ + 3--6 jets + \met & 20.0 & 19 \\ 
CMS-SUS-13-016 \cite{CMS-PAS-SUS13-016} & OS 2$\ell$ + $\geq$3$b$-jets & 19.7 & 1\\\hline
\end{tabular}
\end{center}
\caption{The experimental analyses used in Ref.~\cite{Drees:2015aeo}. The ATLAS-CONF and CMS-SUS papers are only available as conference proceedings, the others are given by their arXiv number. The middle column corresponds to the final state of the respective search, and the third column shows the total integrated luminosity employed in this analysis. The fourth column gives the total number of signal regions.}
\label{tab:lhc_nsusy_searches} 
\end{table}

Each model point was tested against a number of natural SUSY and inclusive SUSY searches with a total number of 156 signal regions, including two CMS searches, summarized in Table~\ref{tab:lhc_nsusy_searches}. {\tt CheckMATE} determines the signal region with the highest expected sensitivity, as well as the selection efficiency for this particular signal region. Finally, {\tt CheckMATE}  determines if the model point is excluded at the $95\%$ CL, using the CL$_S$ method~\cite{Read:2002hq} by evaluating the ratio,
\begin{equation} \label{r}
r \equiv \frac{S-1.96\cdot\Delta S} {S_{\rm exp.}^{95}}\,, 
\end{equation}
where $S$ is the number of signal events, $\Delta S$ denotes its theoretical uncertainty, and $S_{\rm exp.}^{95}$ is the experimentally determined 95$\%$ confidence level limit on the signal. A statistical error due to the finite MC sample, i.e.\ $\Delta S = \sqrt{S}$, as well as a 10$\%$ systematic error has been assumed. The parameter $r$ is only computed for the best expected signal region, in order to avoid exclusions due to downward fluctuations of the experimental data, which is expected considering the large number of signal regions. {\tt CheckMATE} does
not statistically combine signal regions nor combine different analyses. It considers a parameter point
to be excluded at 95$\%$ CL if $r$ defined in Eq.~(\ref{r})
exceeds $1.0$. However, the authors followed a more conservative approach.
If the $r$ value was below $0.67$ the point was considered allowed; if it was above $1.5$ it was excluded. All other points were removed from the analysis. 

\begin{figure}[t]
    \centering
    \begin{subfigure}[b]{0.49\textwidth}
        \includegraphics[width=\textwidth]{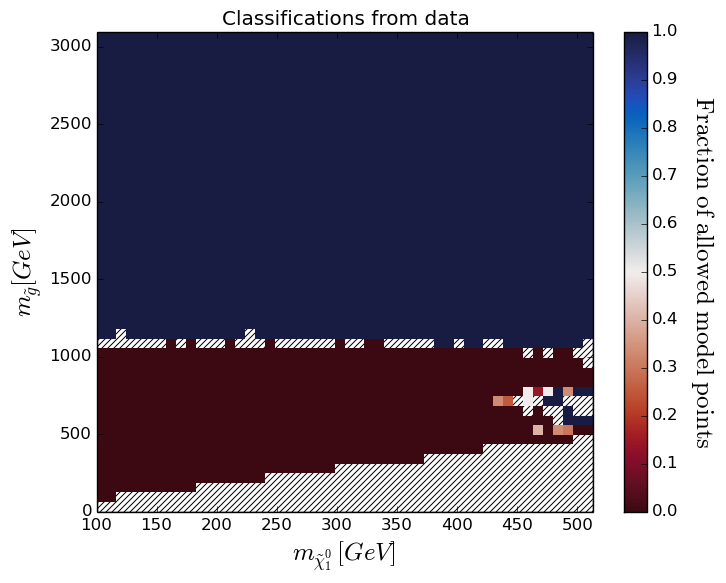}
	\caption{True classification from Ref.~\cite{Drees:2015aeo}.\label{fig:nsusy_clas}}
    \end{subfigure}
    \hfill
    \begin{subfigure}[b]{0.49\textwidth}
        \includegraphics[width=\textwidth]{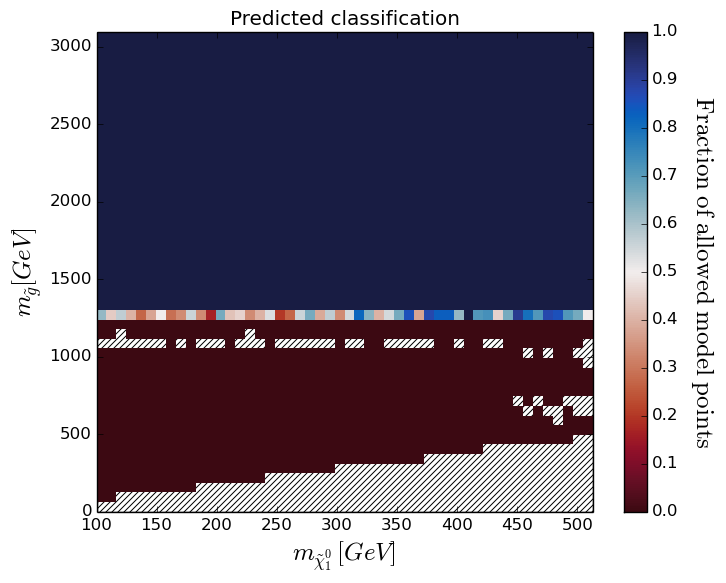}
	\caption{95$\%$ CL\ SUSY-AI classification result.\label{fig:nsusy_pred}}
    \end{subfigure}
    \hfill
    \begin{subfigure}[b]{0.49\textwidth}
        \includegraphics[width=\textwidth]{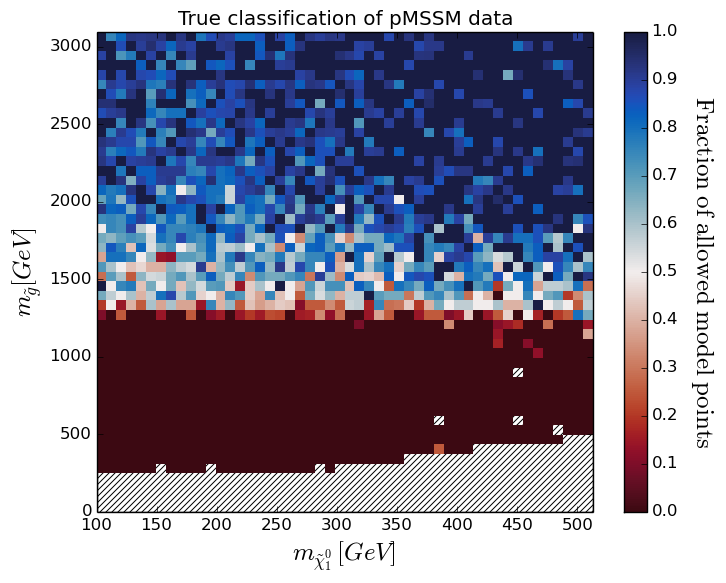}
	\caption{pMSSM data used for training.\label{fig:nsusy_pmssm}}
    \end{subfigure}
    \caption{Results of testing a natural SUSY scenario with the trained classifier in the higgsino LSP and gluino mass plane assuming stop masses larger than 600~GeV. The colors
    indicate the probability that a particular point is not excluded. For reference panel (c) shows the training data
    in the same plane as panels (a) and (b) after applying a constraint on the stop mass to filter out data points mimicking natural SUSY. The dashed bins contain no data points. The dashed stripe  in panels (a) and (b) corresponds
    to the points that were outside the 95$\%$ CL boundaries of \texttt{CheckMATE}, see text. }
\end{figure}

Figure~\ref{fig:nsusy_clas} shows the exclusion limit in the gluino--LSP-mass plane assuming $m_{\tilde t_1}\ge600$~GeV from Ref.~\cite{Drees:2015aeo}. 
Here, the red (blue) shaded areas indicate excluded (allowed) regions of parameter space, while the fraction of allowed points is shown by the color intensity according to a color bar. The limit is
essentially driven by the production of gluino pairs. Hence, a clear separation between the allowed and
excluded regions can be observed in the figure.

\begin{figure}[t]
    \centering
    \includegraphics[width=0.6\textwidth]{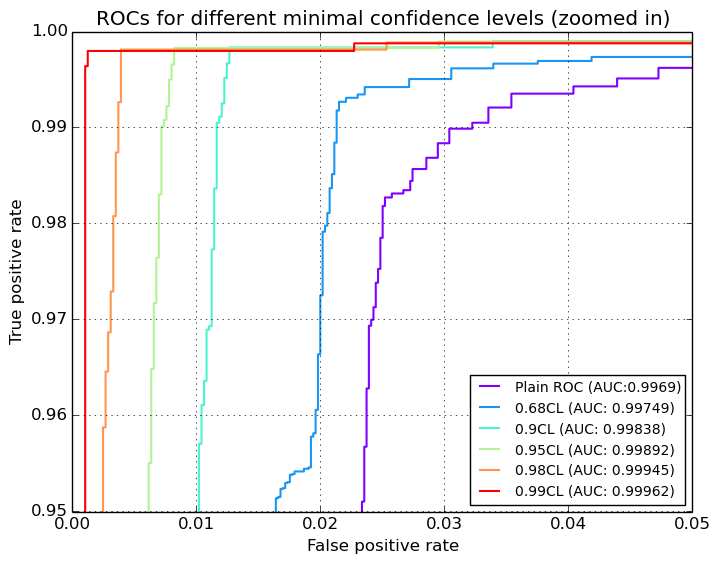}
    \caption{{Several ROCs for the pMSSM-trained classifier with varying CL cuts when tested on the natural SUSY sample.}}
    \label{fig:nsusy_rocs}
\end{figure}

Figure~\ref{fig:nsusy_pred} shows the result from the prediction with the 95\% CL cut. The classification by SUSY-AI reproduces the results from Ref.~\cite{Drees:2015aeo} very well. It produces slightly better limits, since the classifier was trained using more recent searches and a larger number of analyses, showing that the procedure of \texttt{CheckMATE} is conservative. Figure~\ref{fig:nsusy_pmssm} shows the same plot using the pMSSM training data and its true classification. This confirms that the location of the decision boundary in Figure~\ref{fig:nsusy_pred} is indeed learned from the training data and not an artifact of the natural SUSY data sample.

Again a series of ROC curves were generated. These are plotted in Figure \ref{fig:nsusy_rocs}. A large part of the model points is classified correctly; however, there remains a small number of false negatives (assuming the \texttt{CheckMATE} classification to be correct). This can be deduced from the spacing between the $\mathrm{TPR} = 1.0$ line and the ROC curves in Figure~\ref{fig:nsusy_rocs}. Nevertheless, the pMSSM-trained classifier provides a reliable classification, especially when a confidence level cut is applied, resulting in AUC of about $0.997$ for the full data set.

\subsection{Performance in a pMSSM submodel: the 5-dimensional constrained MSSM \label{sec:cmssm}}
\begin{figure}
    \centering
    \begin{subfigure}[b]{0.475\textwidth}
        \includegraphics[width=\textwidth]{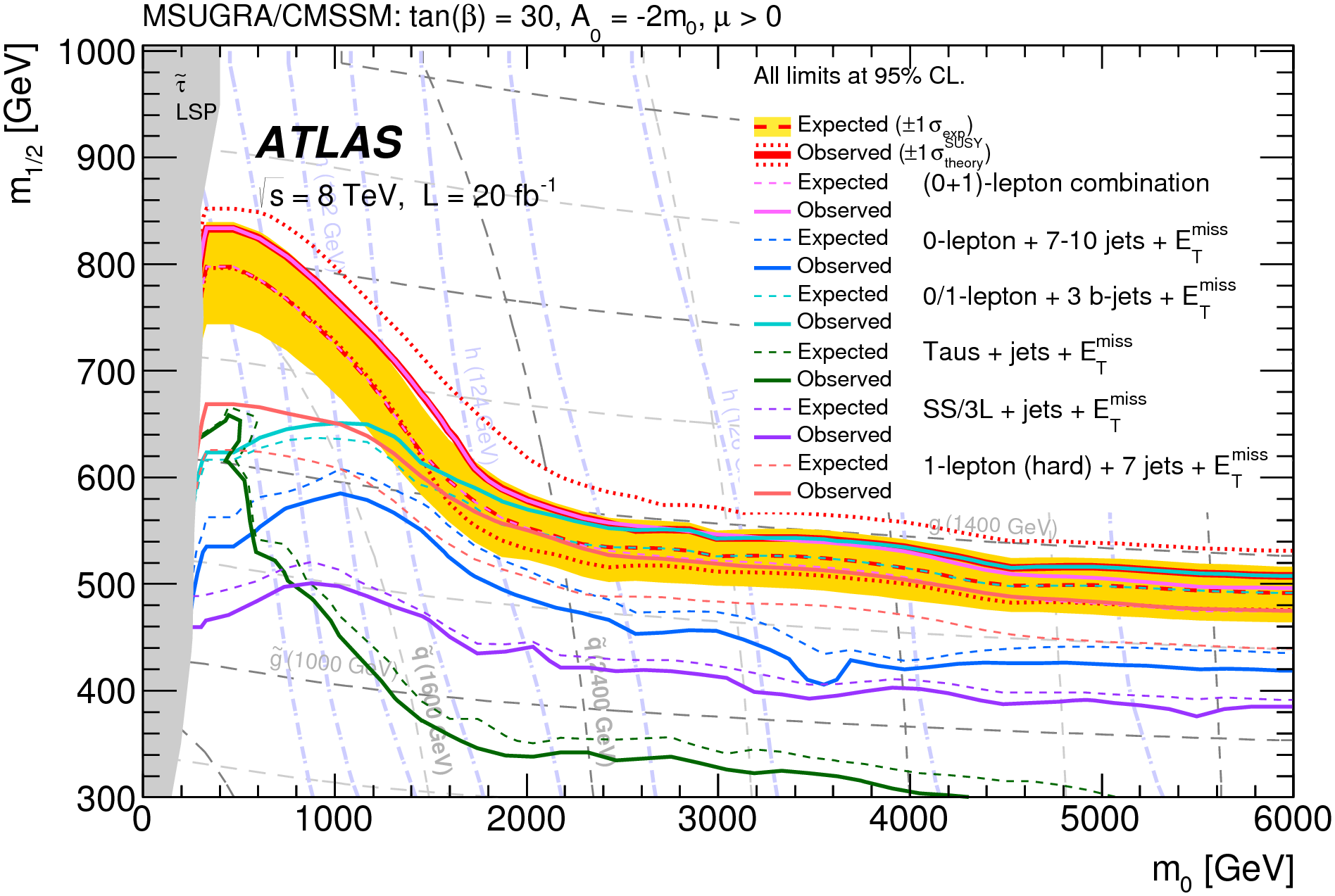}
	\caption{Limits set by ATLAS \cite{Aad:2015baa} on mSUGRA parameter space.}
	\label{fig:msugra_atlas}
    \end{subfigure}
    \hfill
    \begin{subfigure}[b]{0.475\textwidth}
        \includegraphics[width=\textwidth]{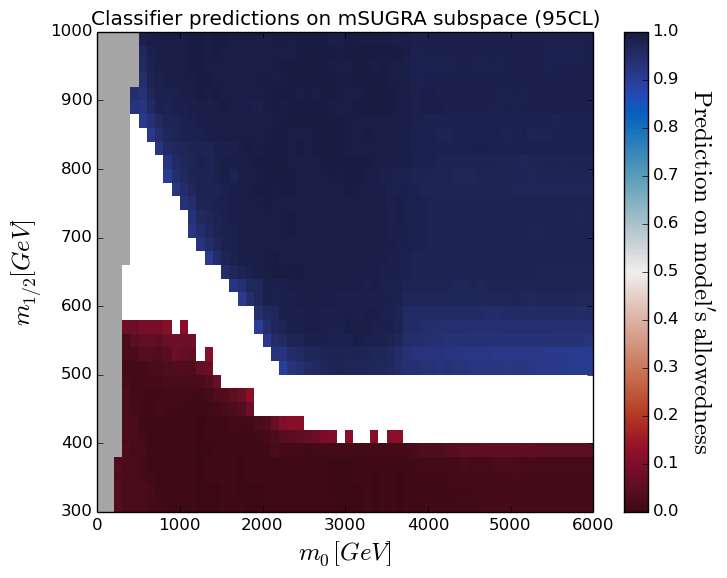}
	\caption{The 95$\%$ CL result of the SUSY-AI classifier.}
	\label{fig:msugra}
    \end{subfigure}
    \caption{Results of testing cMSSM with the trained classifier. The colors in Figure \ref{fig:msugra} indicate the probability of the single point not being excluded. The white band in Figure~\ref{fig:msugra} corresponds to the points that were outside the 95$\%$ CL boundaries.}\label{fig:nsusy_pure}
\end{figure}

A second test was performed on the constrained MSSM (cMSSM or mSUGRA)~\cite{Chamseddine:1982jx,Barbieri:1982eh,
Hall:1983iz}. The MSSM with a particularly popular choice of the universal boundary conditions for the soft breaking terms at the grand unification scale
is called the cMSSM. It is defined in terms of five parameters: common scalar ($m_0$), gaugino ($m_{1/2}$)  and
trilinear ($A_0$) mass parameters (all specified at the GUT scale) plus the ratio $\tan \beta$ of Higgs vacuum
expectation values and sign($\mu$), where $\mu$  is the Higgs/higgsino mass parameter whose square is computed from the
conditions of radiative electroweak symmetry breaking.                            
For this model, ATLAS has set constraints shown in Figure \ref{fig:msugra_atlas} \cite{Aad:2015iea}. Using {\tt SuSpect}
\cite{Djouadi:2002ze}, the same slice of parameter space was sampled randomly following an uniform distribution over parameter space and classified using the tested
classifier. In this scan, we set $\tan\beta=30$, $A_0=2m_{1/2}$ and the sign of $\mu$ to +1 in order to facilitate the comparison
with ATLAS results. {In this search no further constraints were imposed, for example on the Higgs mass or from dark matter measurements.} The result of the classification on the data can be seen in Figure~\ref{fig:msugra}, in which
similarities with Figure~\ref{fig:msugra_atlas} can be observed. 

In this plot, only the data points within the 95$\%$ CL are shown. The white band, therefore, corresponds to the parameter points that could not be classified within 95$\%$ CL. All data points that lay outside of the sampling range as specified in Table~\ref{tab:softbreaking_pmssm} (or close to the border) were relocated  into the sampling region in order to reduce boundary effects of the classifier. In particular, for the points with $m_0>4$~TeV the masses of the scalars where moved back to values approximately 4~TeV. This has a small effect on physics since the heavy scalars have masses outside of the sensitivity of the LHC at 8~TeV. %\BS{The grey band on the left side of the plot only contained model points with a no-neutrino LSP. Since SUSY-AI was only trained on neutrino LSP models, these model points can not be expected to be classified correctly.}

\subsection{Effects of limited training data and applicability range\label{sec:limitations}}
%\subsection{Limitations of SUSY-AI}
In the previous subsections, we have shown that SUSY-AI indeed performs very well despite the fact that the training sample is relatively small. However, here we want to discuss the current limitations of SUSY-AI. Some regions of parameter space of the pMSSM-19 are poorly sampled by the ATLAS data since in these corners of the parameter space it is difficult to satisfy all phenomenological constraints. For example, there are only a few parameter points with very light stops since this would require the maximal mixing scenario with a very heavy $\tilde t_2$ in order to obtain a sufficiently heavy SM-like Higgs boson. In these corners, however, the lack of training data translates to a lower value for the confidence level. This effect can be observed in the plots in the previous chapters. Although the initial prediction may be incorrect, applying a confidence level cut removes almost all incorrectly classified data points from the tested sample.

\begin{table}[t]
\begin{center}
\begin{tabular}{ll}
\hline
Parameter & Range \\
\hline
$m_{\tilde L_1}$&  \small{$[600 \ {\rm GeV},\, 4 \ {\rm TeV}]$}\\
$m_{\tilde E_1}$ &  \small{$[600 \ {\rm GeV},\, 4 \ {\rm TeV}]$}\\
$m_{\tilde L_3}$&  \small{$[600 \ {\rm GeV},\, 4 \ {\rm TeV}]$}\\
$m_{\tilde E_3}$ &  \small{$[600 \ {\rm GeV},\, 4 \ {\rm TeV}]$}\\
$m_{\tilde Q_1}$&\small{$[1200 \ {\rm GeV},\, 4 \ {\rm TeV}]$}\\
$m_{\tilde U_1}$ &  \small{$[1200 \ {\rm GeV},\, 4 \ {\rm TeV}]$}\\
$m_{\tilde D_1}$ & \small{$[1200 \ {\rm GeV},\, 4 \ {\rm TeV}]$}\\
$m_{\tilde Q_3}$&  \small{$[1200 \ {\rm GeV},\, 4 \ {\rm TeV}]$}\\
$m_{\tilde U_3}$ & \small{$[100 \ {\rm GeV},\, 4 \ {\rm TeV}]$}\\
$m_{\tilde D_3}$ &  \small{$[100 \ {\rm GeV},\, 4 \ {\rm TeV}]$}\\
\hline
$A_{t}$ & \small{$[-8 \ {\rm TeV},\, 8 \ {\rm TeV}]$}\\
$A_{b}$ & \small{$[-4 \ {\rm TeV},\, 4 \ {\rm TeV}]$}\\
$A_{\tau}$ &  \small{$[-4 \ {\rm TeV},\, 4 \ {\rm TeV}]$}\\
\hline
$|\mu|$ &  \small{$[80 \ {\rm GeV},\, 4\ {\rm TeV}]$}\\
$|M_1|$ & \small{$[600 \ {\rm TeV},\, 4 \ {\rm TeV}]$}\\
$|M_2|$ & \small{$[600 \ {\rm GeV},\, 4 \ {\rm TeV}]$}\\
$M_3$ & \small{$[1300 \ {\rm GeV},\, 4 \ {\rm TeV}]$}\\
\hline
$M_A$ &  \small{$[600 \ {\rm GeV},\, 4 \ {\rm TeV}]$}\\
$\tan\beta$ & \small{$[1, \, 60]$} \\
\hline
\end{tabular}\hspace*{1cm}
\begin{tabular}{ll}
\hline
Parameter & Range \\
\hline
$m_{\tilde L_1}$&  \small{$[700 \ {\rm GeV},\, 4 \ {\rm TeV}]$}\\
$m_{\tilde E_1}$ &  \small{$[700 \ {\rm GeV},\, 4 \ {\rm TeV}]$}\\
$m_{\tilde L_3}$&  \small{$[700 \ {\rm GeV},\, 4 \ {\rm TeV}]$}\\
$m_{\tilde E_3}$ &  \small{$[700 \ {\rm GeV},\, 4 \ {\rm TeV}]$}\\
$m_{\tilde Q_1}$&\small{$[1200 \ {\rm GeV},\, 4 \ {\rm TeV}]$}\\
$m_{\tilde U_1}$ &  \small{$[1200 \ {\rm GeV},\, 4 \ {\rm TeV}]$}\\
$m_{\tilde D_1}$ & \small{$[1200 \ {\rm GeV},\, 4 \ {\rm TeV}]$}\\
$m_{\tilde Q_3}$&  \small{$[1200 \ {\rm GeV},\, 4 \ {\rm TeV}]$}\\
$m_{\tilde U_3}$ & \small{$[1200 \ {\rm GeV},\, 4 \ {\rm TeV}]$}\\
$m_{\tilde D_3}$ &  \small{$[1200 \ {\rm GeV},\, 4 \ {\rm TeV}]$}\\
\hline
$A_{t}$ & \small{$[-8 \ {\rm TeV},\, 8 \ {\rm TeV}]$}\\
$A_{b}$ & \small{$[-4 \ {\rm TeV},\, 4 \ {\rm TeV}]$}\\
$A_{\tau}$ &  \small{$[-4 \ {\rm TeV},\, 4 \ {\rm TeV}]$}\\
\hline
$|\mu|$ &  \small{$[80 \ {\rm GeV},\, 4\ {\rm TeV}]$}\\
$|M_1|$ & \small{$[0 \ {\rm TeV},\, 4 \ {\rm TeV}]$}\\
$|M_2|$ & \small{$[70 \ {\rm GeV},\, 4 \ {\rm TeV}]$}\\
$M_3$ & \small{$[1300 \ {\rm GeV},\, 4 \ {\rm TeV}]$}\\
\hline
$M_A$ &  \small{$[700 \ {\rm GeV},\, 4 \ {\rm TeV}]$}\\
$\tan\beta$ &\small{$[1, \, 60]$} \\
\hline
\end{tabular}
\end{center}
\caption{Input parameters of the pMSSM subspace in the light stop (left) and the electroweakino (right) scenarios.
 \label{tab:softbreaking_pmssm_EW}}
\end{table}

The lack of data points, but also the improvement on the difference between the true classification and the predicted classification, can be observed in Figure~\ref{fig:mstop_mlsp_nsusy}, which shows density projections on the stop--LSP-mass plane of the total number of parameter points used for testing, their true classification, prediction from the SUSY-AI classifier and the fraction of misclassified points. Here we show a subspace in the general pMSSM-19 parameter space summarized in Table~\ref{tab:softbreaking_pmssm_EW} (left), which corresponds to a subset of the pMSSM-19 resembling a natural-SUSY scenario with relatively light stops but heavy sleptons, and first and second generation squarks. The figure shows the classification for all points as well as for points satisfying the 95\% CL and 99\% CL limit, respectively. As expected, with an increased CL level the misclassification ratio consistently decreases, as can be seen in the right column. In the bottom left corner of the stop--LSP-mass plane many light stop points are excluded if no CL cut is demanded. As can be seen in the left column of this figure, however, this corner was relatively poorly trained due to the lack of data points in that region. It is because of this that the number of data points left after a confidence level cut decreases for increasingly higher cuts, which is consistent with our discussion of the performance of the classifier.

%\begin{figure}[hp]
\afterpage{
    \begin{sidewaysfigure}
	\centering
	\begin{subfigure}[b]{0.03\textwidth}
	    \hspace{0.5cm}
	\end{subfigure}
	\hfill
	\begin{subfigure}[b]{0.22\textwidth}
	    \begin{center}
		Number of model points
	    \end{center}
	    \vspace{0.5cm}
	\end{subfigure}
	\hfill
	\begin{subfigure}[b]{0.22\textwidth}
	    \begin{center}
		True classification
	    \end{center}
	    \vspace{0.5cm}
	\end{subfigure}
	\hfill
	\begin{subfigure}[b]{0.22\textwidth}
	    \begin{center}
		Prediction by classifier
	    \end{center}
	    \vspace{0.5cm}
	\end{subfigure}
	\hfill
	\begin{subfigure}[b]{0.22\textwidth}
	    \begin{center}
		Ratio of misclassified points
	    \end{center}
	    \vspace{0.5cm}
	\end{subfigure}
	\hfill

	\begin{subfigure}[b]{0.03\textwidth}
      \rotatebox{90}{\hspace{2cm}All data}
      \end{subfigure}
	\hfill
	\begin{subfigure}[b]{0.23\textwidth}
	    \includegraphics[width=\textwidth]{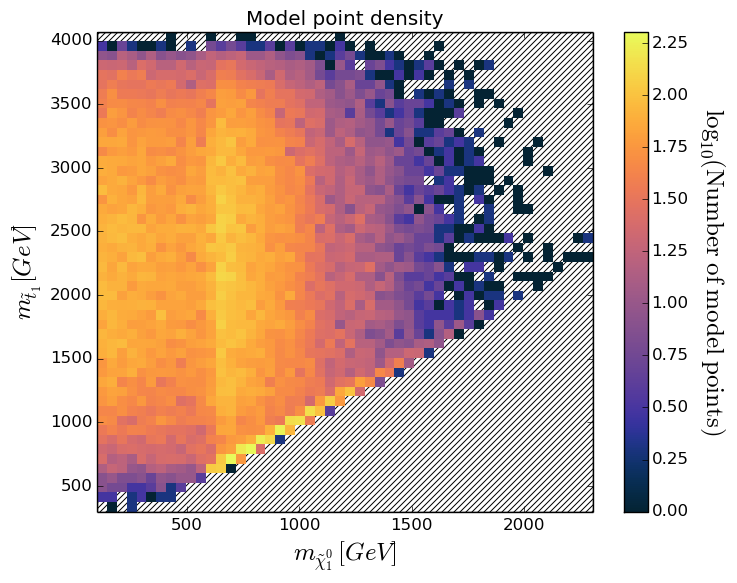}
	\end{subfigure}
	\hfill
	\begin{subfigure}[b]{0.23\textwidth}
	    \includegraphics[width=\textwidth]{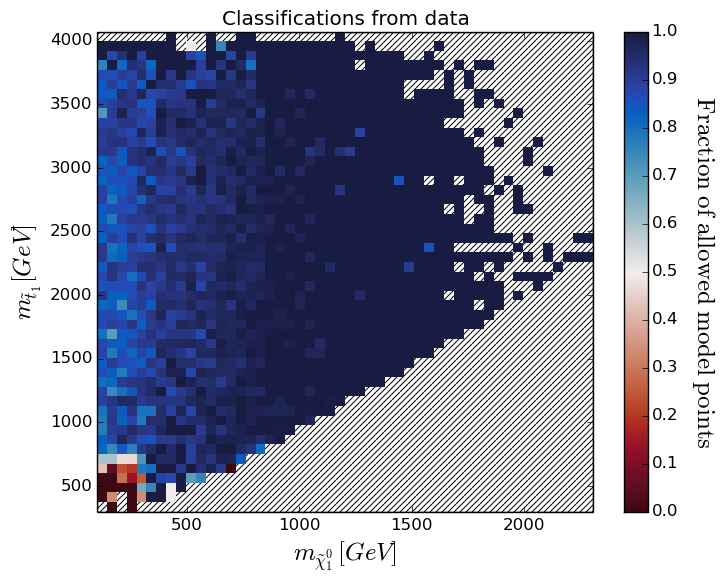}
	\end{subfigure}
	\hfill
	\begin{subfigure}[b]{0.23\textwidth}
	    \includegraphics[width=\textwidth]{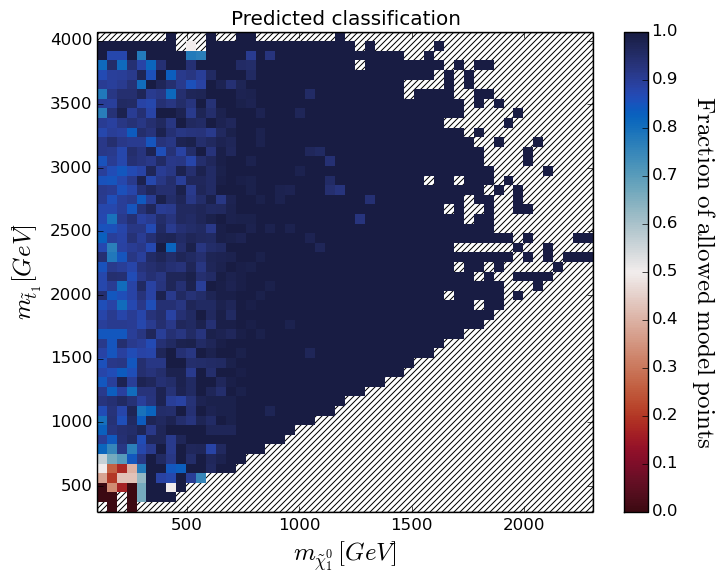}
	\end{subfigure}
	\hfill
	\begin{subfigure}[b]{0.23\textwidth}
	    \includegraphics[width=\textwidth]{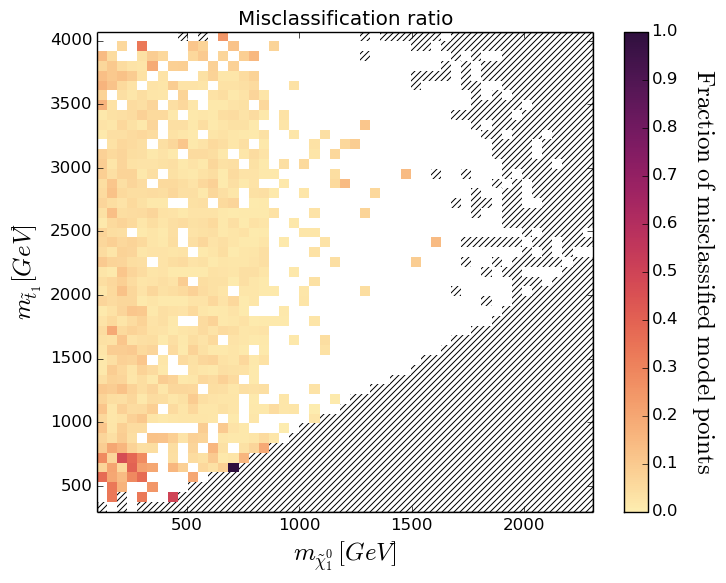}
	\end{subfigure}
	
	\begin{subfigure}[b]{0.03\textwidth}
	    \rotatebox{90}{\hspace{2.15cm}95\-CL}
	\end{subfigure}
	\hfill
	\begin{subfigure}[b]{0.23\textwidth}
	    \includegraphics[width=\textwidth]{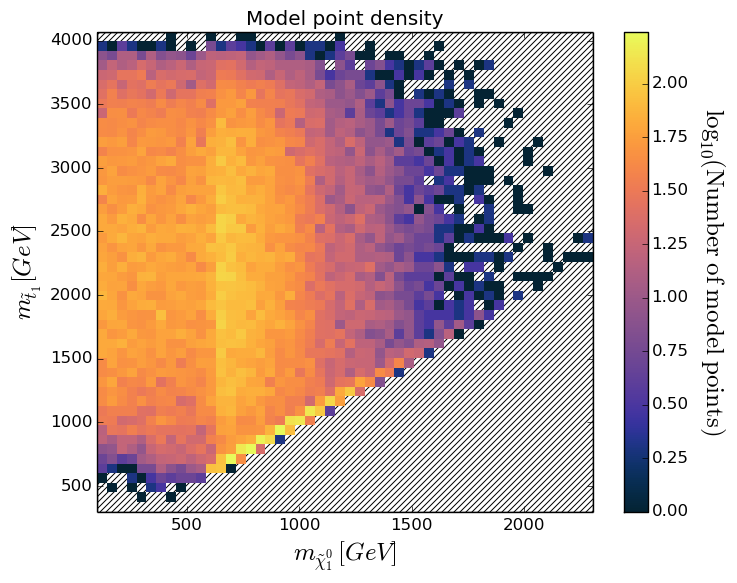}
	\end{subfigure}
	\hfill
	\begin{subfigure}[b]{0.23\textwidth}
	    \includegraphics[width=\textwidth]{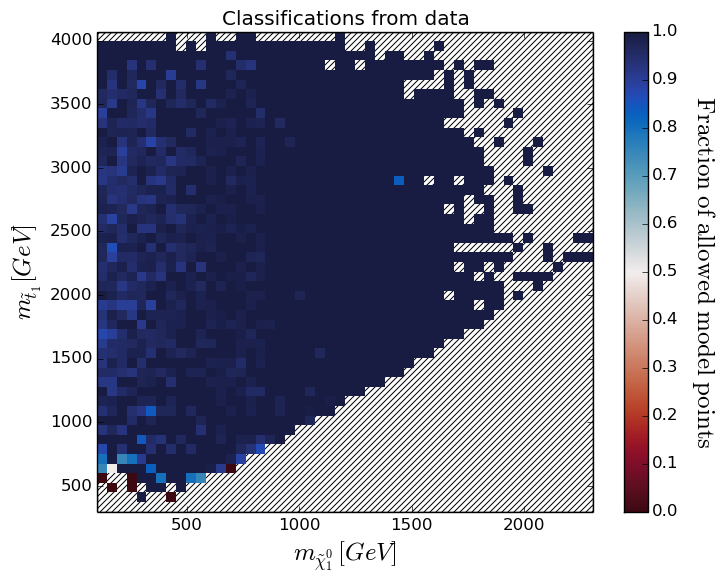}
	\end{subfigure}
	\hfill
	\begin{subfigure}[b]{0.23\textwidth}
	    \includegraphics[width=\textwidth]{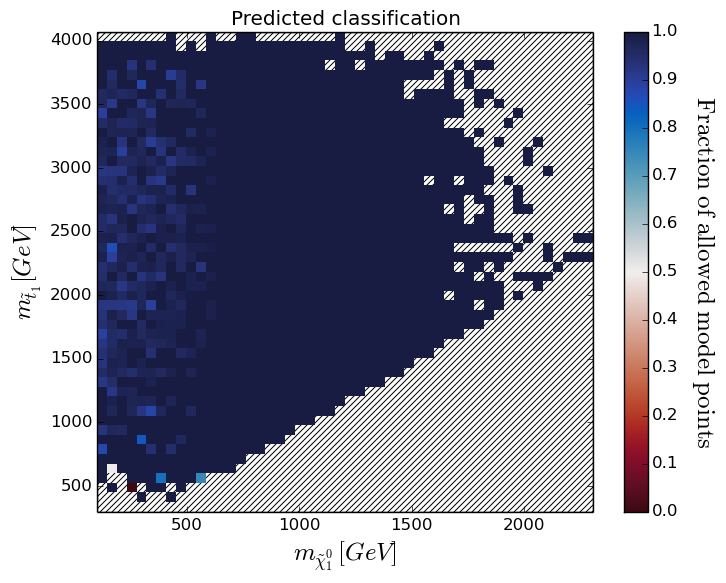}
	\end{subfigure}
	\hfill
	\begin{subfigure}[b]{0.23\textwidth}
	    \includegraphics[width=\textwidth]{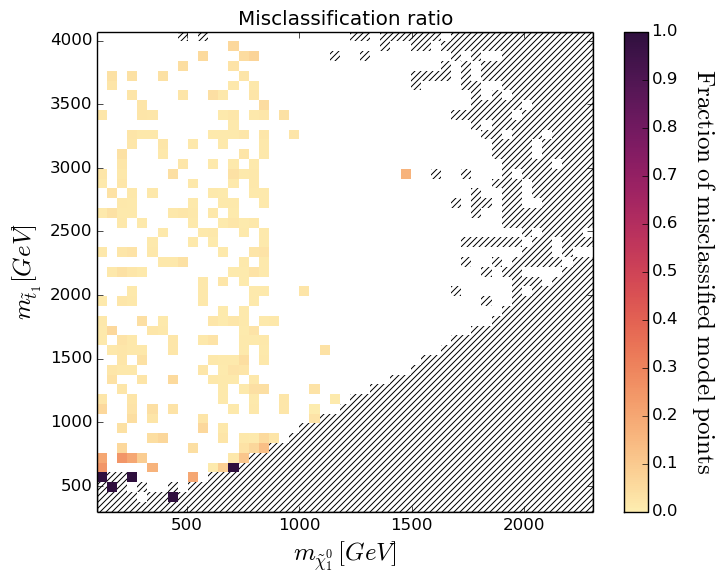}
	\end{subfigure}  
	
	\begin{subfigure}[b]{0.03\textwidth}
	    \rotatebox{90}{\hspace{2.15cm}99\-CL}
	\end{subfigure}
	\hfill
	\begin{subfigure}[b]{0.23\textwidth}
	    \includegraphics[width=\textwidth]{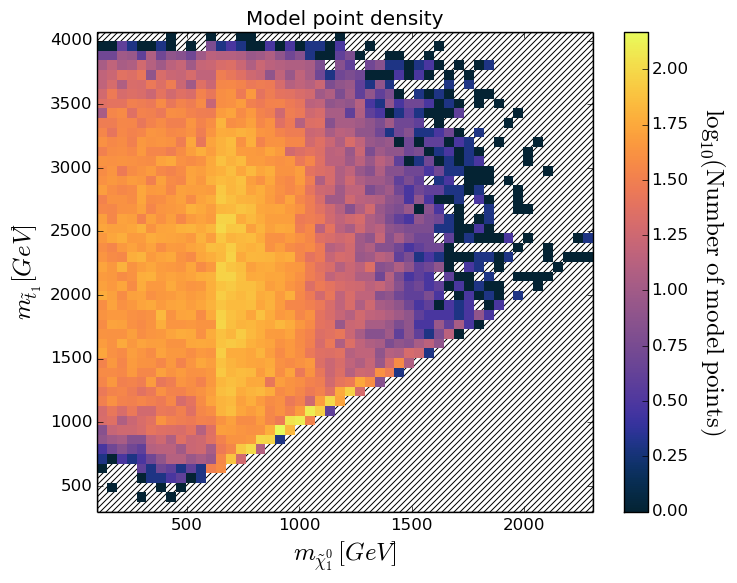}
	\end{subfigure}
	\hfill
	\begin{subfigure}[b]{0.23\textwidth}
	    \includegraphics[width=\textwidth]{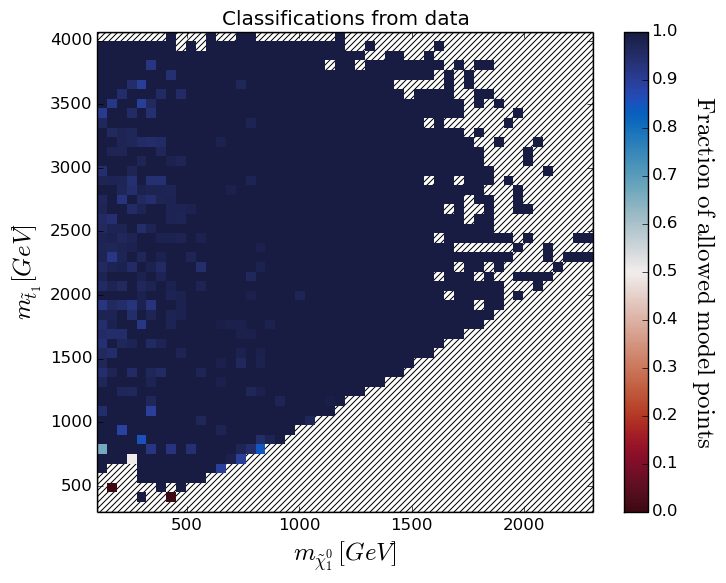}
	\end{subfigure}
	\hfill
	\begin{subfigure}[b]{0.23\textwidth}
	    \includegraphics[width=\textwidth]{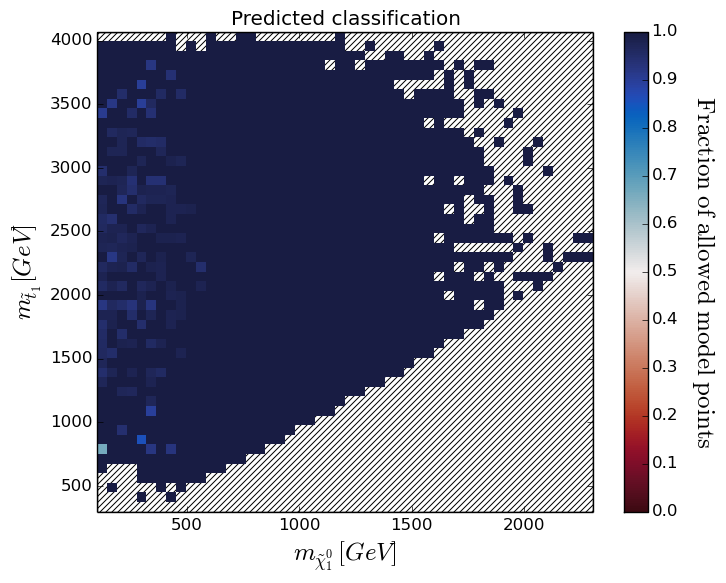}
	\end{subfigure}
	\hfill
	\begin{subfigure}[b]{0.23\textwidth}
	    \includegraphics[width=\textwidth]{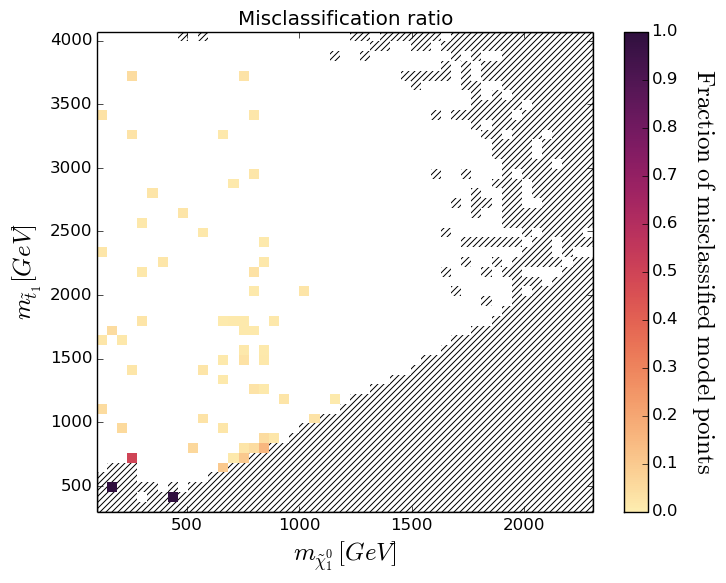}
	\end{subfigure}
	\caption{Color histograms for a projection of the 19-dimensional pMSSM parameter space on the $m_{\tilde{t}_1}$--$m_{\tilde{\chi}^0_1}$ plane after imposing the constraints on the soft breaking parameters summarized in Table~\ref{tab:softbreaking_pmssm_EW} (left). The color in the second and third column indicates the fraction of allowed data points. The last column shows the fraction of misclassified points. The dashed bins contain no data points.}
	\label{fig:mstop_mlsp_nsusy}
    \end{sidewaysfigure}
    \clearpage
}
%\end{figure}

%\begin{figure}[hp]
\afterpage{
    \begin{sidewaysfigure}
	\centering
	\begin{subfigure}[b]{0.03\textwidth}
	    \hspace{0.5cm}
	\end{subfigure}
	\hfill
	\begin{subfigure}[b]{0.22\textwidth}
	    \begin{center}
		Number of model points
	    \end{center}
	    \vspace{0.5cm}
	\end{subfigure}
	\hfill
	\begin{subfigure}[b]{0.22\textwidth}
	    \begin{center}
		True classification
	    \end{center}
	    \vspace{0.5cm}
	\end{subfigure}
	\hfill
	\begin{subfigure}[b]{0.22\textwidth}
	    \begin{center}
		Prediction by classifier
	    \end{center}
	    \vspace{0.5cm}
	\end{subfigure}
	\hfill
	\begin{subfigure}[b]{0.22\textwidth}
	    \begin{center}
		Ratio of misclassified points
	    \end{center}
	    \vspace{0.5cm}
	\end{subfigure}
	\hfill

	\begin{subfigure}[b]{0.03\textwidth}
      \rotatebox{90}{\hspace{2cm}All data}
      \end{subfigure}
	\hfill
	\begin{subfigure}[b]{0.23\textwidth}
	    \includegraphics[width=\textwidth]{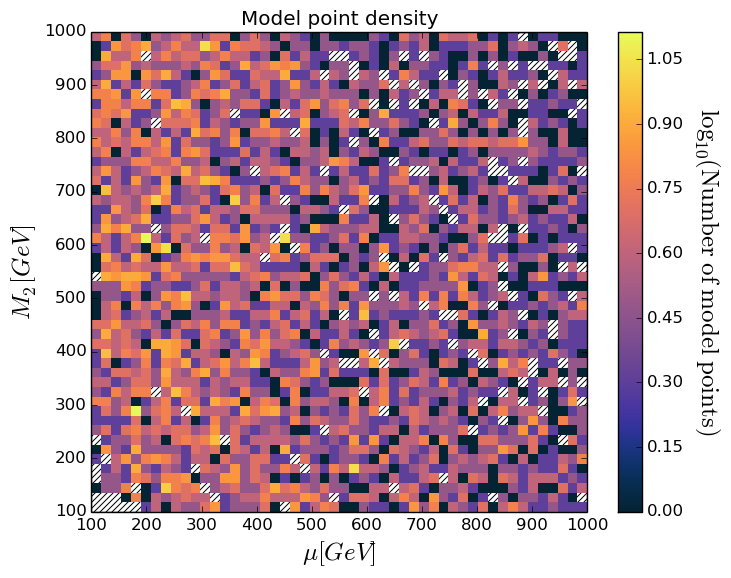}
	\end{subfigure}
	\hfill
	\begin{subfigure}[b]{0.23\textwidth}
	    \includegraphics[width=\textwidth]{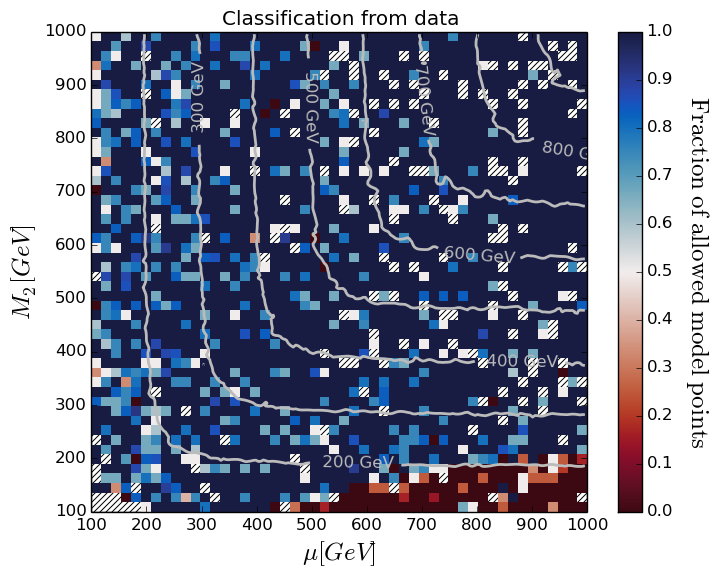}
	\end{subfigure}
	\hfill
	\begin{subfigure}[b]{0.23\textwidth}
	    \includegraphics[width=\textwidth]{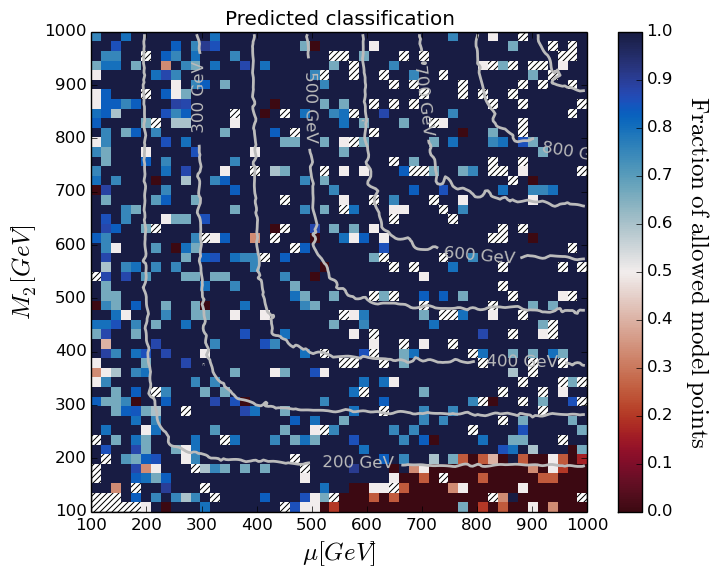}
	\end{subfigure}
	\hfill
	\begin{subfigure}[b]{0.23\textwidth}
	    \includegraphics[width=\textwidth]{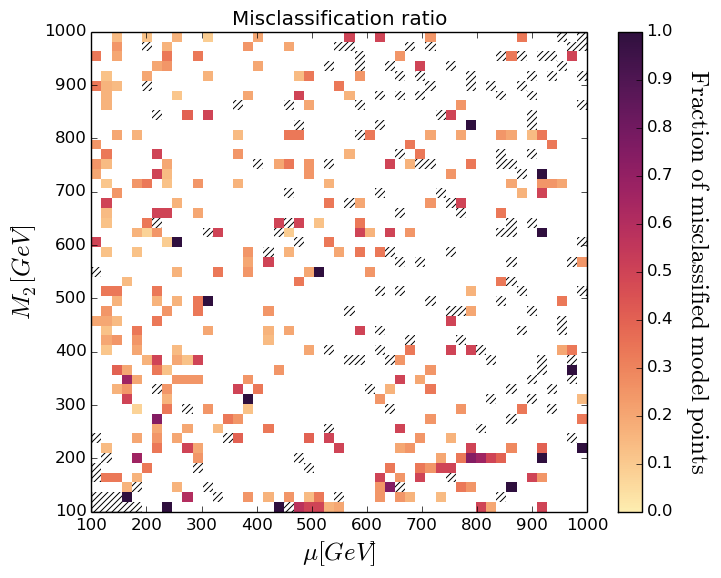}
	\end{subfigure}
	
	\begin{subfigure}[b]{0.03\textwidth}
	    \rotatebox{90}{\hspace{2.15cm}95\-CL}
	\end{subfigure}
	\hfill
	\begin{subfigure}[b]{0.23\textwidth}
	    \includegraphics[width=\textwidth]{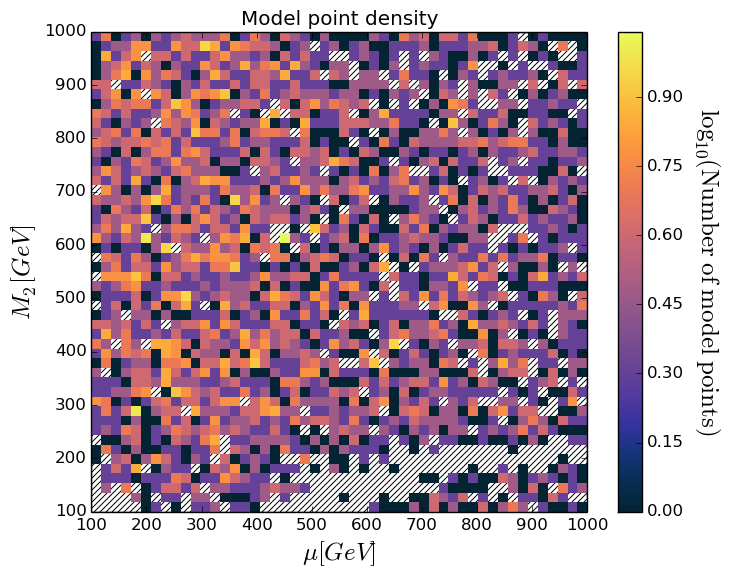}
	\end{subfigure}
	\hfill
	\begin{subfigure}[b]{0.23\textwidth}
	    \includegraphics[width=\textwidth]{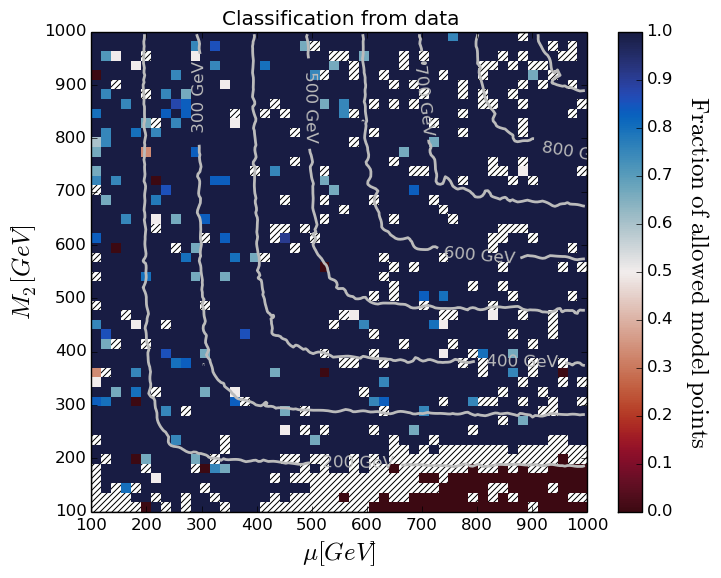}
	\end{subfigure}
	\hfill
	\begin{subfigure}[b]{0.23\textwidth}
	    \includegraphics[width=\textwidth]{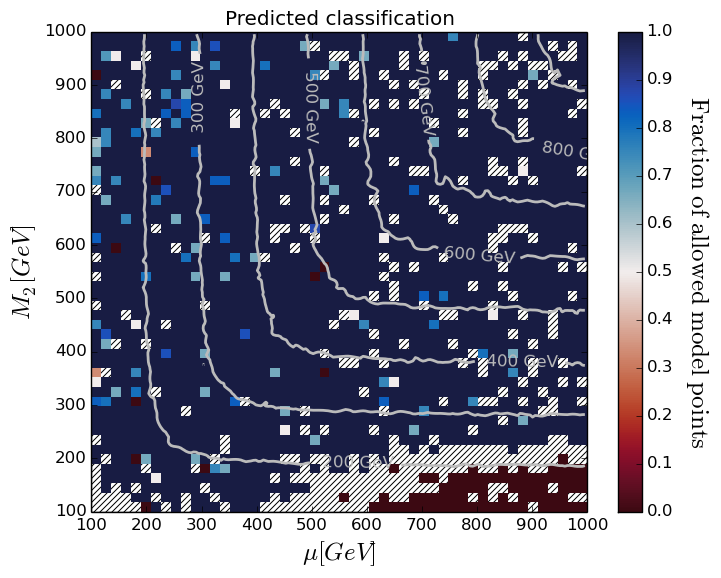}
	\end{subfigure}
	\hfill
	\begin{subfigure}[b]{0.23\textwidth}
	    \includegraphics[width=\textwidth]{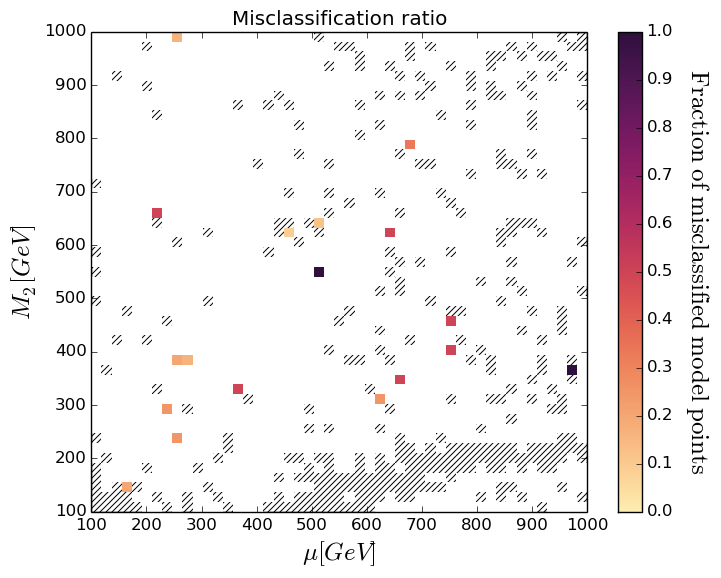}
	\end{subfigure}  
	
	\begin{subfigure}[b]{0.03\textwidth}
	    \rotatebox{90}{\hspace{2.15cm}99\-CL}
	\end{subfigure}
	\hfill
	\begin{subfigure}[b]{0.23\textwidth}
	    \includegraphics[width=\textwidth]{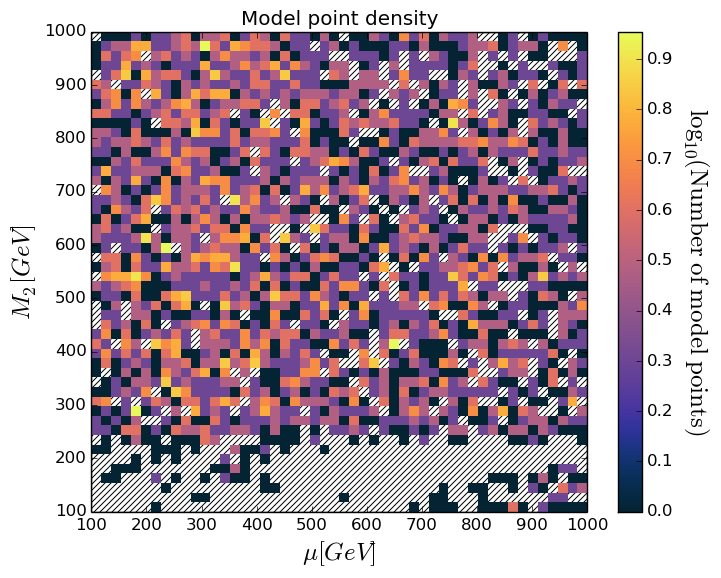}
	\end{subfigure}
	\hfill
	\begin{subfigure}[b]{0.23\textwidth}
	    \includegraphics[width=\textwidth]{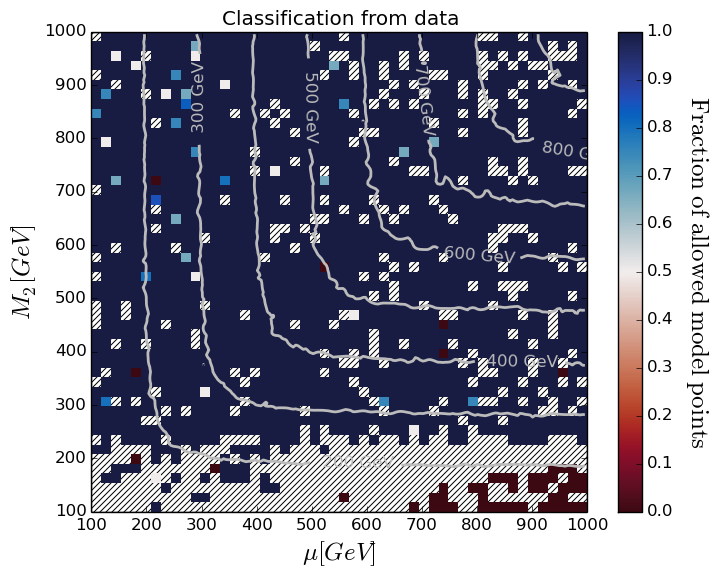}
	\end{subfigure}
	\hfill
	\begin{subfigure}[b]{0.23\textwidth}
	    \includegraphics[width=\textwidth]{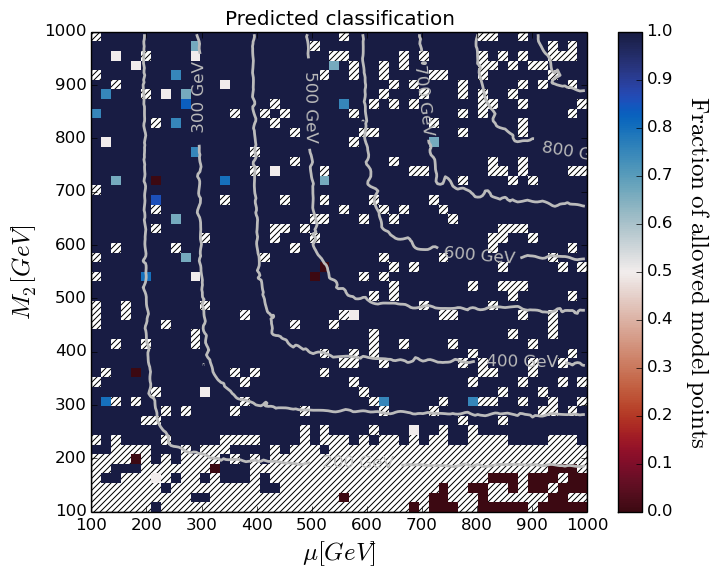}
	\end{subfigure}
	\hfill
	\begin{subfigure}[b]{0.23\textwidth}
	    \includegraphics[width=\textwidth]{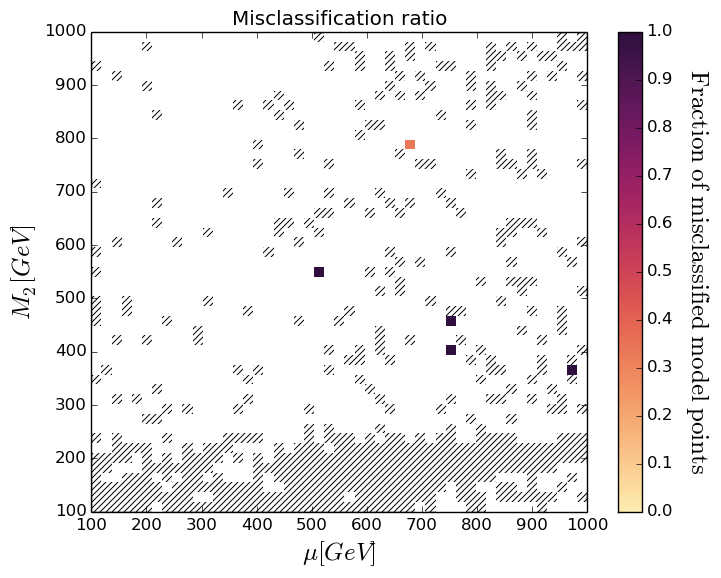}
	\end{subfigure}
	\caption{Color histograms for a projection of the 19-dimensional pMSSM parameter space on the $M_2$--$\mu$ plane after imposing the constraints on the soft breaking parameters summarized in Table~\ref{tab:softbreaking_pmssm_EW} (right). The contours denote the mass of $\chi^0_1$. The color in the second and third column indicates the fraction of allowed data points. The last column shows the fraction of misclassified points. The dashed bins contain no data points.}
	\label{fig:softbreaking_pmssm_EW}
    \end{sidewaysfigure}
    \clearpage
}
%\end{figure}

Figure~\ref{fig:softbreaking_pmssm_EW} shows a second example in the $M_2$--$\mu$ plane. This subset of the pMSSM-19 is defined in Table~\ref{tab:softbreaking_pmssm_EW} (right) and it resembles an electroweakino scenario with severe restrictions on the parameter space. As a result, the $M_2$--$\mu$ plane is sparsely populated in the training sample. One can again observe a corner in the parameter space that is excluded if no CL cut is imposed. In particular, the pure wino LSP scenario is excluded due to long-lived sparticle searches. However, without any cut the misclassification ratio is non-negligible. With increasing CL cuts, however, the points with lower CL are removed and the misclassification ratio is significantly reduced. This demonstrates that the CL assignment fulfills its role: it reveals the `uncertain' points that require a more detailed assessment.  

Although introducing a cut on the confidence level removes data points on which a prediction can be made from a testing sample, both Figures~\ref{fig:mstop_mlsp_nsusy} and \ref{fig:softbreaking_pmssm_EW} show an increase in the quality of the prediction. Using confidence levels in making predictions, therefore, corresponds to removing data points on which the resulted binary prediction was uncertain, automatically removing data points in regions of parameter space with a low density of training data.

{Another limitation of the current version is that it only uses the combined classification from all searches, without making a distinction which particular analysis excluded a given parameter point. While this may underpower some of the analyses, e.g.\ electroweak searches, the validation plots in Figures~\ref{fig:appplot01}--\ref{fig:appplot04} show nevertheless a good sensitivity to light electroweakinos and sleptons. The future versions will aim to also use this additional information in order to improve performance in this region of the parameter space. }

The pMSSM sample used for training of SUSY-AI meets the set of constraints discussed in Section~\ref{sec:atlasandpmssm}. As we showed in various validation plots, the code performs well on points that belong to the to the tested subspace of the MSSM. The CMSSM example further demonstrates that some of the constraints (e.g.\ Higgs mass or dark matter relic density) can be relaxed. A user who wishes to use SUSY-AI on samples that are outside the ranges of the ATLAS sample or do not fulfill some of the constraints 
should first perform revalidation of the code. The CL measure that we introduced can greatly assist in this process. 
Generally speaking, a clear sign that a classification cannot be trusted would be a high fraction of points with low CL scores for the sample being tested. Another method would be to compare SUSY-AI predictions to a small number of fully simulated points (a MC simulation and detector simulation using \texttt{CheckMATE} would suffice) for which one can clearly conclude about their exclusion status. We advise against using SUSY-AI for models that have significantly different phenomenology from the training pMSSM sample, for example including $R$-parity violation or the gravitino LSP.  Finally, as an additional functionality, SUSY-AI issues and automated warning when a tested point lies outside the limits specified in Table~\ref{tab:softbreaking_pmssm}. When this is the case the point can be automatically moved within the limits and the decision is left to the user if the prediction can be trusted.

%%%%%%%%%%%%%%%%%%%%%%%%%%%%%%%%%%%%%%%%%%%%%%%%%%%%%%%%

\section{Conclusions \label{sec:conclusions}}
A random forest classifier has been trained on over $310\,000$ data points of the pMSSM. We demonstrate that it provides a reliable classification with an accuracy of 93.8$\%$. The reliability can be improved by demanding a minimum confidence level for the prediction. The trained classifier, SUSY-AI, is tested on the 19-dimensional pMSSM, the 6-dimensional natural SUSY model and on the 5-dimensional constrained MSSM. All these tests yield results that confirm reliable classification.

SUSY-AI will be continuously updated with future LHC results as a part of the BSM-AI project. When possible, the publicly available ATLAS and CMS data will be used as in the current work. Additionally, we plan to generate our own MC data samples and recast them to produce limits using \texttt{CheckMATE} based on the existing and future LHC analyses. 
Classifiers and regressors for other models of new physics are also planned so that the whole project could cover a broad range of theories. 

SUSY-AI can be downloaded from the web page 
\begin{center}
\url{http://susyai.hepforge.org}
\end{center}
where we also provide installation instructions, more detailed technical information, frequently asked questions, example codes and updates of SUSY-AI using results of the ongoing Run 2, currently based on Refs.~\cite{Barr:2016inz,Barr:2016sho}.

%%%%%%%%%%%%%%%%%%%%%%%%%%%%%%%%%%%%%%%%%%%%%%%%%%%%%%%%

%\bigskip
\paragraph{Acknowledgments} 
J.S.K.\ wants to thank Ulik Kim for the support while part of this manuscript was prepared and A.~Kodewitz for discussions. S.C.\ and B.S.\ would like to thank Tom Heskes for insightful discussion of machine learning methodology.  

{\small{R.RdA.\ is supported by the Ram\'on y Cajal program of the Spanish MICINN and also thanks the support by the ``SOM Sabor y origen de la Materia" (FPA2011-29678), the ``Fenomenologia y Cosmologia de la Fisica mas alla del Modelo Estandar e lmplicaciones Experimentales en la era del LHC" (FPA2010-17747) MEC projects, the Severo Ochoa MINECO project SEV-2014-0398, the Consolider-Ingenio 2010 programme under grant MULTIDARK CSD2009-00064 and by the Invisibles European ITN project FP7-PEOPLE-2011-ITN, PITN-GA-2011-289442-INVISIBLES. J.S.K.\ and K.R.\ have been partially supported by the MINECO (Spain) under contract FPA2013-44773-P;  Consolider-Ingenio CPAN CSD2007-00042; the Spanish MINECO Centro de excelencia Severo Ochoa  Program under grant SEV-2012-0249; and by JAE-Doc program. K.R.\ was supported by the National Science Centre (Poland) under Grant 2015/19/D/ST2/03136 and the Collaborative Research Center SFB676 of the DFG, ``Particles, Strings, and the Early Universe''. B.S and S.C acknowledge support within the iDark project by the Netherlands eScience Center.}}

%%%%%%%%%%%%%%%%%%%%%%%%%%%%%%%%%%%%%%%%%%%%%%%%%%%%%%%%

%\newpage
%\newpage
\begin{appendices}
\section{Comparison of out-of-bag estimation with train:test split} \label{app:splitting}

The SUSY-AI classifier was validated using out-of-bag estimation, i.e.\ using the full available data set. In this appendix we compare SUSY-AI to an identically configured classifier    trained on a subset of the data, so that is could be validated using the remaining data. Although this is the standard way of classifier validation, it has a drawback that the classifier does not fully exploit the available data.

A comparison of the out-of-bag method and the splitting of the dataset is shown in Tables~\ref{tab:comparison_oob} and \ref{tab:comparison_split}. The column labels are defined as follows: 
\begin{equation} \label{eq:measures}.
\begin{split}
\textrm{accuracy} \quad =& \quad {\frac{\textrm{TP + TN}}{\textrm{TP} + \textrm{FP} + \textrm{FN} + \textrm{TN}}}\,, \\
\textrm{precision} \quad =& \quad {\frac{\textrm{TP}}{\textrm{TP} + \textrm{FP}}}\,, \\
\textrm{sensitivity} \quad =& \quad {\frac{\textrm{TP}}{\textrm{TP} + \textrm{FN}}}\,, \\
\textrm{specificity} \quad =& \quad {\frac{\textrm{TN}}{\textrm{TN} + \textrm{FP}}}\,, \\
\textrm{ negative\ predictive\ value\ (NPV)} \quad =& \quad {\frac{\textrm{TN}}{\textrm{TN} + \textrm{FN}}}\,,
\end{split}
\end{equation}
while different tags for each point are assigned following the rules in Table~\ref{tab:hypothesistesting}.
The comparison yields similar results with the out-of-bag method performing better on accuracy, precision and sensitivity. 

%\newpage
\begin{minipage}{\textwidth}
\centering
\Large{Out-of-bag}
%\begin{table}[t]
%\centering
\small{
\begin{tabular}{lccccccr}
\hline
{CL} & {\#} & {\# / total} & {Accuracy} & {Precision} & {Sensitivity} & {NPV} & {Specificity}\\
\hline
0.0 & $310\,324$ & 1.0000 & 0.93226 & 0.93951 & 0.94665 & 0.92152 & 0.91133 \\
0.68 & $289\,371$ & 0.93248 & 0.95735 & 0.96072 & 0.96835 & 0.95222 & 0.94094 \\
0.95 & $219\,233$ & 0.70646 & 0.99094 & 0.99092 & 0.99426 & 0.99096 & 0.98573 \\
0.98 & $184\,230$ & 0.59367 & 0.99543 & 0.99573 & 0.99672 & 0.99496 & 0.99346 \\
0.99 & $160\,034$ & 0.51570 & 0.99708 & 0.99747 & 0.99764 & 0.99649 & 0.99624 \\
\hline
\end{tabular}
}
\captionof{table}{Results of the validation of the SUSY-AI classifier with out-of-bag estimation.}
\label{tab:comparison_oob}
%\end{table}
\end{minipage}

\begin{minipage}{\textwidth}
\Large{Dataset splitting train:test = 75:25}
%\begin{table}[h]
\centering
\small{
\begin{tabular}{lccccccr}
\hline
{CL} & {\#} & {\# / total} & {Accuracy} & {Precision} & {Sensitivity} & {NPV} & {Specificity}\\
\hline
0.0 & $77\,581$ & 1.0000 & 0.92271 & 0.91653 & 0.93049 & 0.92912 & 0.91491 \\
0.68 & $70\,375$ & 0.90712 & 0.9545 & 0.95516 & 0.95302 & 0.95386 & 0.95595 \\
0.95 & $48\,900$ & 0.63031 & 0.99022 & 0.99047 & 0.9893 & 0.99 & 0.99109 \\
0.98 & $39\,815$ & 0.51321 & 0.99485 & 0.99559 & 0.99353 & 0.99419 & 0.99604 \\
0.99 & $34\,004$ & 0.43830 & 0.99644 & 0.99685 & 0.99554 & 0.99608 & 0.99724 \\
\hline
\end{tabular}
}
\captionof{table}{Results of the validation of the RF classifier with a split dataset (0.75 training, 0.25 testing).}
\label{tab:comparison_split}
%\end{table}
\end{minipage}

\begin{table}[h]\renewcommand{\arraystretch}{1.2}
  \centering
  \small{
  \begin{tabular}{>{\centering\arraybackslash}m{0.8in} >{\centering\arraybackslash}m{0.7in} || >{\centering\arraybackslash}m{1.4in} | >{\centering\arraybackslash}m{1.4in}}
    & & \multicolumn{2}{c}{True classification} \\
    & & Positive & Negative \\
    \hline
    \hline
    \multirow{2}{*}{Prediction} & {Positive} &{True positive (TP)} & False positive (FP) \\
%    & & & \footnotesize{\textit{Error of first kind}} \\
    \cline{2-4}
    & {Negative} & False negative (FN) &{True negative (TN)} \\
%    & & \footnotesize{\textit{Error of second kind}} & \\
  \end{tabular}
  }
  \caption{Classification of events following from comparison of true classification and prediction~\cite{Witten:2005}.}
  \label{tab:hypothesistesting}
\end{table}
\section{Projections of the pMSSM} \label{app:projections}

In this appendix we provide additional validation plots demonstrating the performance of SUSY-AI. They are presented in a similar manner to Figures~\ref{fig:mstop_mlsp_nsusy} and \ref{fig:softbreaking_pmssm_EW} as various two-dimensional projections of the 19-dimensional pMSSM parameter space. We show the following projections: $m_{\tilde{b}_1}$--$m_{\tilde{\chi}_1^0}$ plane in Figure~\ref{fig:appplot01}, $m_{\tilde{\ell}_L}$--$m_{\tilde{\chi}_1^0}$ plane in Figure~\ref{fig:appplot02}, $m_{\tilde{\chi}_2^0}$--$m_{\tilde{\chi}_1^0}$ plane in Figure~\ref{fig:appplot03}, $m_{\tilde{\chi}_1^\pm}$--$m_{\tilde{\chi}_1^0}$ plane in Figure~\ref{fig:appplot04}, $m_{A^0}$--$\tan\beta$ plane in Figure~\ref{fig:appplot05}, $\mu$--$M_2$  plane in Figure~\ref{fig:appplot06}, $M_3$--$m_{\widetilde{Q}_1}$ plane in Figure~\ref{fig:appplot07}, and $m_{\widetilde{Q}_1}$--$m_{\widetilde{D}_1}$ plane in Figure~\ref{fig:appplot08}.

\afterpage{
    \begin{sidewaysfigure}
	\centering
	\begin{subfigure}[b]{0.03\textwidth}
	    \hspace{0.5cm}
	\end{subfigure}
	\hfill
	\begin{subfigure}[b]{0.22\textwidth}
	    \begin{center}
		Number of model points
	    \end{center}
	    \vspace{0.5cm}
	\end{subfigure}
	\hfill
	\begin{subfigure}[b]{0.22\textwidth}
	    \begin{center}
		True classification
	    \end{center}
	    \vspace{0.5cm}
	\end{subfigure}
	\hfill
	\begin{subfigure}[b]{0.22\textwidth}
	    \begin{center}
		Prediction by classifier
	    \end{center}
	    \vspace{0.5cm}
	\end{subfigure}
	\hfill
	\begin{subfigure}[b]{0.22\textwidth}
	    \begin{center}
		Ratio of misclassified points
	    \end{center}
	    \vspace{0.5cm}
	\end{subfigure}
	\hfill

	\begin{subfigure}[b]{0.03\textwidth}
      \rotatebox{90}{\hspace{2cm}All data}
      \end{subfigure}
	\hfill
	\begin{subfigure}[b]{0.23\textwidth}
	    \includegraphics[width=\textwidth]{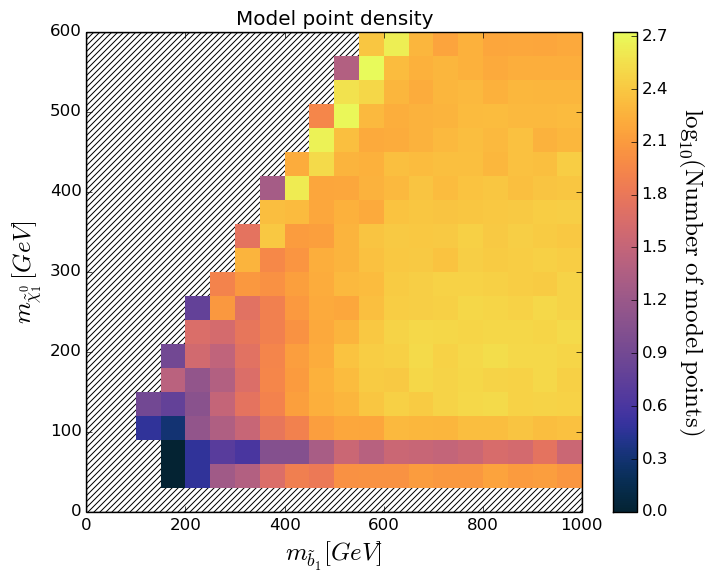}
	\end{subfigure}
	\hfill
	\begin{subfigure}[b]{0.23\textwidth}
	    \includegraphics[width=\textwidth]{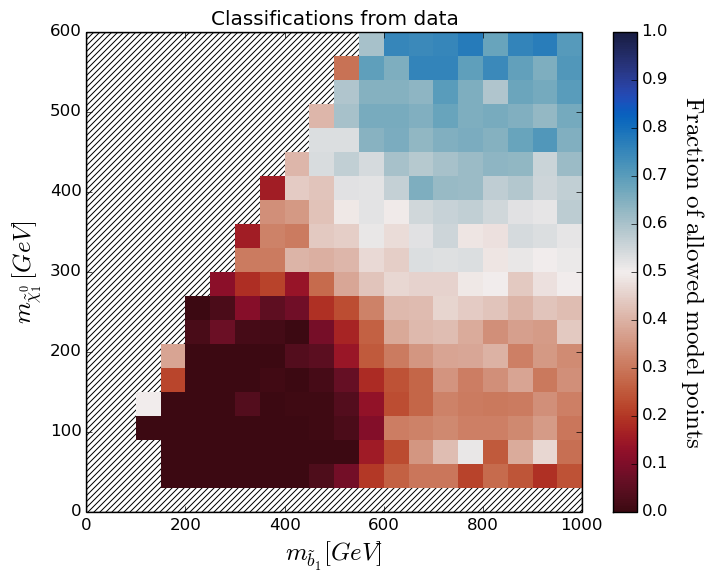}
	\end{subfigure}
	\hfill
	\begin{subfigure}[b]{0.23\textwidth}
	    \includegraphics[width=\textwidth]{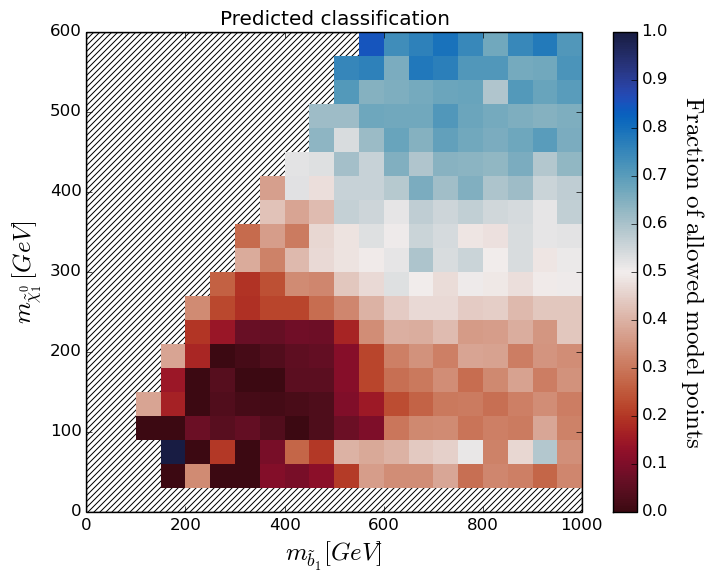}
	\end{subfigure}
	\hfill
	\begin{subfigure}[b]{0.23\textwidth}
	    \includegraphics[width=\textwidth]{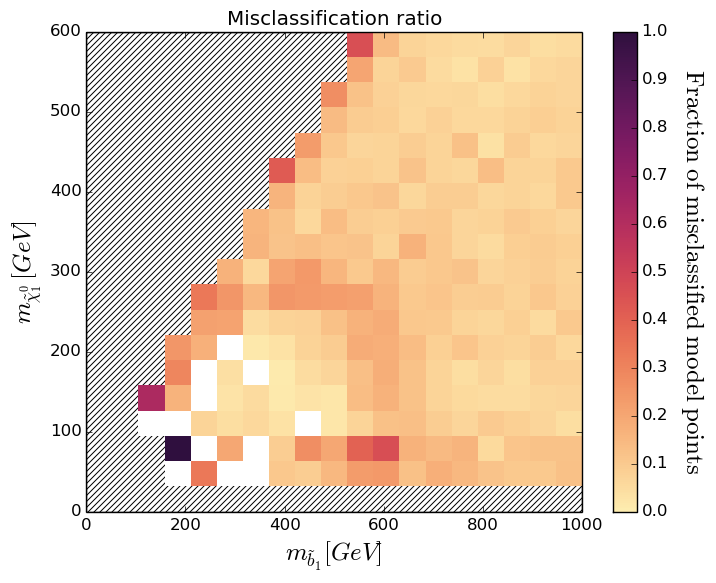}
	\end{subfigure}
	
	\begin{subfigure}[b]{0.03\textwidth}
	    \rotatebox{90}{\hspace{2.15cm}95\-CL}
	\end{subfigure}
	\hfill
	\begin{subfigure}[b]{0.23\textwidth}
	    \includegraphics[width=\textwidth]{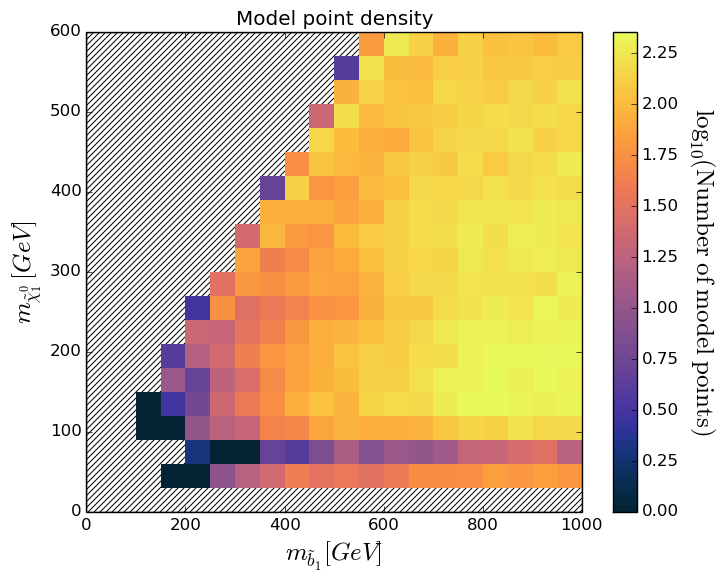}
	\end{subfigure}
	\hfill
	\begin{subfigure}[b]{0.23\textwidth}
	    \includegraphics[width=\textwidth]{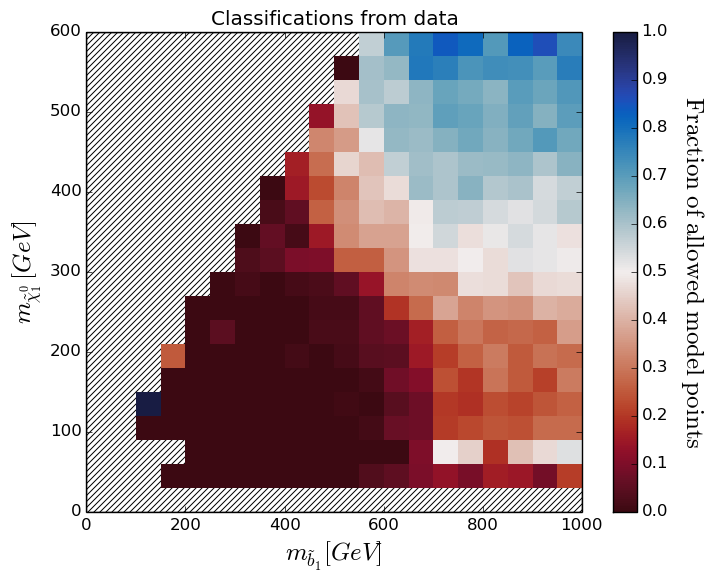}
	\end{subfigure}
	\hfill
	\begin{subfigure}[b]{0.23\textwidth}
	    \includegraphics[width=\textwidth]{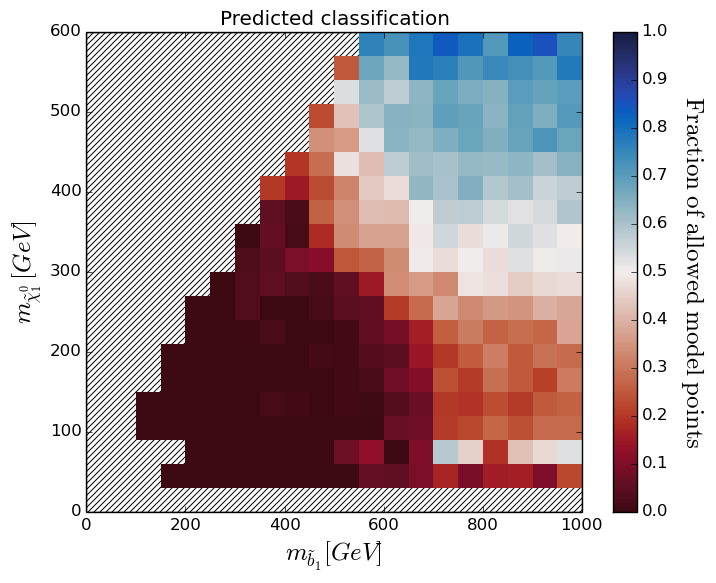}
	\end{subfigure}
	\hfill
	\begin{subfigure}[b]{0.23\textwidth}
	    \includegraphics[width=\textwidth]{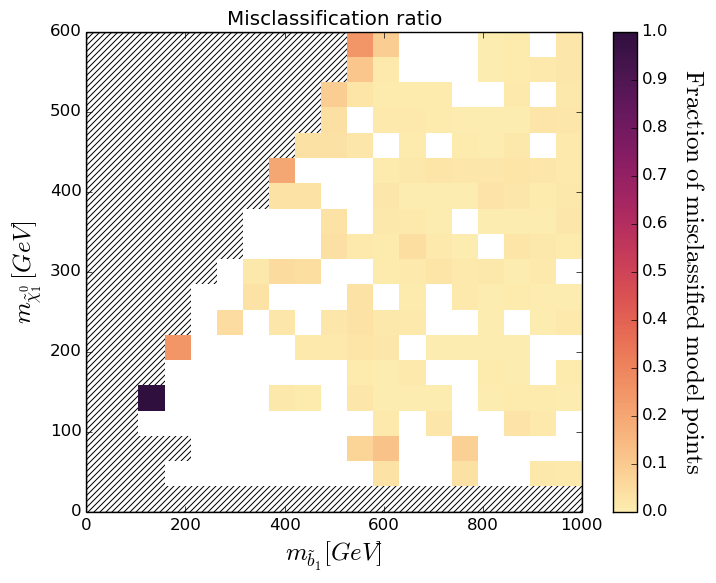}
	\end{subfigure}  
	
	\begin{subfigure}[b]{0.03\textwidth}
	    \rotatebox{90}{\hspace{2.15cm}99\-CL}
	\end{subfigure}
	\hfill
	\begin{subfigure}[b]{0.23\textwidth}
	    \includegraphics[width=\textwidth]{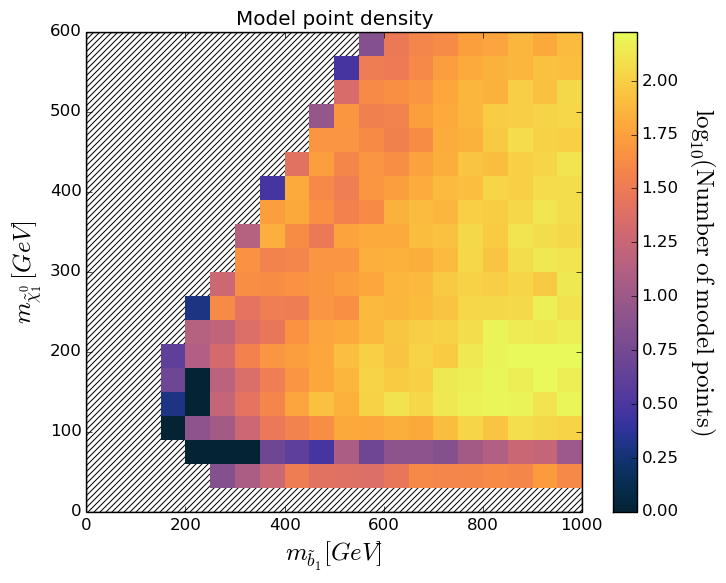}
	\end{subfigure}
	\hfill
	\begin{subfigure}[b]{0.23\textwidth}
	    \includegraphics[width=\textwidth]{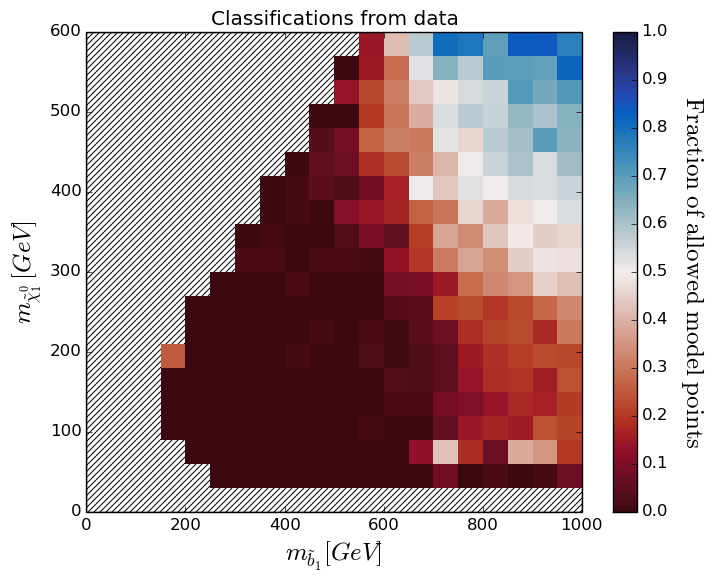}
	\end{subfigure}
	\hfill
	\begin{subfigure}[b]{0.23\textwidth}
	    \includegraphics[width=\textwidth]{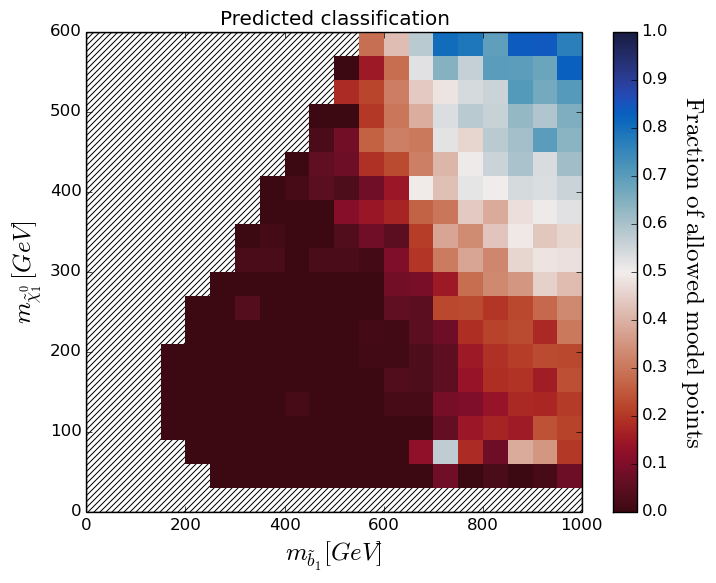}
	\end{subfigure}
	\hfill
	\begin{subfigure}[b]{0.23\textwidth}
	    \includegraphics[width=\textwidth]{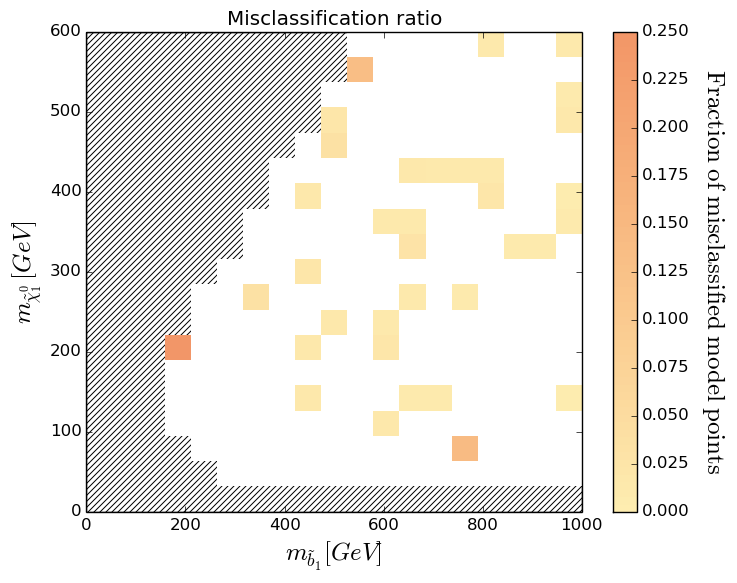}
	\end{subfigure}
	\caption{Color histograms for a projection of the 19-dimensional pMSSM parameter space on the $m_{\tilde{b}_1}$--$m_{\tilde{\chi}_1^0}$ plane. The color in the second and third column indicates the fraction of allowed data points for the true classification and the out-of-bag prediction, respectively. The last column shows the fraction of misclassified model points. The dashed bins contain no data points. Cf.\ Figure~7(a) of Ref.~\cite{Aad:2015baa}.\label{fig:appplot01}}
    \end{sidewaysfigure}    
   }

\afterpage{
     \begin{sidewaysfigure}
	\centering
	\begin{subfigure}[b]{0.03\textwidth}
	    \hspace{0.5cm}
	\end{subfigure}
	\hfill
	\begin{subfigure}[b]{0.22\textwidth}
	    \begin{center}
		Number of model points
	    \end{center}
	    \vspace{0.5cm}
	\end{subfigure}
	\hfill
	\begin{subfigure}[b]{0.22\textwidth}
	    \begin{center}
		True classification
	    \end{center}
	    \vspace{0.5cm}
	\end{subfigure}
	\hfill
	\begin{subfigure}[b]{0.22\textwidth}
	    \begin{center}
		Prediction by classifier
	    \end{center}
	    \vspace{0.5cm}
	\end{subfigure}
	\hfill
	\begin{subfigure}[b]{0.22\textwidth}
	    \begin{center}
		Ratio of misclassified points
	    \end{center}
	    \vspace{0.5cm}
	\end{subfigure}
	\hfill

	\begin{subfigure}[b]{0.03\textwidth}
      \rotatebox{90}{\hspace{2cm}All data}
      \end{subfigure}
	\hfill
	\begin{subfigure}[b]{0.23\textwidth}
	    \includegraphics[width=\textwidth]{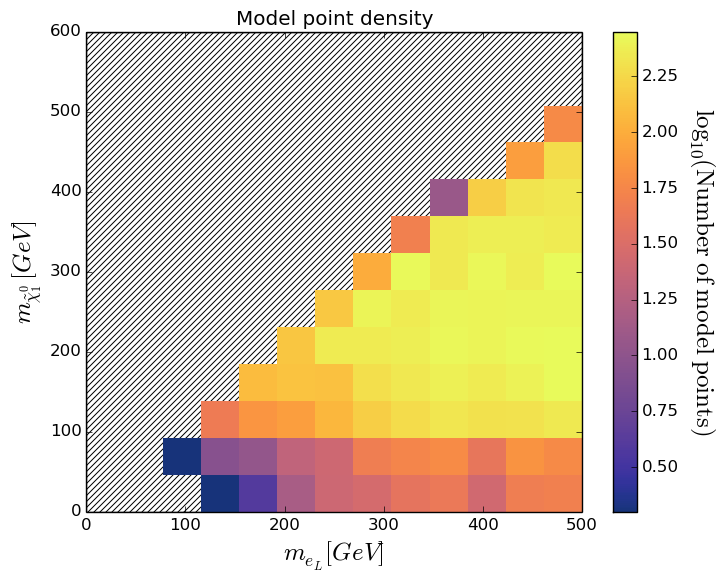}
	\end{subfigure}
	\hfill
	\begin{subfigure}[b]{0.23\textwidth}
	    \includegraphics[width=\textwidth]{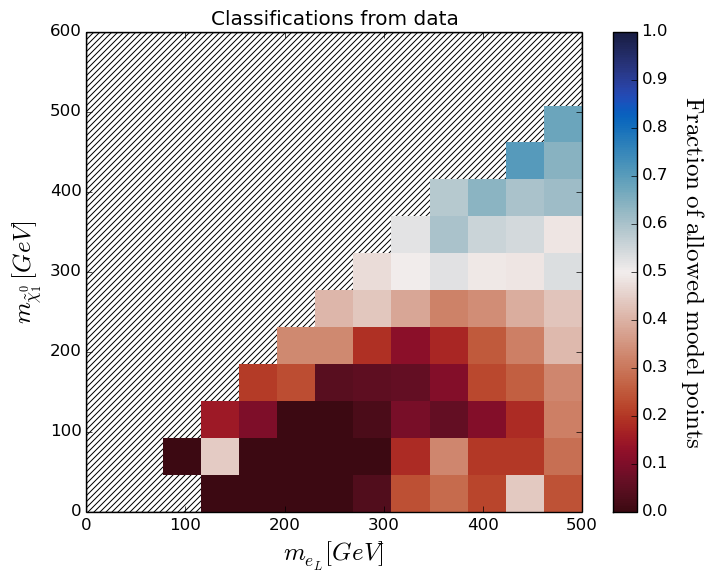}
	\end{subfigure}
	\hfill
	\begin{subfigure}[b]{0.23\textwidth}
	    \includegraphics[width=\textwidth]{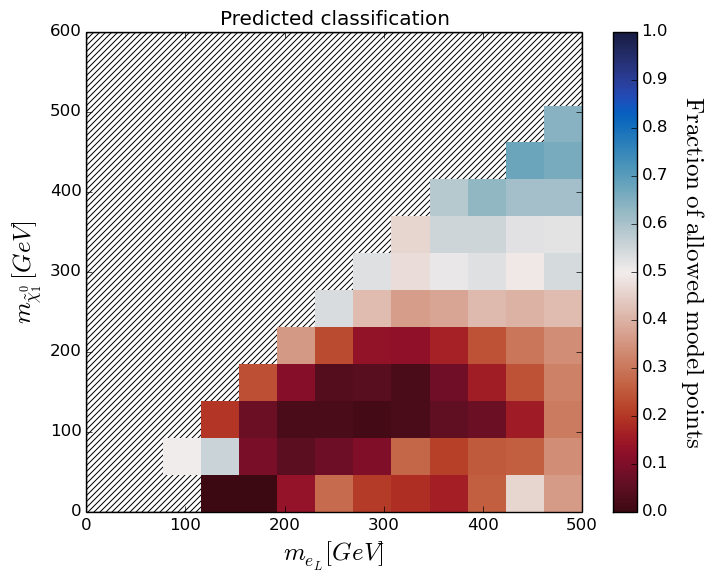}
	\end{subfigure}
	\hfill
	\begin{subfigure}[b]{0.23\textwidth}
	    \includegraphics[width=\textwidth]{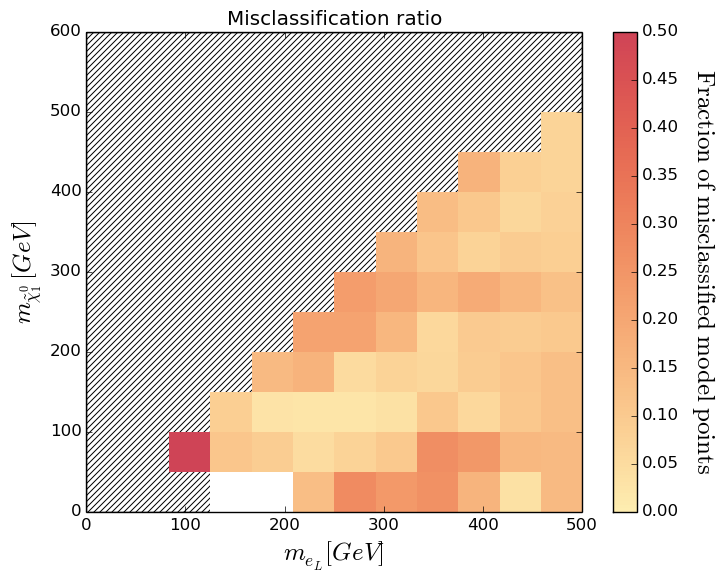}
	\end{subfigure}
	
	\begin{subfigure}[b]{0.03\textwidth}
	    \rotatebox{90}{\hspace{2.15cm}95\-CL}
	\end{subfigure}
	\hfill
	\begin{subfigure}[b]{0.23\textwidth}
	    \includegraphics[width=\textwidth]{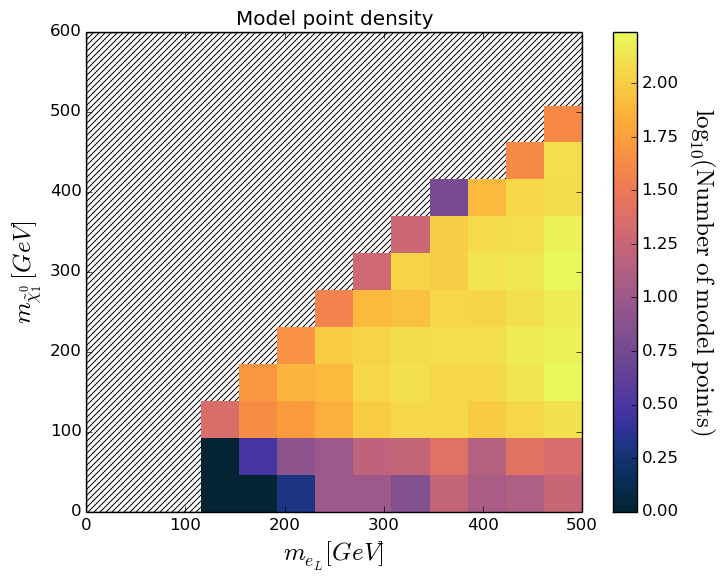}
	\end{subfigure}
	\hfill
	\begin{subfigure}[b]{0.23\textwidth}
	    \includegraphics[width=\textwidth]{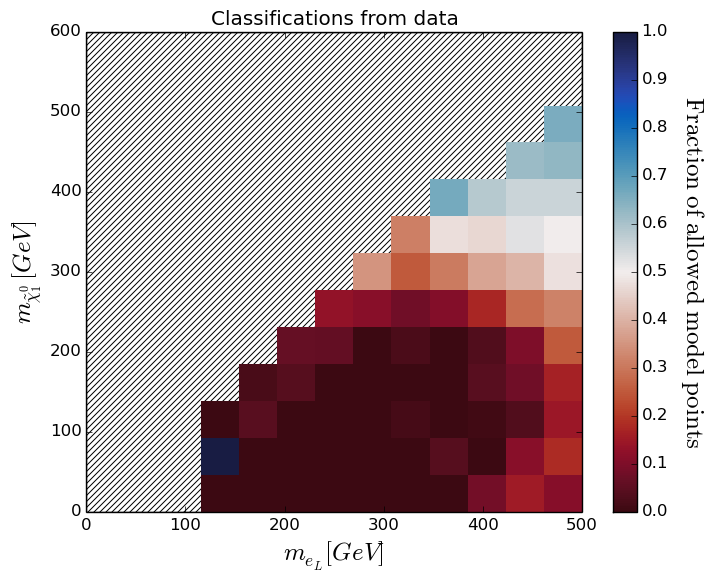}
	\end{subfigure}
	\hfill
	\begin{subfigure}[b]{0.23\textwidth}
	    \includegraphics[width=\textwidth]{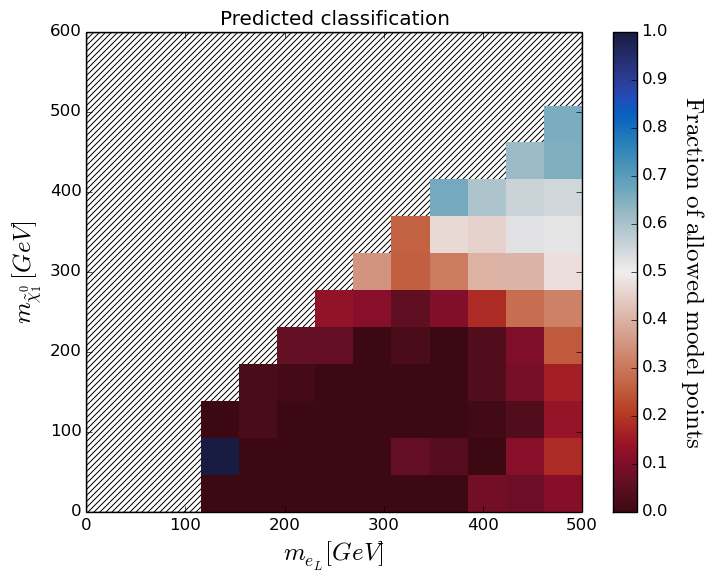}
	\end{subfigure}
	\hfill
	\begin{subfigure}[b]{0.23\textwidth}
	    \includegraphics[width=\textwidth]{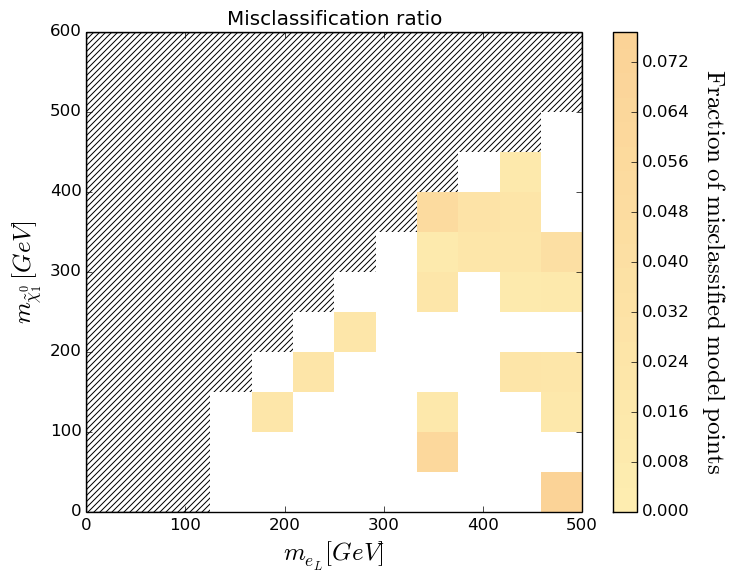}
	\end{subfigure}  
	
	\begin{subfigure}[b]{0.03\textwidth}
	    \rotatebox{90}{\hspace{2.15cm}99\-CL}
	\end{subfigure}
	\hfill
	\begin{subfigure}[b]{0.23\textwidth}
	    \includegraphics[width=\textwidth]{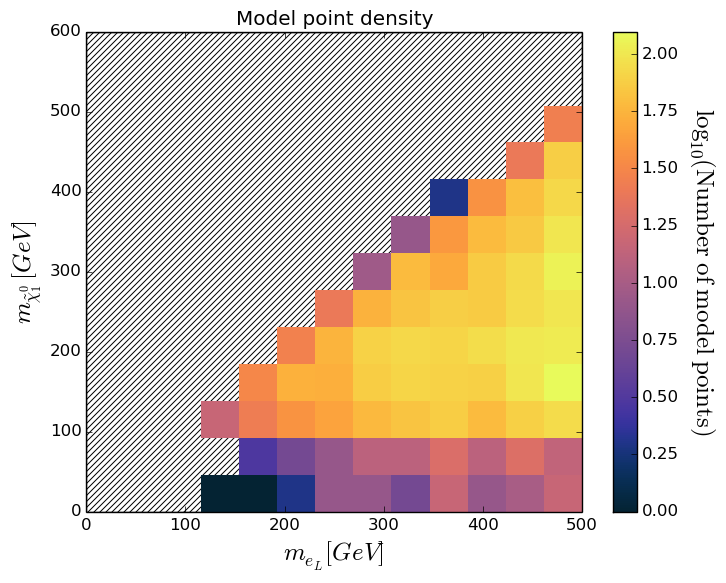}
	\end{subfigure}
	\hfill
	\begin{subfigure}[b]{0.23\textwidth}
	    \includegraphics[width=\textwidth]{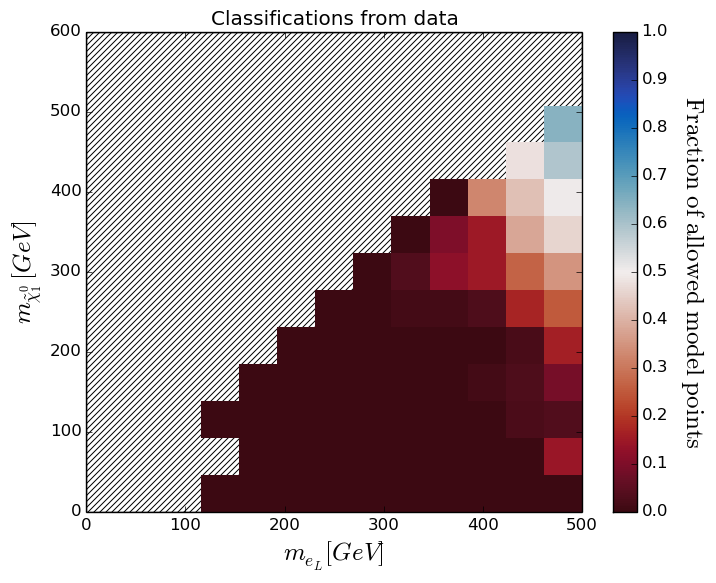}
	\end{subfigure}
	\hfill
	\begin{subfigure}[b]{0.23\textwidth}
	    \includegraphics[width=\textwidth]{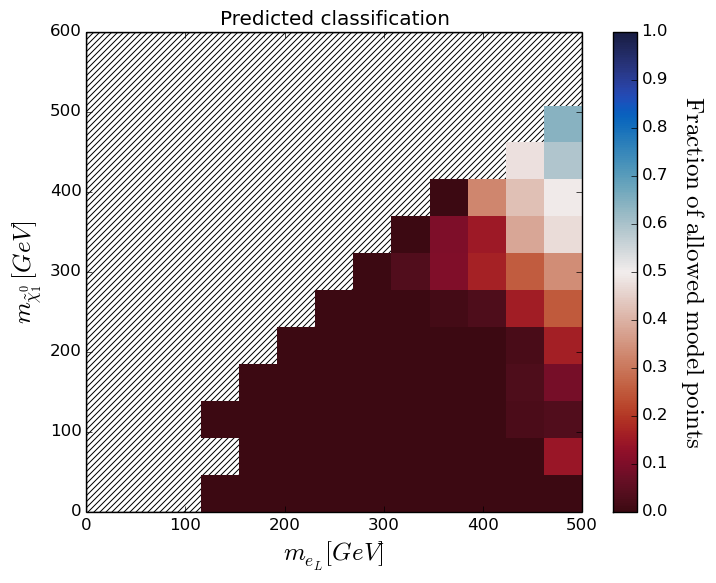}
	\end{subfigure}
	\hfill
	\begin{subfigure}[b]{0.23\textwidth}
	    \includegraphics[width=\textwidth]{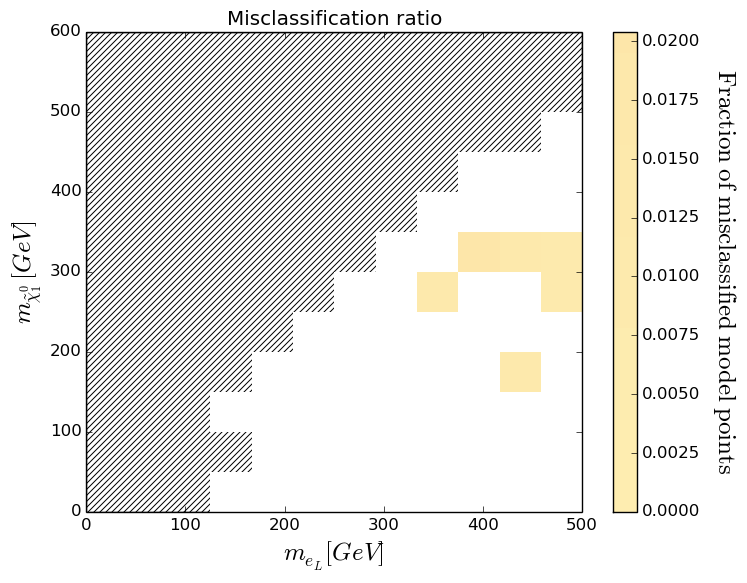}
	\end{subfigure}
	\caption{Color histograms for a projection of the 19-dimensional pMSSM parameter space on the $m_{\tilde{\ell}_L}$--$m_{\tilde{\chi}_1^0}$ plane. The color in the second and third column indicates the fraction of allowed data points for the true classification and the out-of-bag prediction, respectively. The last column shows the fraction of misclassified model points. The dashed bins contain no data points. Cf.\ Figure~9(a) of Ref.~\cite{Aad:2015baa}; however, note that the plots here take into account all searches.\label{fig:appplot02}}
    \end{sidewaysfigure}  
    }
   
\afterpage{
     \begin{sidewaysfigure}
	\centering
	\begin{subfigure}[b]{0.03\textwidth}
	    \hspace{0.5cm}
	\end{subfigure}
	\hfill
	\begin{subfigure}[b]{0.22\textwidth}
	    \begin{center}
		Number of model points
	    \end{center}
	    \vspace{0.5cm}
	\end{subfigure}
	\hfill
	\begin{subfigure}[b]{0.22\textwidth}
	    \begin{center}
		True classification
	    \end{center}
	    \vspace{0.5cm}
	\end{subfigure}
	\hfill
	\begin{subfigure}[b]{0.22\textwidth}
	    \begin{center}
		Prediction by classifier
	    \end{center}
	    \vspace{0.5cm}
	\end{subfigure}
	\hfill
	\begin{subfigure}[b]{0.22\textwidth}
	    \begin{center}
		Ratio of misclassified points
	    \end{center}
	    \vspace{0.5cm}
	\end{subfigure}
	\hfill

	\begin{subfigure}[b]{0.03\textwidth}
      \rotatebox{90}{\hspace{2cm}All data}
      \end{subfigure}
	\hfill
	\begin{subfigure}[b]{0.23\textwidth}
	    \includegraphics[width=\textwidth]{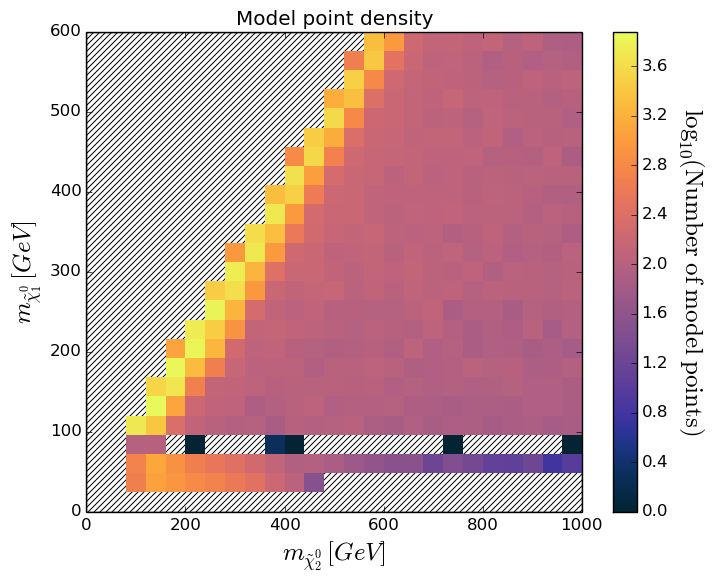}
	\end{subfigure}
	\hfill
	\begin{subfigure}[b]{0.23\textwidth}
	    \includegraphics[width=\textwidth]{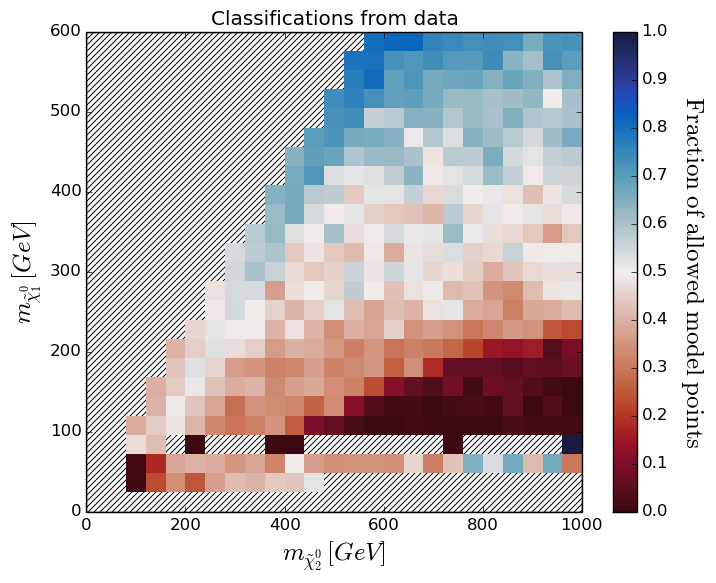}
	\end{subfigure}
	\hfill
	\begin{subfigure}[b]{0.23\textwidth}
	    \includegraphics[width=\textwidth]{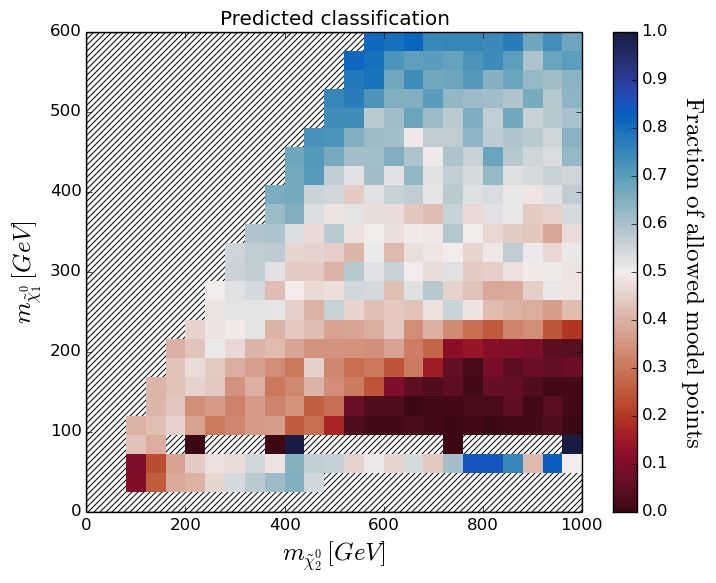}
	\end{subfigure}
	\hfill
	\begin{subfigure}[b]{0.23\textwidth}
	    \includegraphics[width=\textwidth]{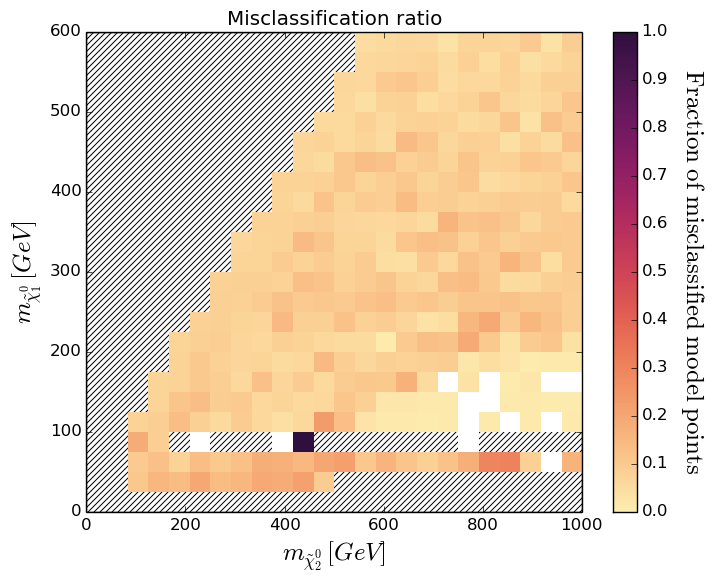}
	\end{subfigure}
	
	\begin{subfigure}[b]{0.03\textwidth}
	    \rotatebox{90}{\hspace{2.15cm}95\-CL}
	\end{subfigure}
	\hfill
	\begin{subfigure}[b]{0.23\textwidth}
	    \includegraphics[width=\textwidth]{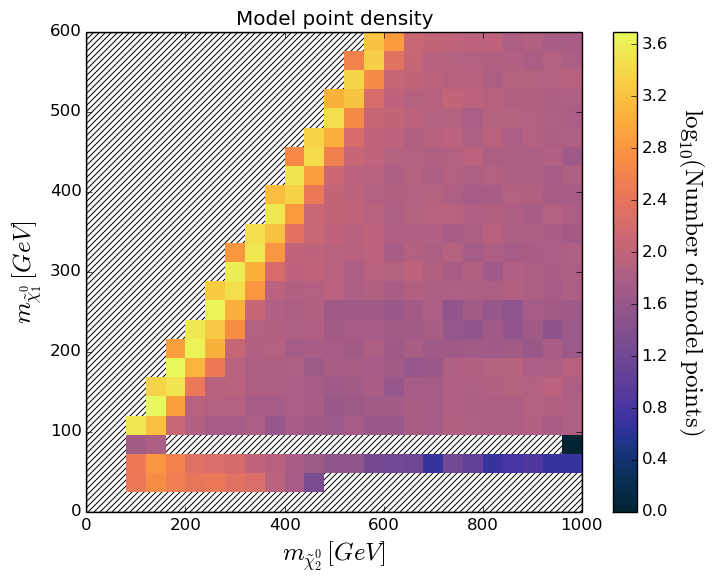}
	\end{subfigure}
	\hfill
	\begin{subfigure}[b]{0.23\textwidth}
	    \includegraphics[width=\textwidth]{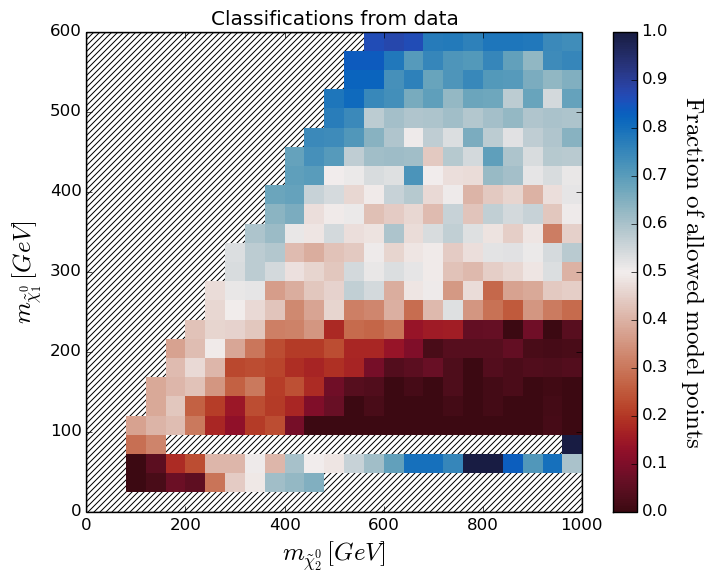}
	\end{subfigure}
	\hfill
	\begin{subfigure}[b]{0.23\textwidth}
	    \includegraphics[width=\textwidth]{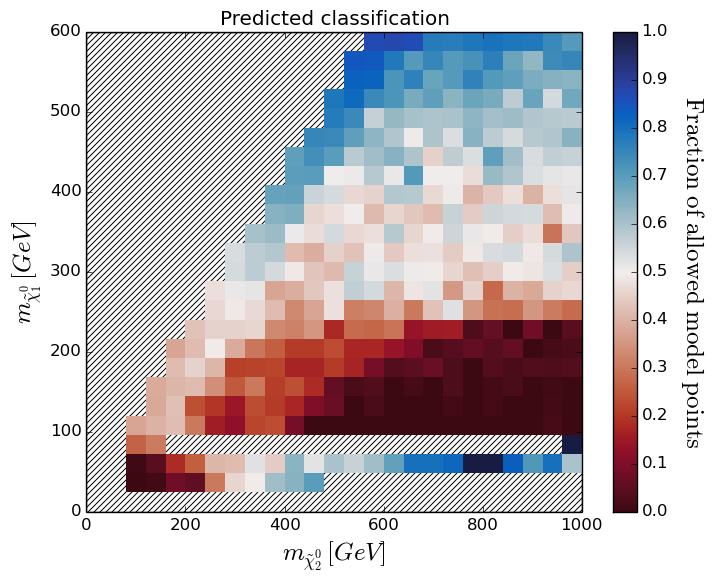}
	\end{subfigure}
	\hfill
	\begin{subfigure}[b]{0.23\textwidth}
	    \includegraphics[width=\textwidth]{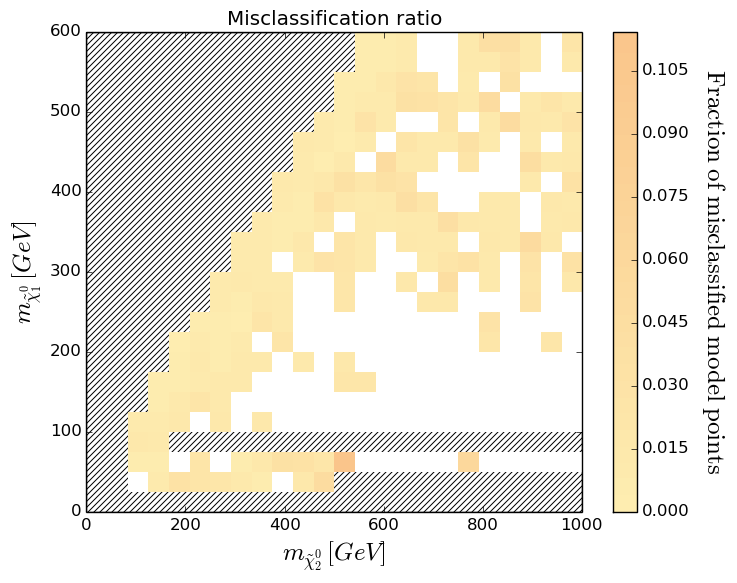}
	\end{subfigure}  
	
	\begin{subfigure}[b]{0.03\textwidth}
	    \rotatebox{90}{\hspace{2.15cm}99\-CL}
	\end{subfigure}
	\hfill
	\begin{subfigure}[b]{0.23\textwidth}
	    \includegraphics[width=\textwidth]{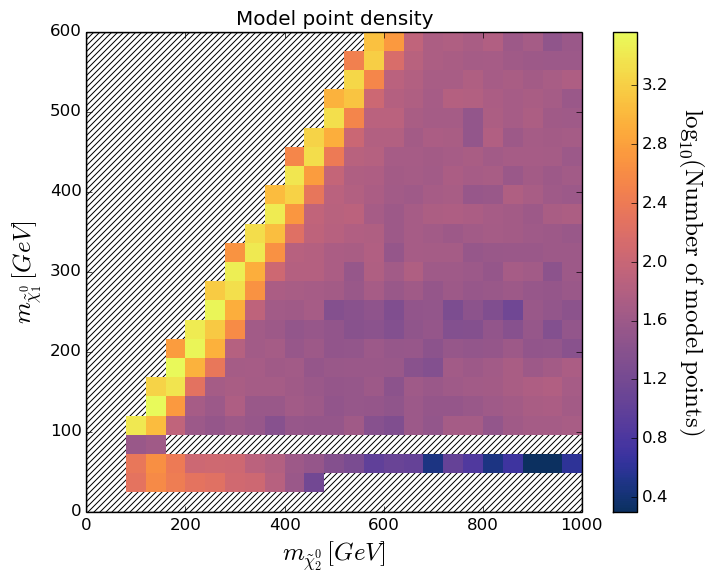}
	\end{subfigure}
	\hfill
	\begin{subfigure}[b]{0.23\textwidth}
	    \includegraphics[width=\textwidth]{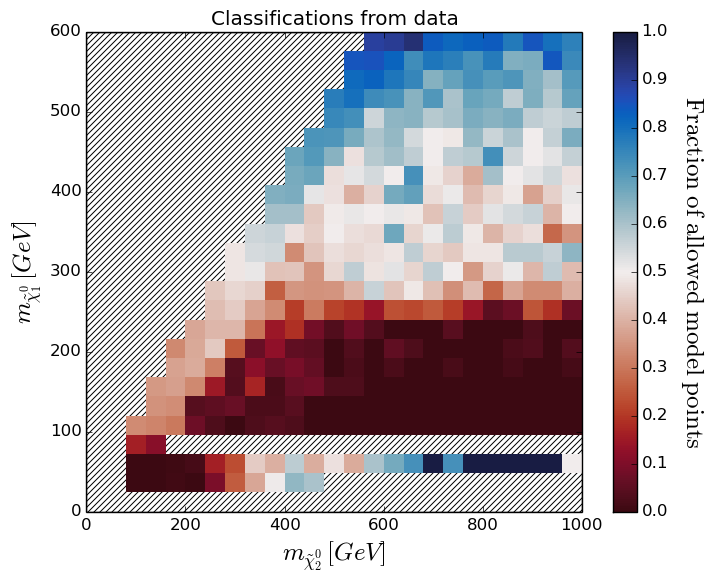}
	\end{subfigure}
	\hfill
	\begin{subfigure}[b]{0.23\textwidth}
	    \includegraphics[width=\textwidth]{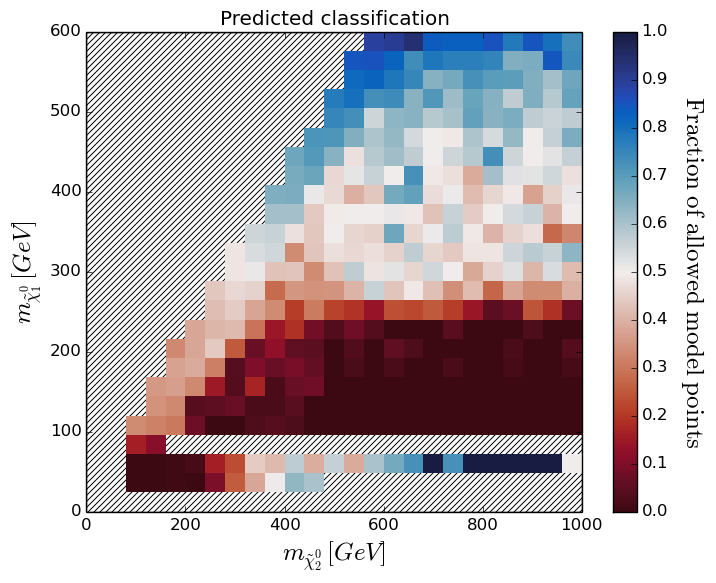}
	\end{subfigure}
	\hfill
	\begin{subfigure}[b]{0.23\textwidth}
	    \includegraphics[width=\textwidth]{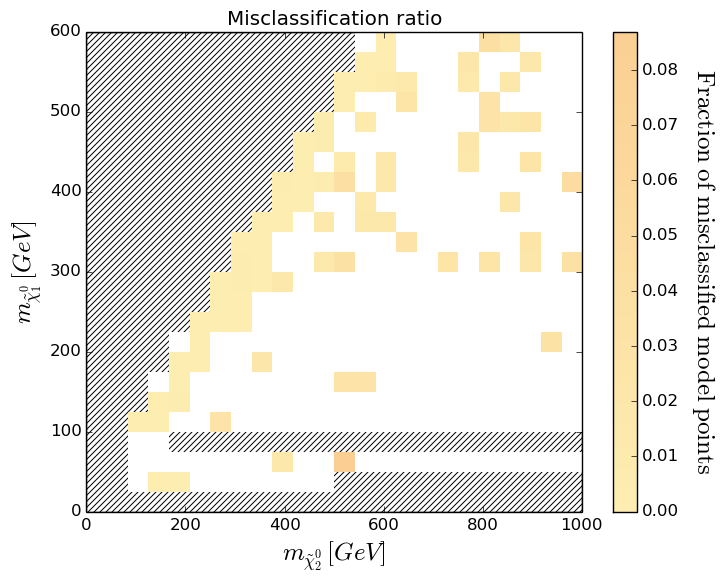}
	\end{subfigure}
	\caption{Color histograms for a projection of the 19-dimensional pMSSM parameter space on the $m_{\tilde{\chi}_2^0}$--$m_{\tilde{\chi}_1^0}$ plane. The color in the second and third column indicates the fraction of allowed data points for the true classification and the out-of-bag prediction, respectively. The last column shows the fraction of misclassified model points. The dashed bins contain no data points. Cf.\ Figure~11(a) of Ref.~\cite{Aad:2015baa}; however note that the plots here take into account all searches.\label{fig:appplot03}}
    \end{sidewaysfigure}    
}
   
\afterpage{
     \begin{sidewaysfigure}
	\centering
	\begin{subfigure}[b]{0.03\textwidth}
	    \hspace{0.5cm}
	\end{subfigure}
	\hfill
	\begin{subfigure}[b]{0.22\textwidth}
	    \begin{center}
		Number of model points
	    \end{center}
	    \vspace{0.5cm}
	\end{subfigure}
	\hfill
	\begin{subfigure}[b]{0.22\textwidth}
	    \begin{center}
		True classification
	    \end{center}
	    \vspace{0.5cm}
	\end{subfigure}
	\hfill
	\begin{subfigure}[b]{0.22\textwidth}
	    \begin{center}
		Prediction by classifier
	    \end{center}
	    \vspace{0.5cm}
	\end{subfigure}
	\hfill
	\begin{subfigure}[b]{0.22\textwidth}
	    \begin{center}
		Ratio of misclassified points
	    \end{center}
	    \vspace{0.5cm}
	\end{subfigure}
	\hfill

	\begin{subfigure}[b]{0.03\textwidth}
      \rotatebox{90}{\hspace{2cm}All data}
      \end{subfigure}
	\hfill
	\begin{subfigure}[b]{0.23\textwidth}
	    \includegraphics[width=\textwidth]{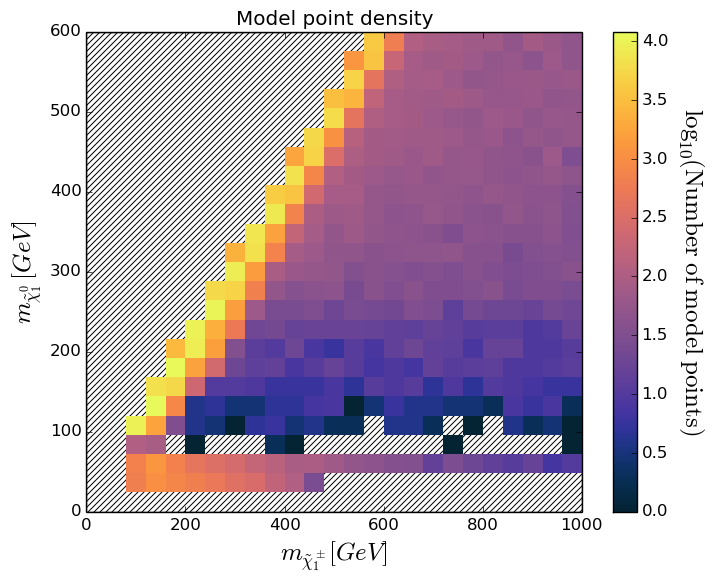}
	\end{subfigure}
	\hfill
	\begin{subfigure}[b]{0.23\textwidth}
	    \includegraphics[width=\textwidth]{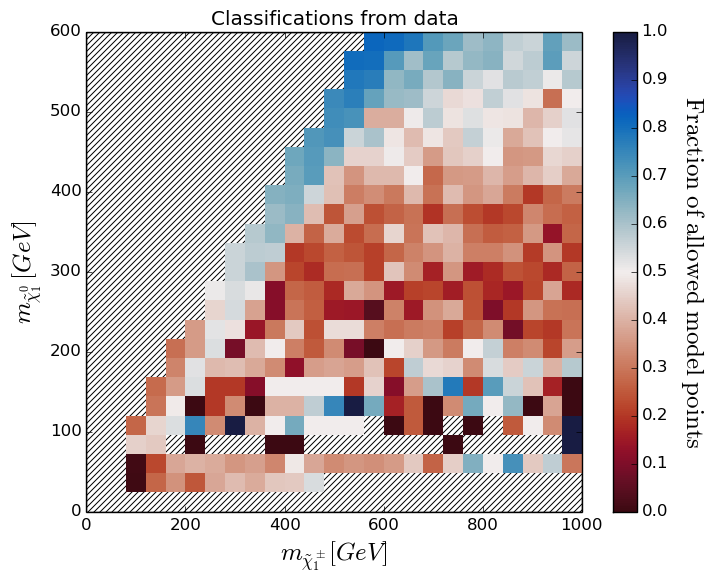}
	\end{subfigure}
	\hfill
	\begin{subfigure}[b]{0.23\textwidth}
	    \includegraphics[width=\textwidth]{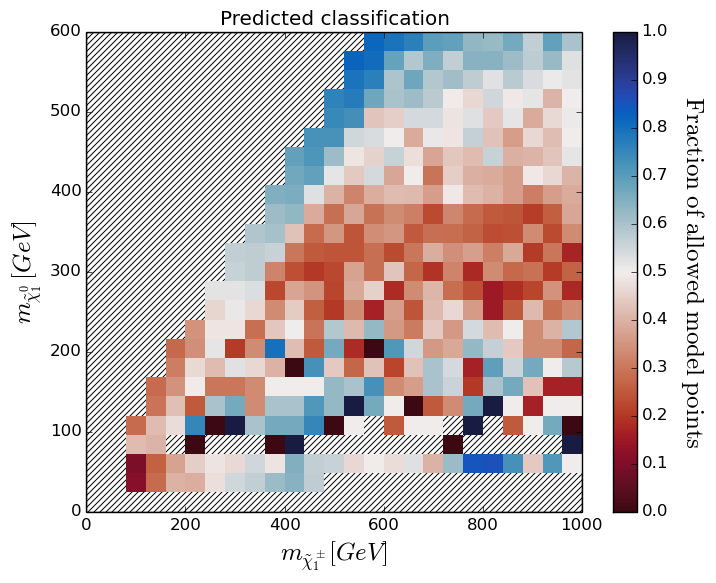}
	\end{subfigure}
	\hfill
	\begin{subfigure}[b]{0.23\textwidth}
	    \includegraphics[width=\textwidth]{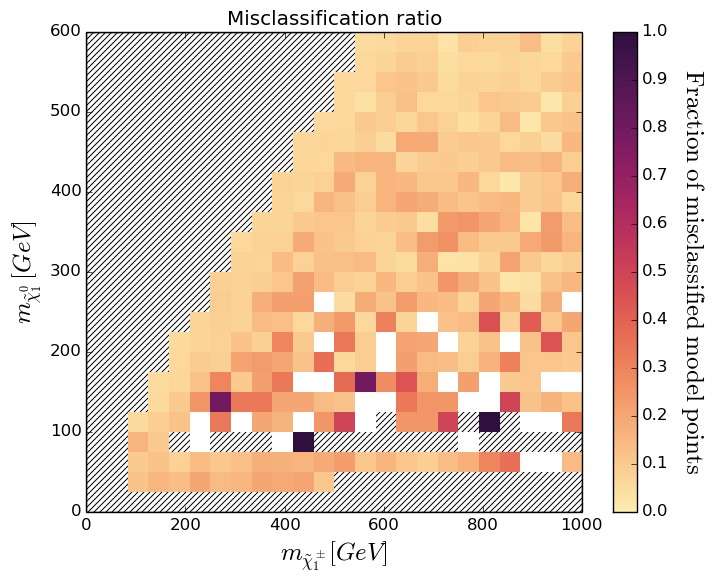}
	\end{subfigure}
	
	\begin{subfigure}[b]{0.03\textwidth}
	    \rotatebox{90}{\hspace{2.15cm}95\-CL}
	\end{subfigure}
	\hfill
	\begin{subfigure}[b]{0.23\textwidth}
	    \includegraphics[width=\textwidth]{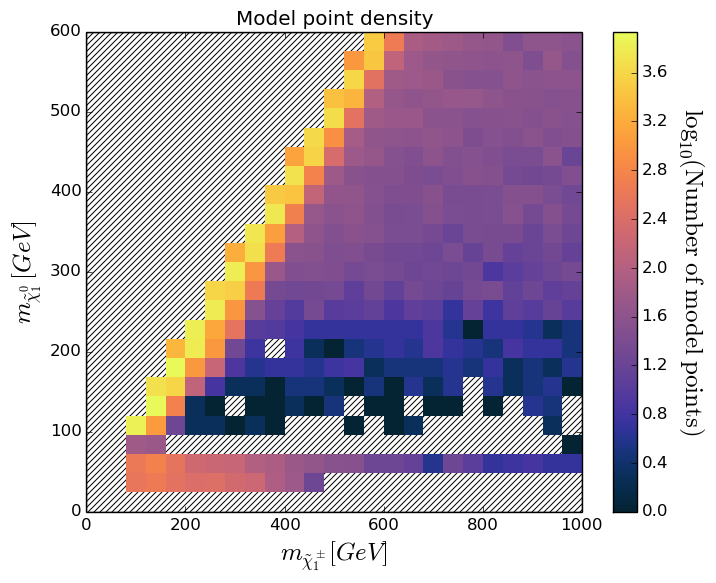}
	\end{subfigure}
	\hfill
	\begin{subfigure}[b]{0.23\textwidth}
	    \includegraphics[width=\textwidth]{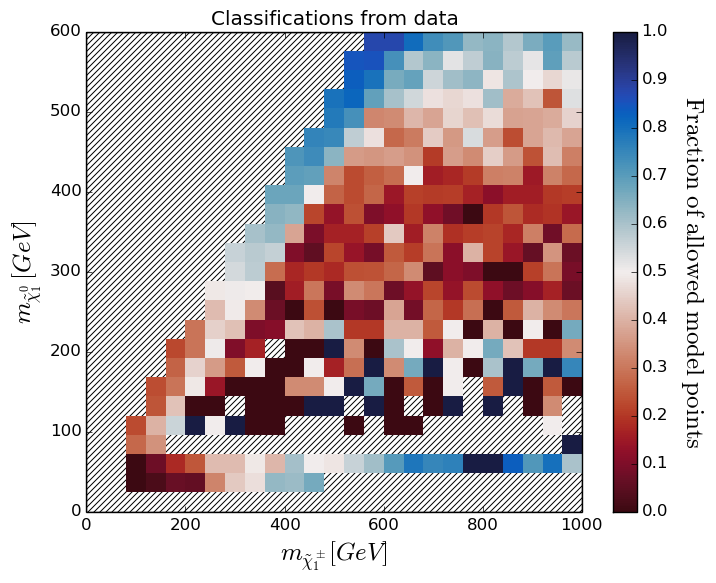}
	\end{subfigure}
	\hfill
	\begin{subfigure}[b]{0.23\textwidth}
	    \includegraphics[width=\textwidth]{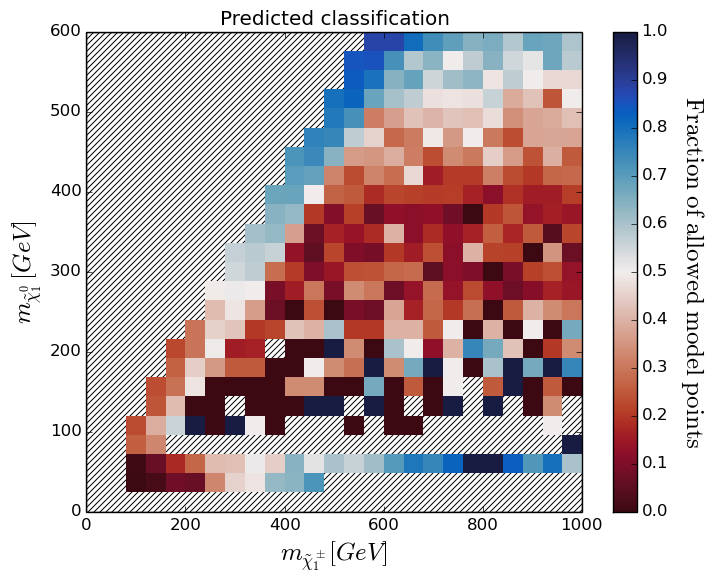}
	\end{subfigure}
	\hfill
	\begin{subfigure}[b]{0.23\textwidth}
	    \includegraphics[width=\textwidth]{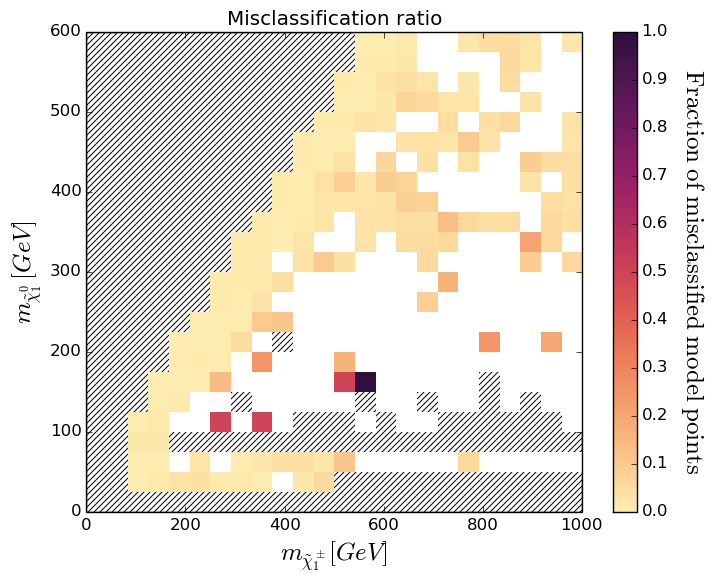}
	\end{subfigure}  
	
	\begin{subfigure}[b]{0.03\textwidth}
	    \rotatebox{90}{\hspace{2.15cm}99\-CL}
	\end{subfigure}
	\hfill
	\begin{subfigure}[b]{0.23\textwidth}
	    \includegraphics[width=\textwidth]{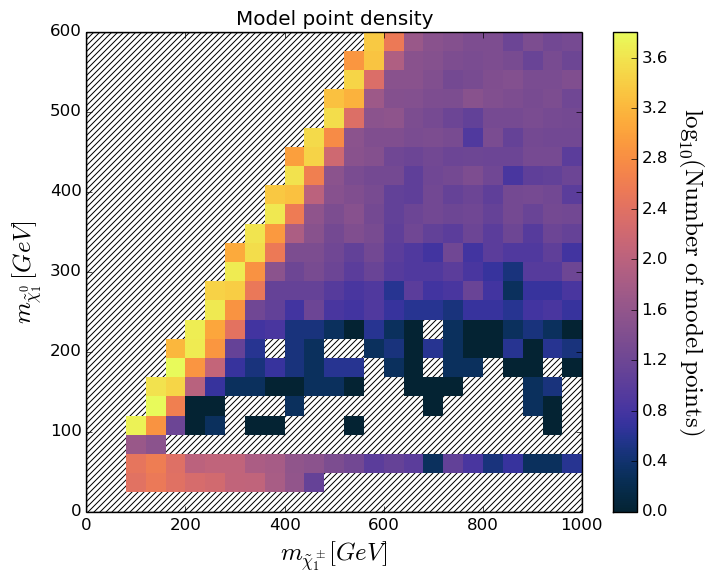}
	\end{subfigure}
	\hfill
	\begin{subfigure}[b]{0.23\textwidth}
	    \includegraphics[width=\textwidth]{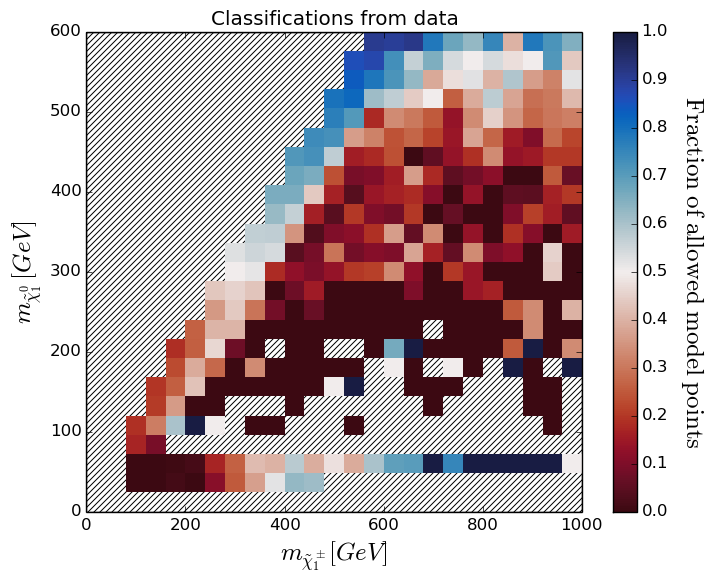}
	\end{subfigure}
	\hfill
	\begin{subfigure}[b]{0.23\textwidth}
	    \includegraphics[width=\textwidth]{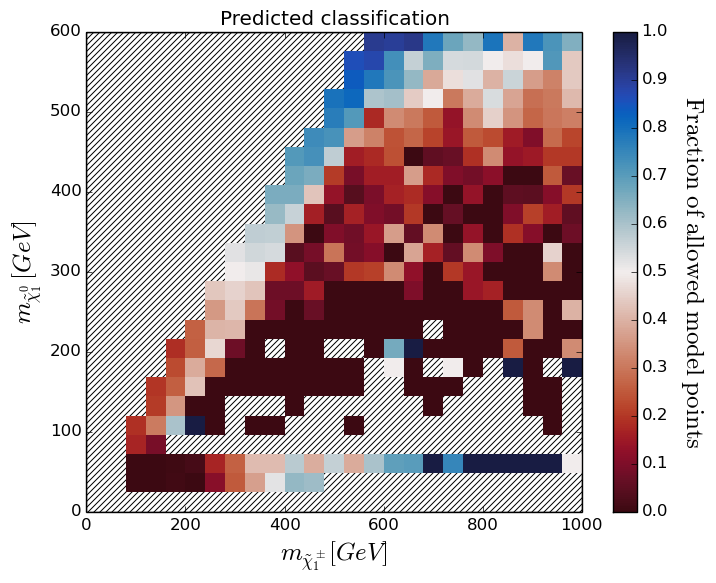}
	\end{subfigure}
	\hfill
	\begin{subfigure}[b]{0.23\textwidth}
	    \includegraphics[width=\textwidth]{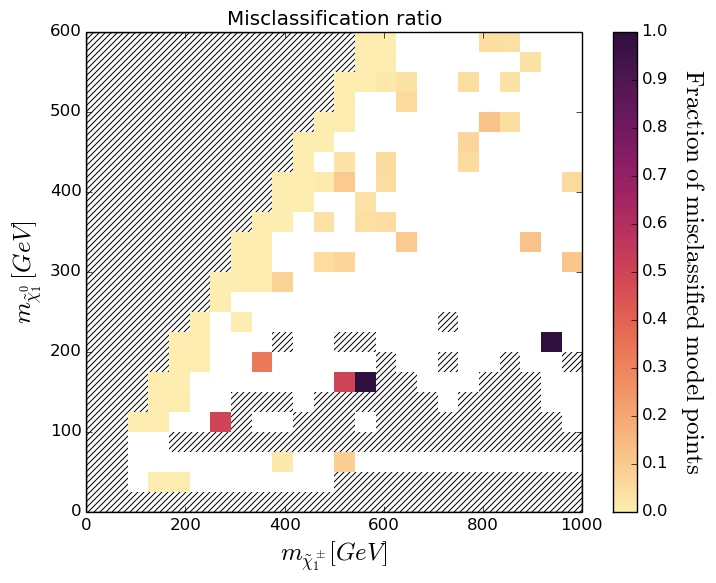}
	\end{subfigure}
	\caption{Color histograms for a projection of the 19-dimensional pMSSM parameter space on the $m_{\tilde{\chi}_1^\pm}$--$m_{\tilde{\chi}_1^0}$ plane. The color in the second and third column indicates the fraction of allowed data points for the true classification and the out-of-bag prediction. respectively. The last column shows the fraction of misclassified model points. The dashed bins contain no data points. Cf.\ Figure~11(b) of Ref.~\cite{Aad:2015baa}; however note that the plots here take into account all searches.\label{fig:appplot04}}
    \end{sidewaysfigure}
    }
    
\afterpage{     
  \begin{sidewaysfigure}
	\centering
	\begin{subfigure}[b]{0.03\textwidth}
	    \hspace{0.5cm}
	\end{subfigure}
	\hfill
	\begin{subfigure}[b]{0.22\textwidth}
	    \begin{center}
		Number of model points
	    \end{center}
	    \vspace{0.5cm}
	\end{subfigure}
	\hfill
	\begin{subfigure}[b]{0.22\textwidth}
	    \begin{center}
		True classification
	    \end{center}
	    \vspace{0.5cm}
	\end{subfigure}
	\hfill
	\begin{subfigure}[b]{0.22\textwidth}
	    \begin{center}
		Prediction by classifier
	    \end{center}
	    \vspace{0.5cm}
	\end{subfigure}
	\hfill
	\begin{subfigure}[b]{0.22\textwidth}
	    \begin{center}
		Ratio of misclassified points
	    \end{center}
	    \vspace{0.5cm}
	\end{subfigure}
	\hfill

	\begin{subfigure}[b]{0.03\textwidth}
      \rotatebox{90}{\hspace{2cm}All data}
      \end{subfigure}
	\hfill
	\begin{subfigure}[b]{0.23\textwidth}
	    \includegraphics[width=\textwidth]{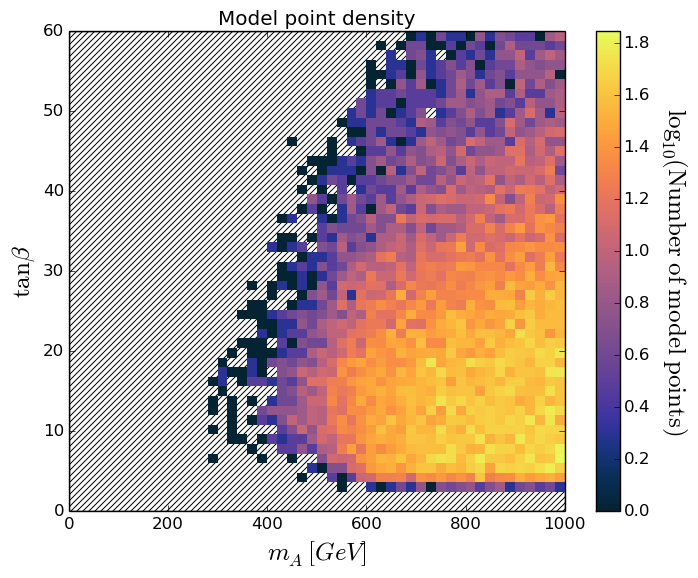}
	\end{subfigure}
	\hfill
	\begin{subfigure}[b]{0.23\textwidth}
	    \includegraphics[width=\textwidth]{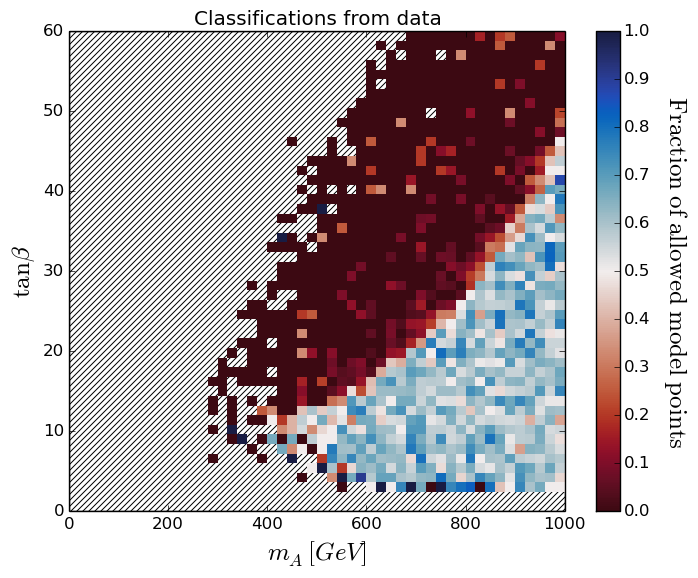}
	\end{subfigure}
	\hfill
	\begin{subfigure}[b]{0.23\textwidth}
	    \includegraphics[width=\textwidth]{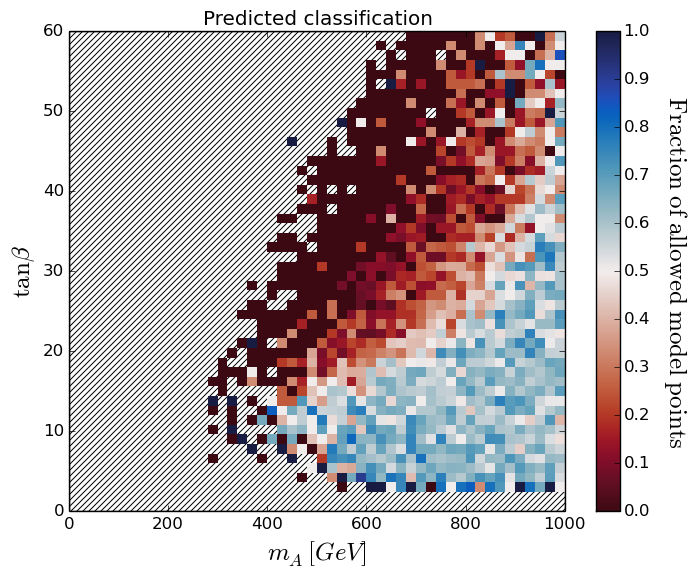}
	\end{subfigure}
	\hfill
	\begin{subfigure}[b]{0.23\textwidth}
	    \includegraphics[width=\textwidth]{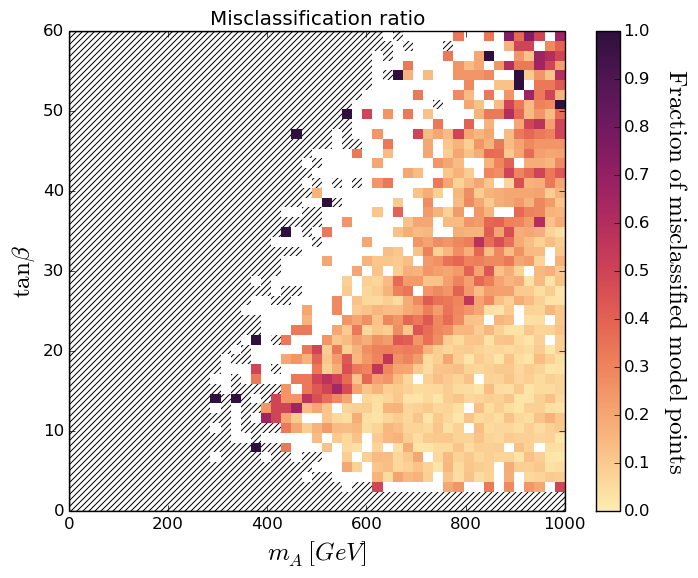}
	\end{subfigure}
	
	\begin{subfigure}[b]{0.03\textwidth}
	    \rotatebox{90}{\hspace{2.15cm}95\-CL}
	\end{subfigure}
	\hfill
	\begin{subfigure}[b]{0.23\textwidth}
	    \includegraphics[width=\textwidth]{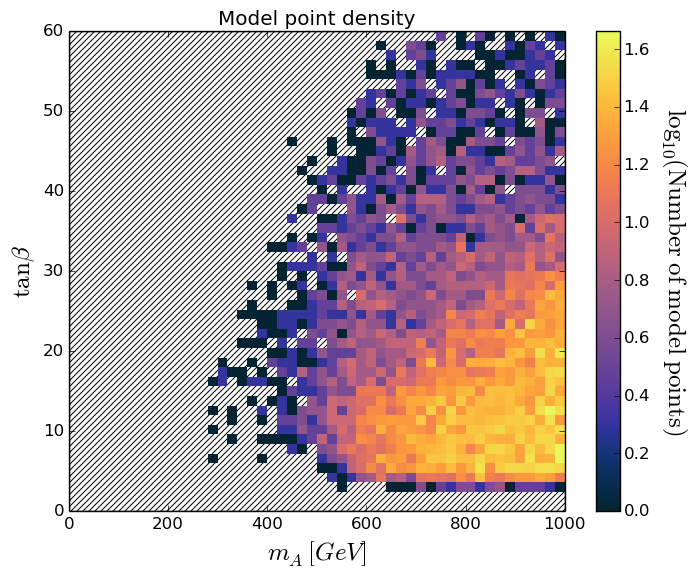}
	\end{subfigure}
	\hfill
	\begin{subfigure}[b]{0.23\textwidth}
	    \includegraphics[width=\textwidth]{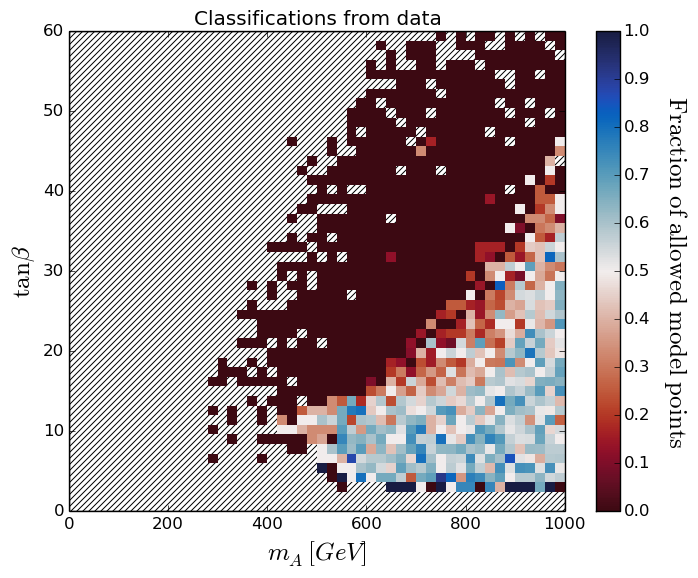}
	\end{subfigure}
	\hfill
	\begin{subfigure}[b]{0.23\textwidth}
	    \includegraphics[width=\textwidth]{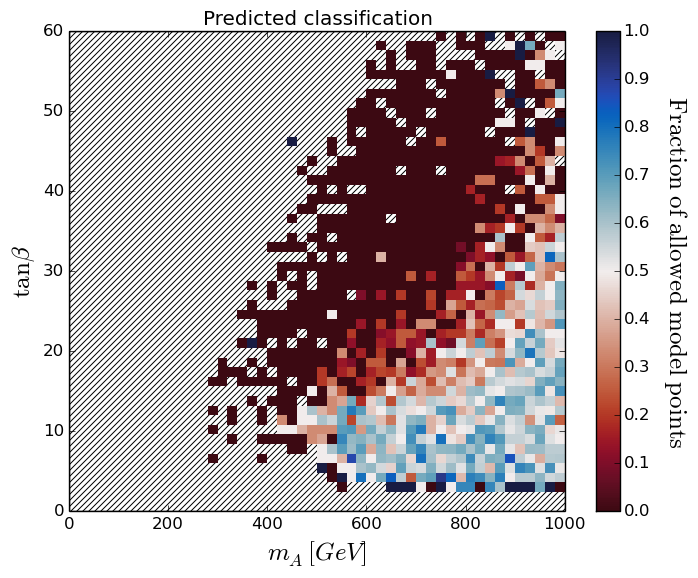}
	\end{subfigure}
	\hfill
	\begin{subfigure}[b]{0.23\textwidth}
	    \includegraphics[width=\textwidth]{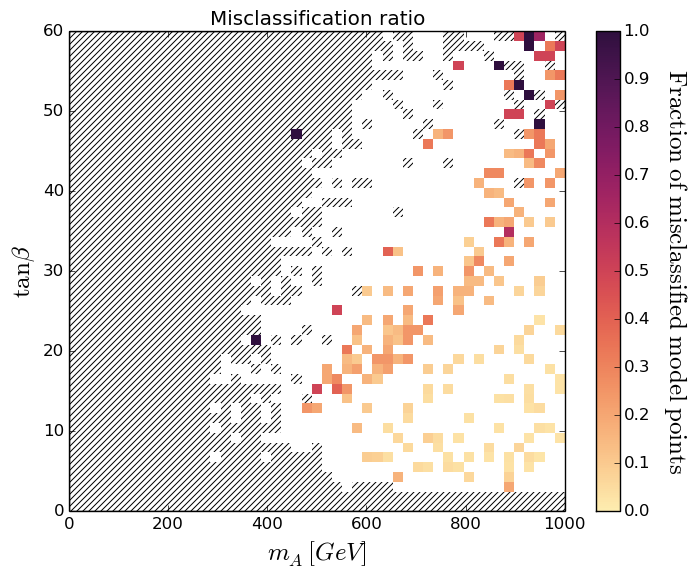}
	\end{subfigure}  
	
	\begin{subfigure}[b]{0.03\textwidth}
	    \rotatebox{90}{\hspace{2.15cm}99\-CL}
	\end{subfigure}
	\hfill
	\begin{subfigure}[b]{0.23\textwidth}
	    \includegraphics[width=\textwidth]{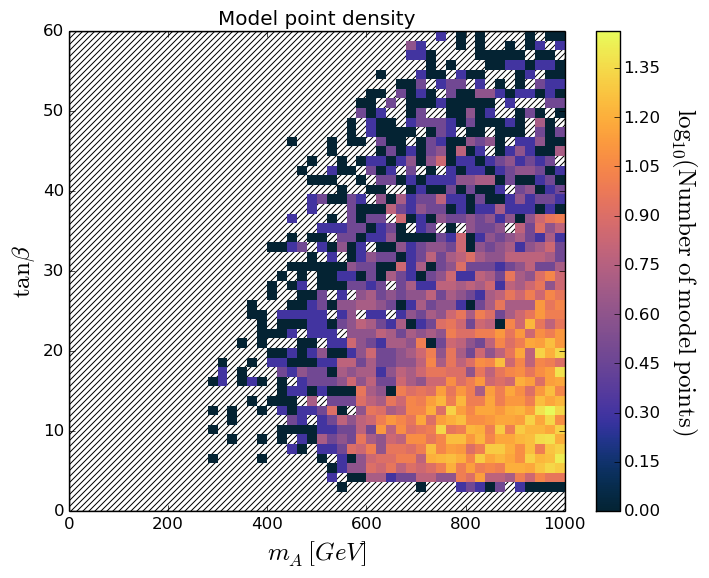}
	\end{subfigure}
	\hfill
	\begin{subfigure}[b]{0.23\textwidth}
	    \includegraphics[width=\textwidth]{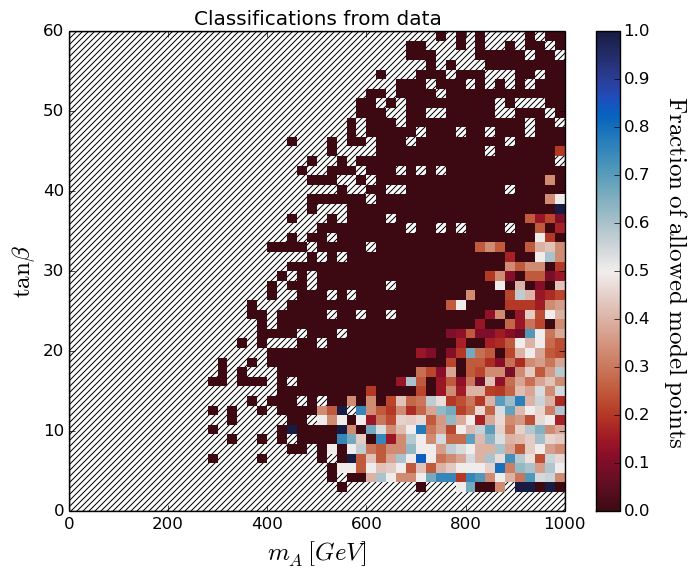}
	\end{subfigure}
	\hfill
	\begin{subfigure}[b]{0.23\textwidth}
	    \includegraphics[width=\textwidth]{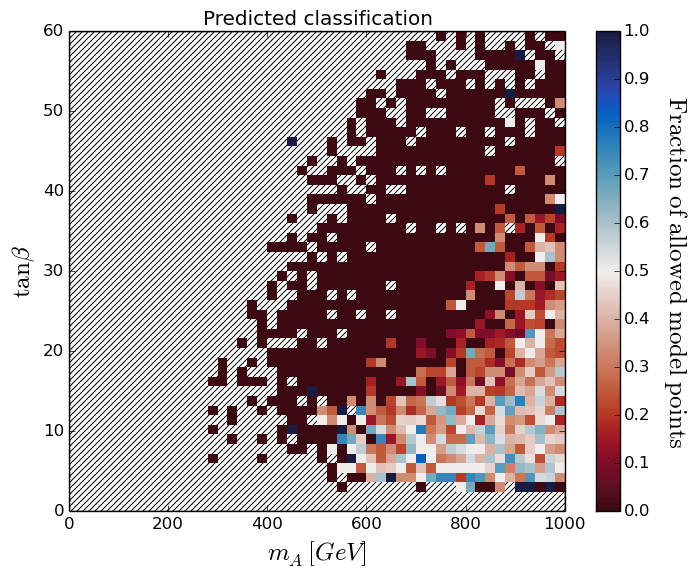}
	\end{subfigure}
	\hfill
	\begin{subfigure}[b]{0.23\textwidth}
	    \includegraphics[width=\textwidth]{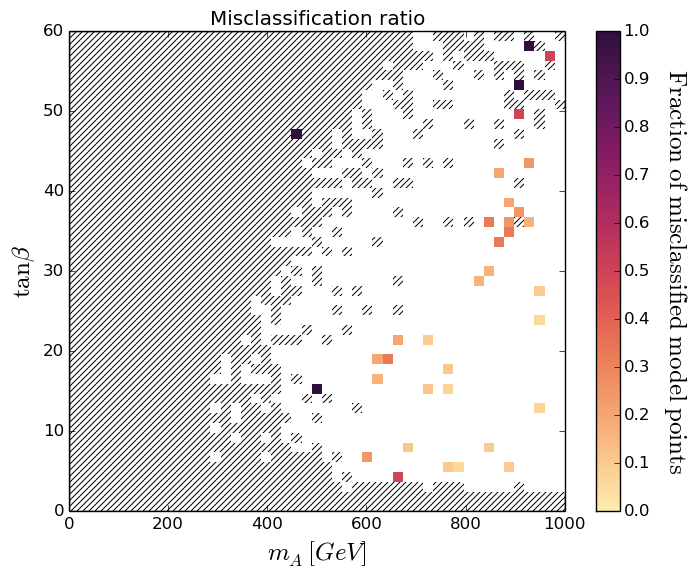}
	\end{subfigure}
	\caption{Color histograms for a projection of the 19-dimensional pMSSM parameter space on the $m_{A^0}$--$\tan\beta$ plane. The color in the second and third column indicates the fraction of allowed data points for the true classification and the out-of-bag prediction, respectively. The last column shows the fraction of misclassified model points. The dashed bins contain no data points. Cf.\ Figure~12 of Ref.~\cite{Aad:2015baa}.\label{fig:appplot05}}
    \end{sidewaysfigure}    
    }

\afterpage{
    \begin{sidewaysfigure}
	\centering
	\begin{subfigure}[b]{0.03\textwidth}
	    \hspace{0.5cm}
	\end{subfigure}
	\hfill
	\begin{subfigure}[b]{0.22\textwidth}
	    \begin{center}
		Number of model points
	    \end{center}
	    \vspace{0.5cm}
	\end{subfigure}
	\hfill
	\begin{subfigure}[b]{0.22\textwidth}
	    \begin{center}
		True classification
	    \end{center}
	    \vspace{0.5cm}
	\end{subfigure}
	\hfill
	\begin{subfigure}[b]{0.22\textwidth}
	    \begin{center}
		Prediction by classifier
	    \end{center}
	    \vspace{0.5cm}
	\end{subfigure}
	\hfill
	\begin{subfigure}[b]{0.22\textwidth}
	    \begin{center}
		Ratio of misclassified points
	    \end{center}
	    \vspace{0.5cm}
	\end{subfigure}
	\hfill

	\begin{subfigure}[b]{0.03\textwidth}
      \rotatebox{90}{\hspace{2cm}All data}
      \end{subfigure}
	\hfill
	\begin{subfigure}[b]{0.23\textwidth}
	    \includegraphics[width=\textwidth]{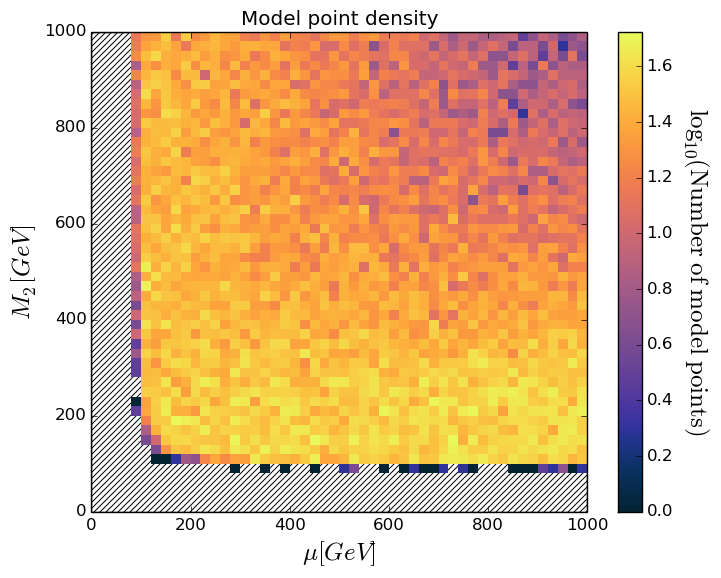}
	\end{subfigure}
	\hfill
	\begin{subfigure}[b]{0.23\textwidth}
	    \includegraphics[width=\textwidth]{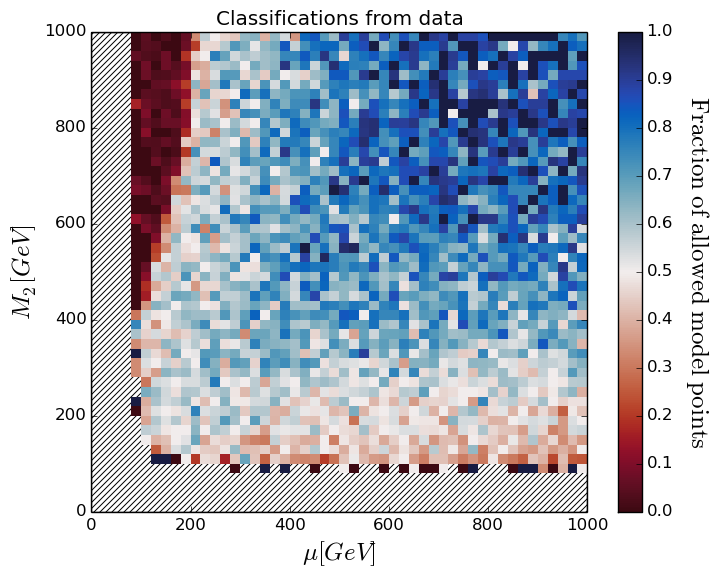}
	\end{subfigure}
	\hfill
	\begin{subfigure}[b]{0.23\textwidth}
	    \includegraphics[width=\textwidth]{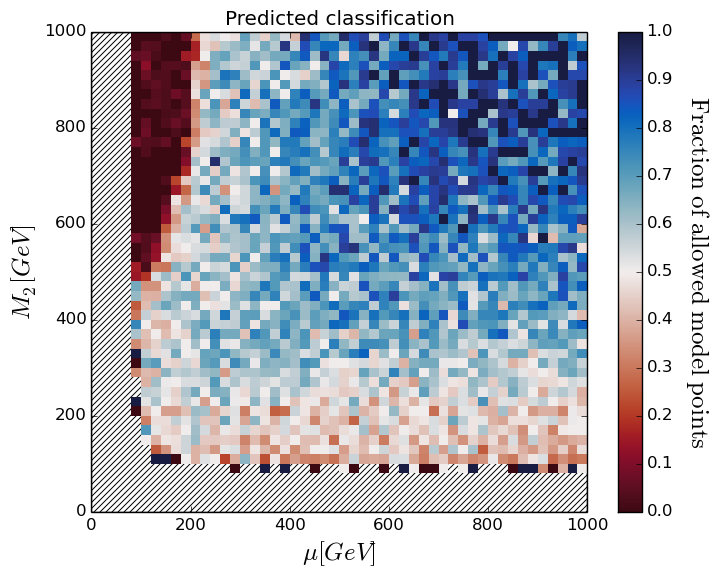}
	\end{subfigure}
	\hfill
	\begin{subfigure}[b]{0.23\textwidth}
	    \includegraphics[width=\textwidth]{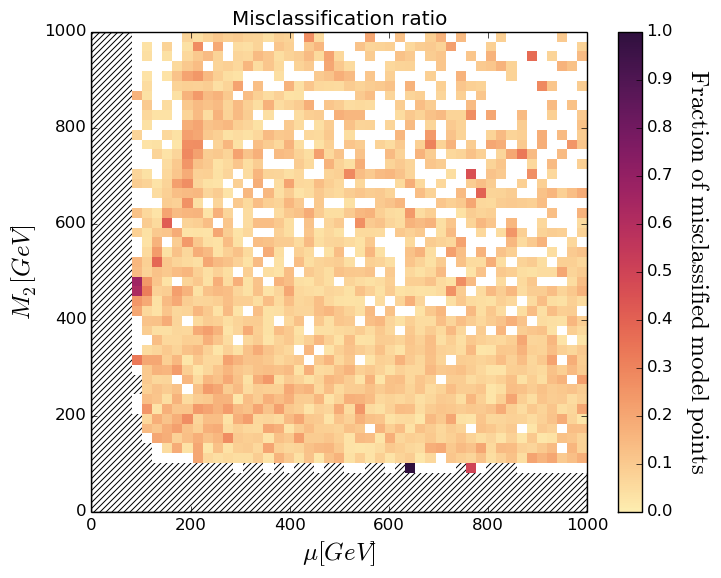}
	\end{subfigure}
	
	\begin{subfigure}[b]{0.03\textwidth}
	    \rotatebox{90}{\hspace{2.15cm}95\-CL}
	\end{subfigure}
	\hfill
	\begin{subfigure}[b]{0.23\textwidth}
	    \includegraphics[width=\textwidth]{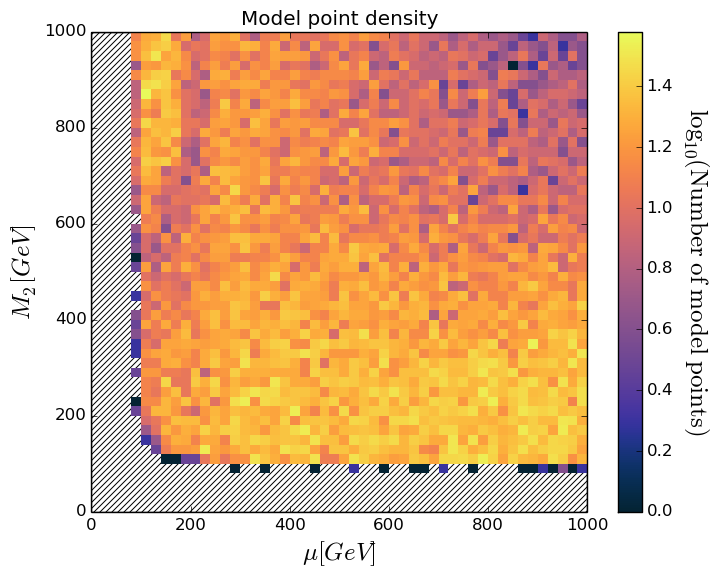}
	\end{subfigure}
	\hfill
	\begin{subfigure}[b]{0.23\textwidth}
	    \includegraphics[width=\textwidth]{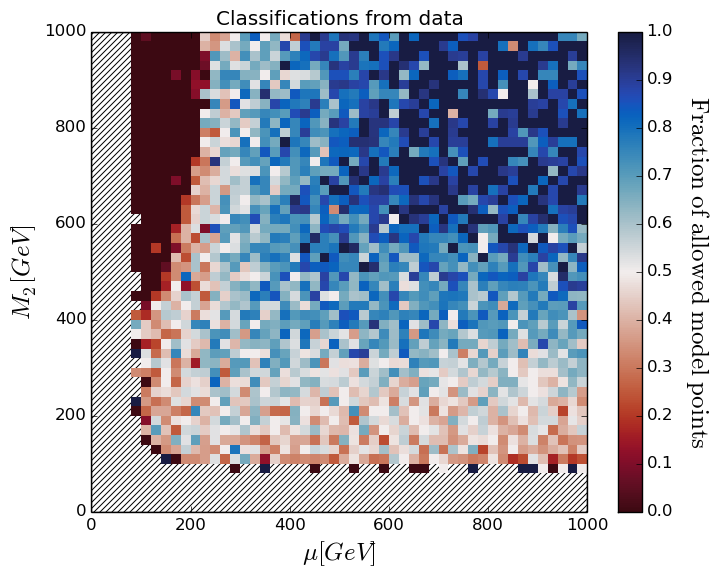}
	\end{subfigure}
	\hfill
	\begin{subfigure}[b]{0.23\textwidth}
	    \includegraphics[width=\textwidth]{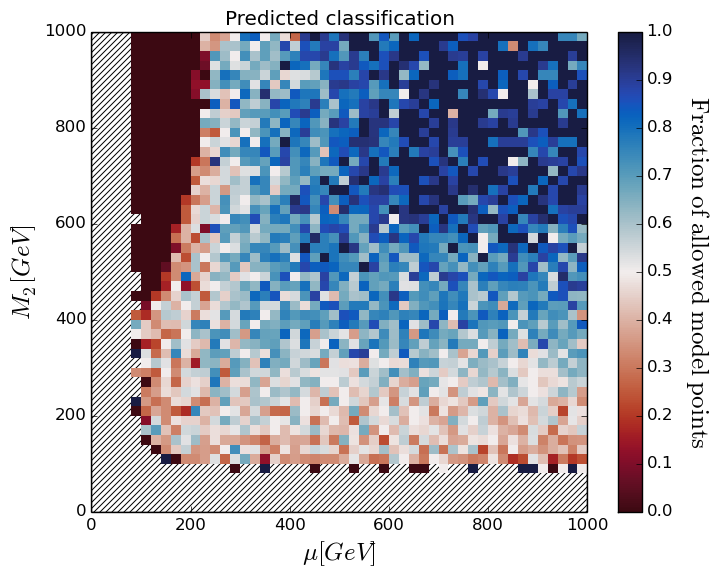}
	\end{subfigure}
	\hfill
	\begin{subfigure}[b]{0.23\textwidth}
	    \includegraphics[width=\textwidth]{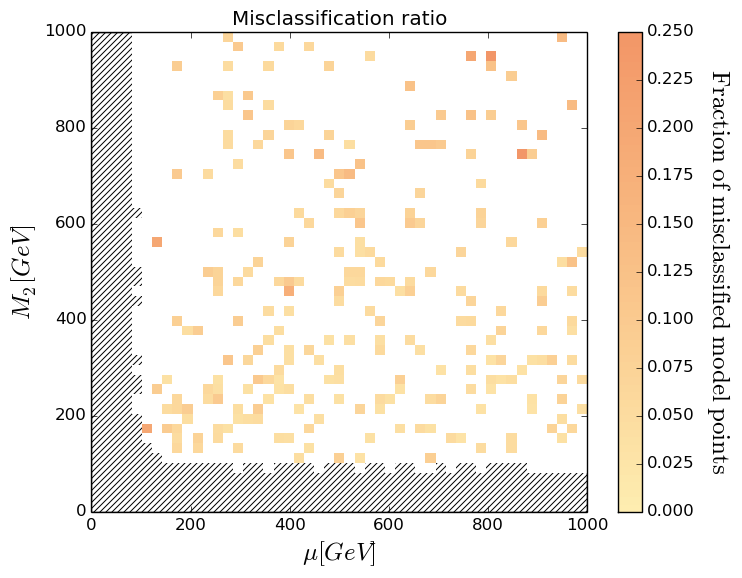}
	\end{subfigure}  
	
	\begin{subfigure}[b]{0.03\textwidth}
	    \rotatebox{90}{\hspace{2.15cm}99\-CL}
	\end{subfigure}
	\hfill
	\begin{subfigure}[b]{0.23\textwidth}
	    \includegraphics[width=\textwidth]{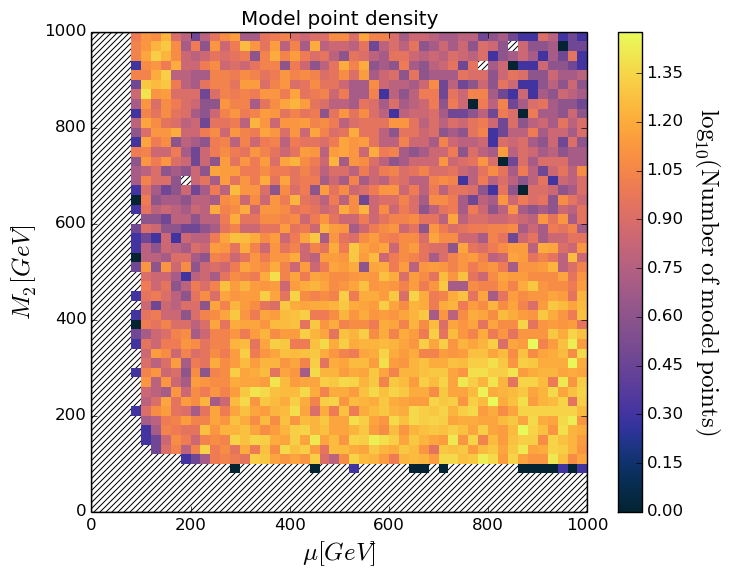}
	\end{subfigure}
	\hfill
	\begin{subfigure}[b]{0.23\textwidth}
	    \includegraphics[width=\textwidth]{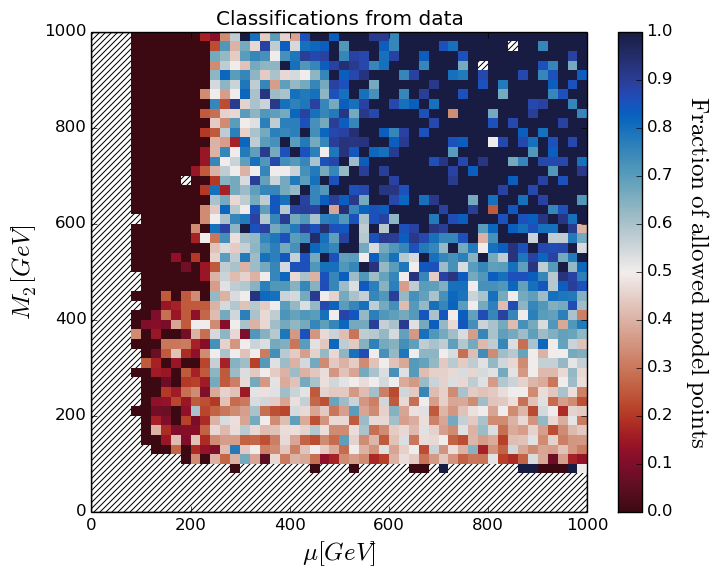}
	\end{subfigure}
	\hfill
	\begin{subfigure}[b]{0.23\textwidth}
	    \includegraphics[width=\textwidth]{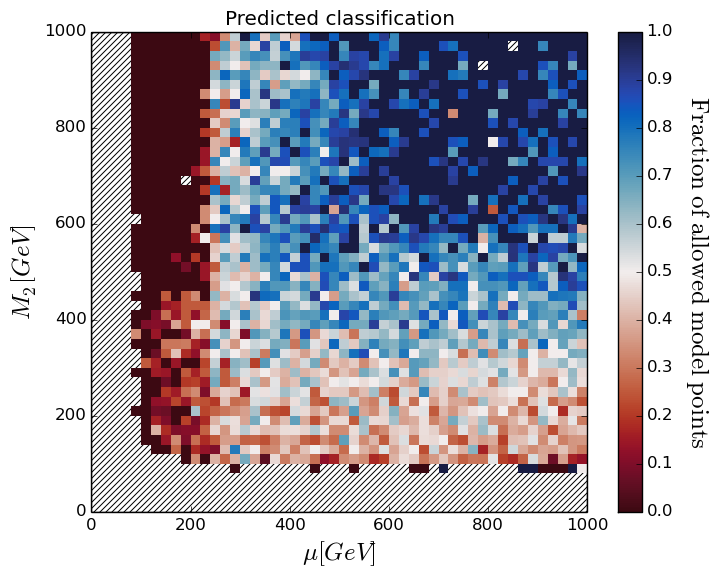}
	\end{subfigure}
	\hfill
	\begin{subfigure}[b]{0.23\textwidth}
	    \includegraphics[width=\textwidth]{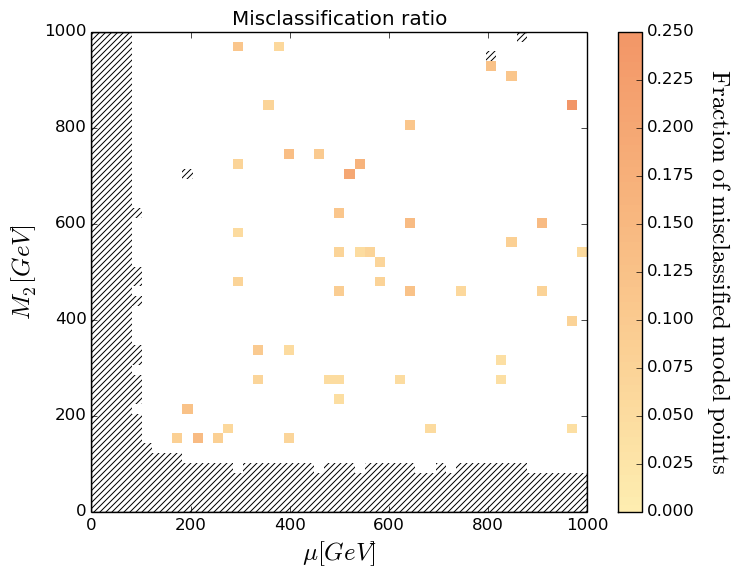}
	\end{subfigure}
	\caption{Color histograms for a projection of the 19-dimensional pMSSM parameter space on the $\mu$--$M_2$ plane. The color in the second and third column indicates the fraction of allowed data points for the true classification and the out-of-bag prediction, respectively. The last column shows the fraction of misclassified model points. The dashed bins contain no data points.}
	\label{fig:appplot06}
    \end{sidewaysfigure}
    \clearpage
}

\afterpage{
    \begin{sidewaysfigure}
	\centering
	\begin{subfigure}[b]{0.03\textwidth}
	    \hspace{0.5cm}
	\end{subfigure}
	\hfill
	\begin{subfigure}[b]{0.22\textwidth}
	    \begin{center}
		Number of model points
	    \end{center}
	    \vspace{0.5cm}
	\end{subfigure}
	\hfill
	\begin{subfigure}[b]{0.22\textwidth}
	    \begin{center}
		True classification
	    \end{center}
	    \vspace{0.5cm}
	\end{subfigure}
	\hfill
	\begin{subfigure}[b]{0.22\textwidth}
	    \begin{center}
		Prediction by classifier
	    \end{center}
	    \vspace{0.5cm}
	\end{subfigure}
	\hfill
	\begin{subfigure}[b]{0.22\textwidth}
	    \begin{center}
		Ratio of misclassified points
	    \end{center}
	    \vspace{0.5cm}
	\end{subfigure}
	\hfill

	\begin{subfigure}[b]{0.03\textwidth}
      \rotatebox{90}{\hspace{2cm}All data}
      \end{subfigure}
	\hfill
	\begin{subfigure}[b]{0.23\textwidth}
	    \includegraphics[width=\textwidth]{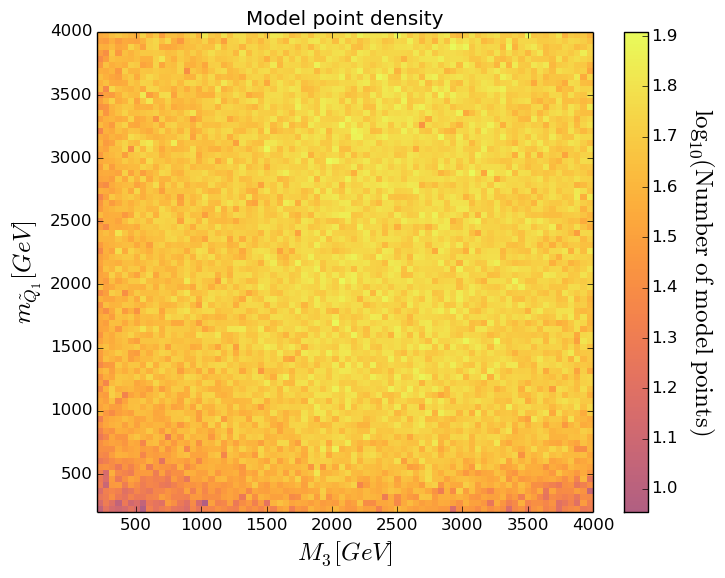}
	\end{subfigure}
	\hfill
	\begin{subfigure}[b]{0.23\textwidth}
	    \includegraphics[width=\textwidth]{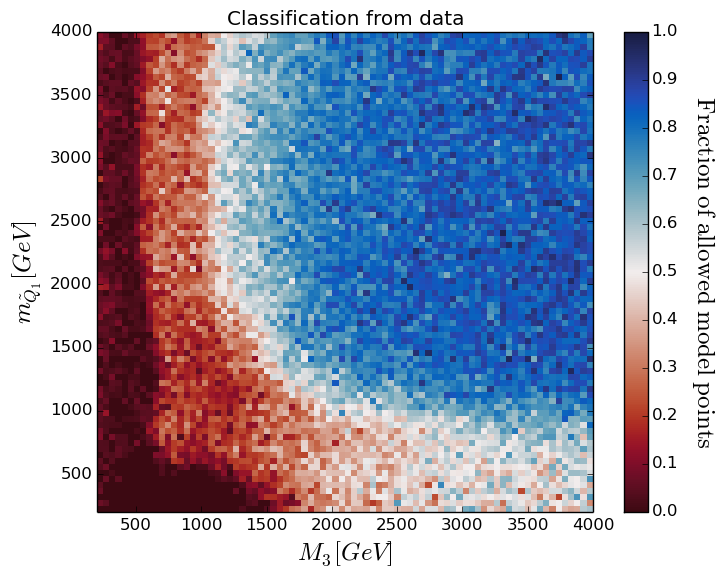}
	\end{subfigure}
	\hfill
	\begin{subfigure}[b]{0.23\textwidth}
	    \includegraphics[width=\textwidth]{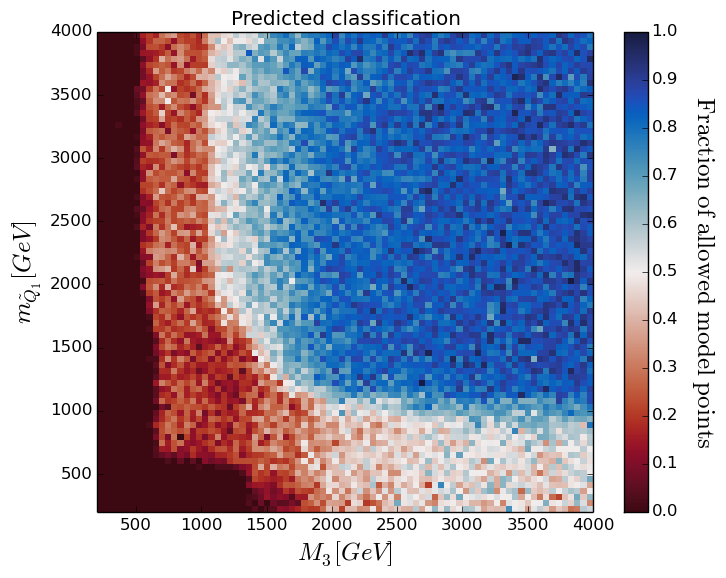}
	\end{subfigure}
	\hfill
	\begin{subfigure}[b]{0.23\textwidth}
	    \includegraphics[width=\textwidth]{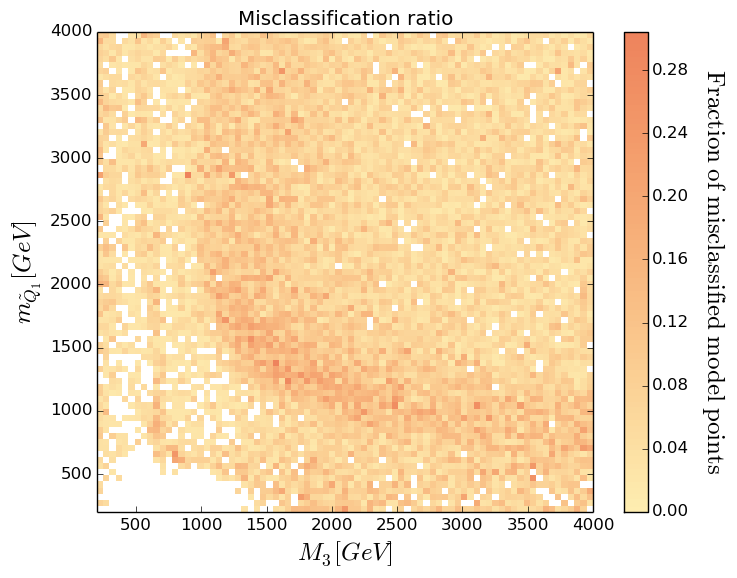}
	\end{subfigure}
	
	\begin{subfigure}[b]{0.03\textwidth}
	    \rotatebox{90}{\hspace{2.15cm}95\-CL}
	\end{subfigure}
	\hfill
	\begin{subfigure}[b]{0.23\textwidth}
	    \includegraphics[width=\textwidth]{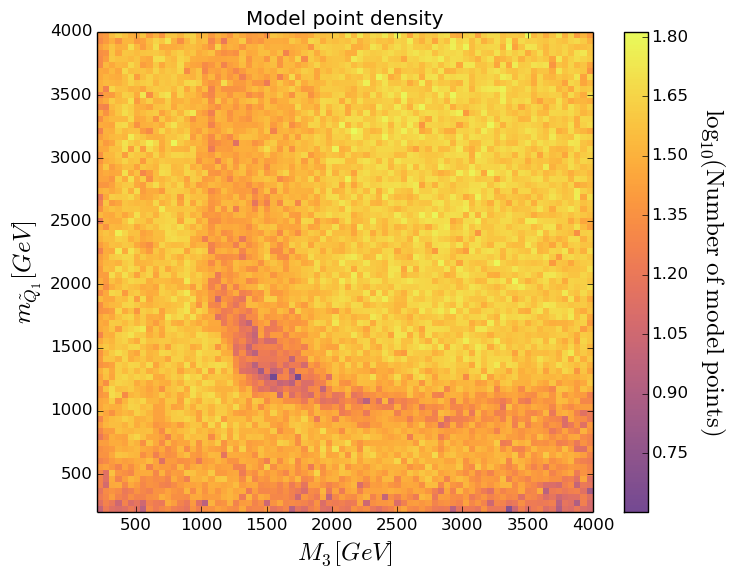}
	\end{subfigure}
	\hfill
	\begin{subfigure}[b]{0.23\textwidth}
	    \includegraphics[width=\textwidth]{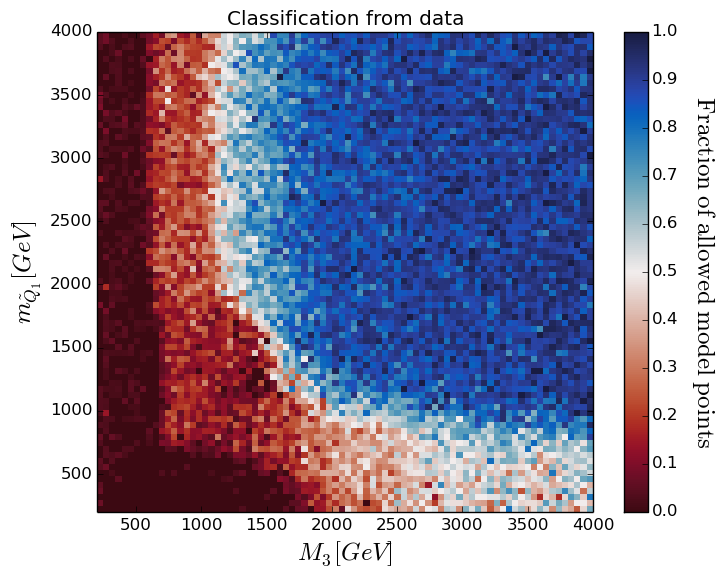}
	\end{subfigure}
	\hfill
	\begin{subfigure}[b]{0.23\textwidth}
	    \includegraphics[width=\textwidth]{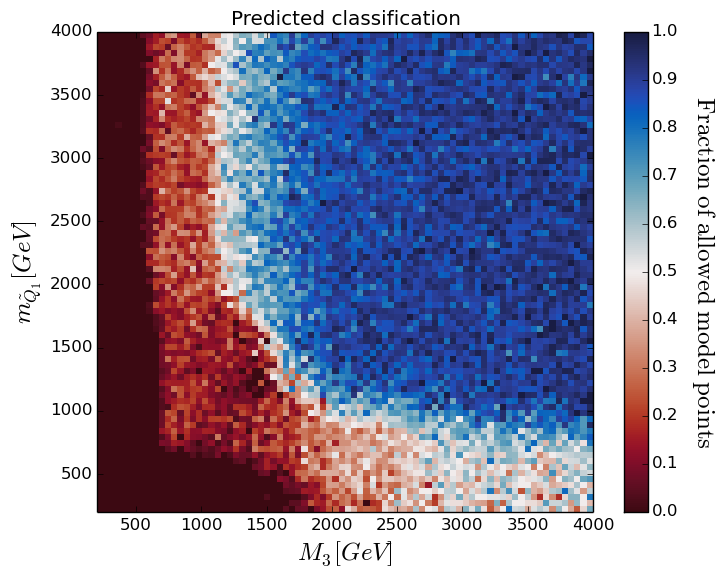}
	\end{subfigure}
	\hfill
	\begin{subfigure}[b]{0.23\textwidth}
	    \includegraphics[width=\textwidth]{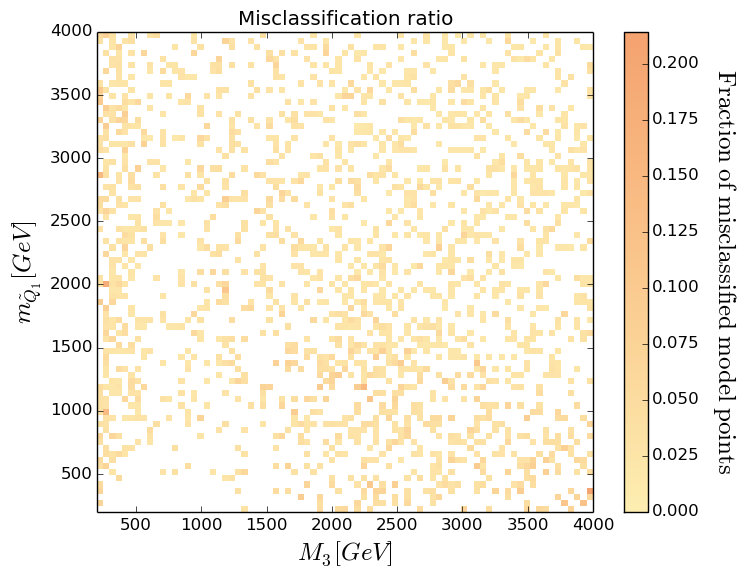}
	\end{subfigure}  
	
	\begin{subfigure}[b]{0.03\textwidth}
	    \rotatebox{90}{\hspace{2.15cm}99\-CL}
	\end{subfigure}
	\hfill
	\begin{subfigure}[b]{0.23\textwidth}
	    \includegraphics[width=\textwidth]{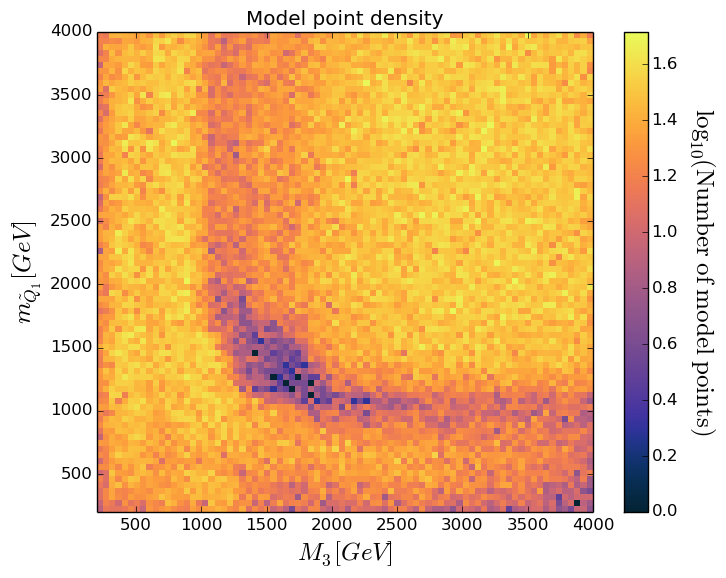}
	\end{subfigure}
	\hfill
	\begin{subfigure}[b]{0.23\textwidth}
	    \includegraphics[width=\textwidth]{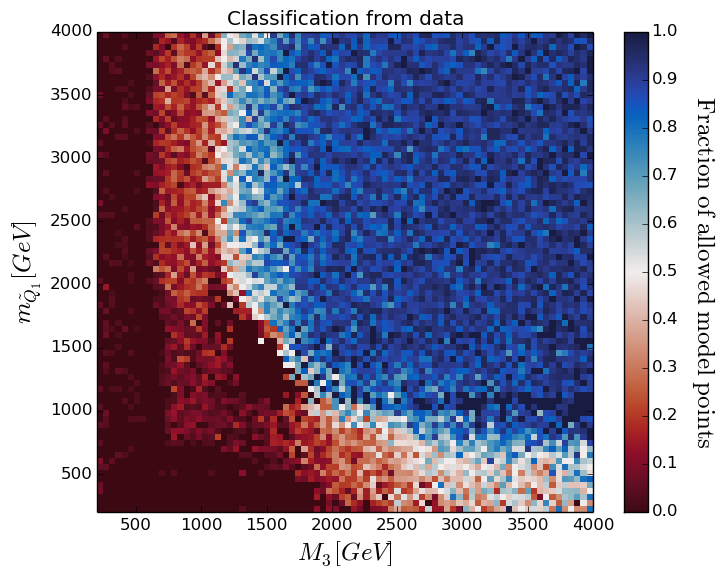}
	\end{subfigure}
	\hfill
	\begin{subfigure}[b]{0.23\textwidth}
	    \includegraphics[width=\textwidth]{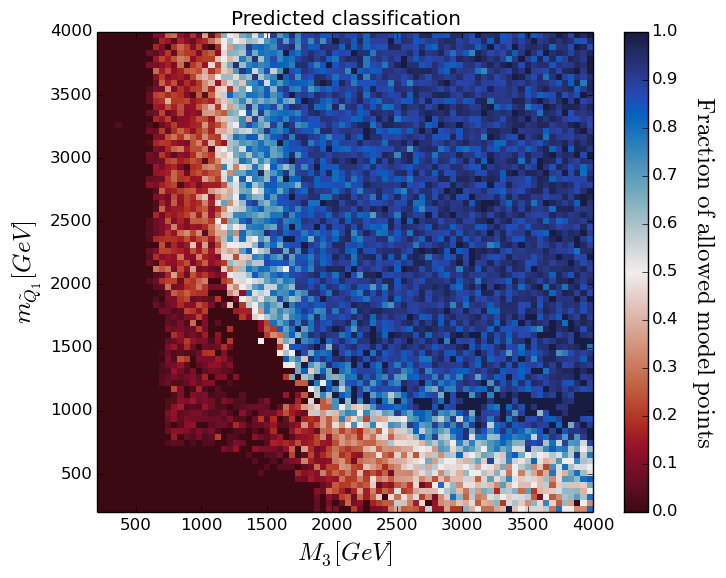}
	\end{subfigure}
	\hfill
	\begin{subfigure}[b]{0.23\textwidth}
	    \includegraphics[width=\textwidth]{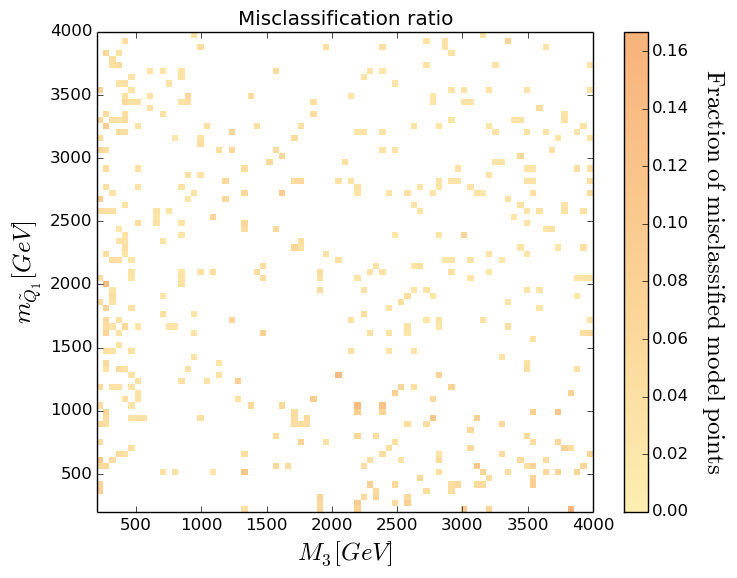}
	\end{subfigure}
	\caption{Color histograms for a projection of the 19-dimensional pMSSM parameter space on the $M_3$--$m_{\widetilde{Q}_1}$ plane. The color in the second and third column indicates the fraction of allowed data points for the true classification and the out-of-bag prediction, respectively. The last column shows the fraction of misclassified model points.}
	\label{fig:appplot07}
    \end{sidewaysfigure}
    \clearpage
}

\afterpage{
    \begin{sidewaysfigure}
	\centering
	\begin{subfigure}[b]{0.03\textwidth}
	    \hspace{0.5cm}
	\end{subfigure}
	\hfill
	\begin{subfigure}[b]{0.22\textwidth}
	    \begin{center}
		Number of model points
	    \end{center}
	    \vspace{0.5cm}
	\end{subfigure}
	\hfill
	\begin{subfigure}[b]{0.22\textwidth}
	    \begin{center}
		True classification
	    \end{center}
	    \vspace{0.5cm}
	\end{subfigure}
	\hfill
	\begin{subfigure}[b]{0.22\textwidth}
	    \begin{center}
		Prediction by classifier
	    \end{center}
	    \vspace{0.5cm}
	\end{subfigure}
	\hfill
	\begin{subfigure}[b]{0.22\textwidth}
	    \begin{center}
		Ratio of misclassified points
	    \end{center}
	    \vspace{0.5cm}
	\end{subfigure}
	\hfill

	\begin{subfigure}[b]{0.03\textwidth}
      \rotatebox{90}{\hspace{2cm}All data}
      \end{subfigure}
	\hfill
	\begin{subfigure}[b]{0.23\textwidth}
	    \includegraphics[width=\textwidth]{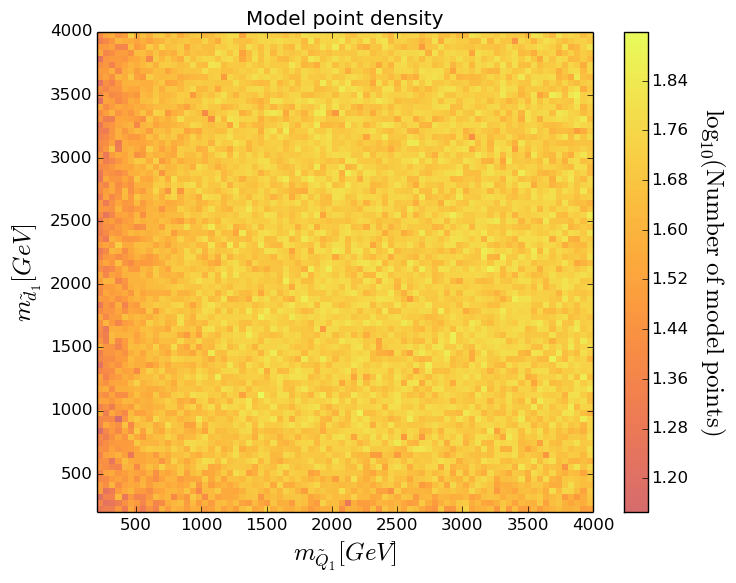}
	\end{subfigure}
	\hfill
	\begin{subfigure}[b]{0.23\textwidth}
	    \includegraphics[width=\textwidth]{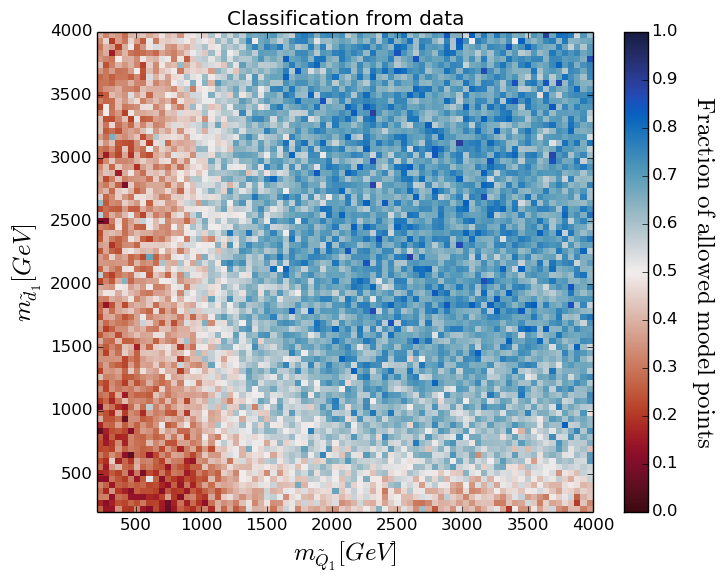}
	\end{subfigure}
	\hfill
	\begin{subfigure}[b]{0.23\textwidth}
	    \includegraphics[width=\textwidth]{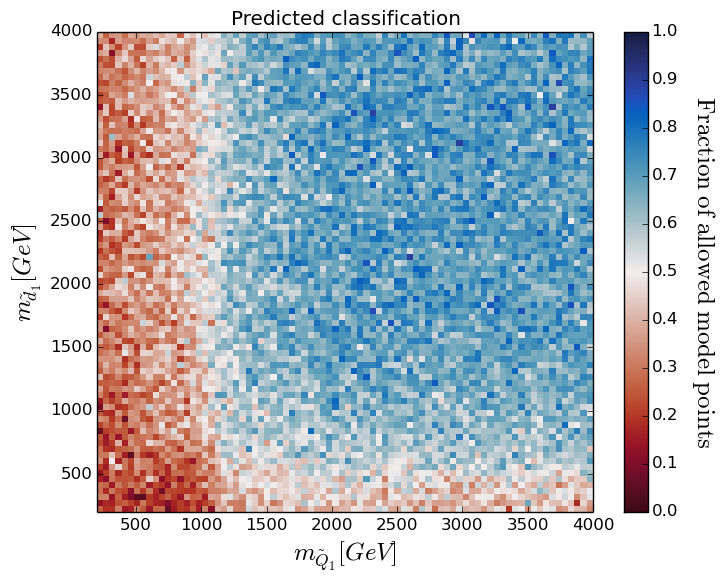}
	\end{subfigure}
	\hfill
	\begin{subfigure}[b]{0.23\textwidth}
	    \includegraphics[width=\textwidth]{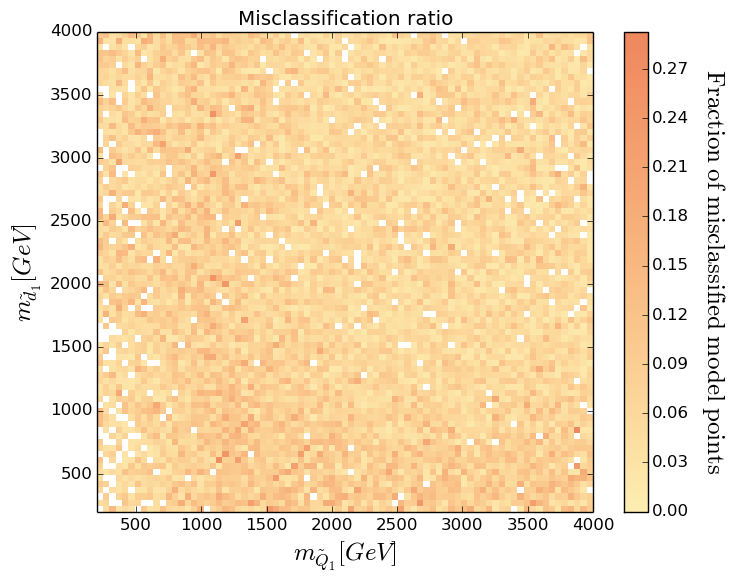}
	\end{subfigure}
	
	\begin{subfigure}[b]{0.03\textwidth}
	    \rotatebox{90}{\hspace{2.15cm}95\-CL}
	\end{subfigure}
	\hfill
	\begin{subfigure}[b]{0.23\textwidth}
	    \includegraphics[width=\textwidth]{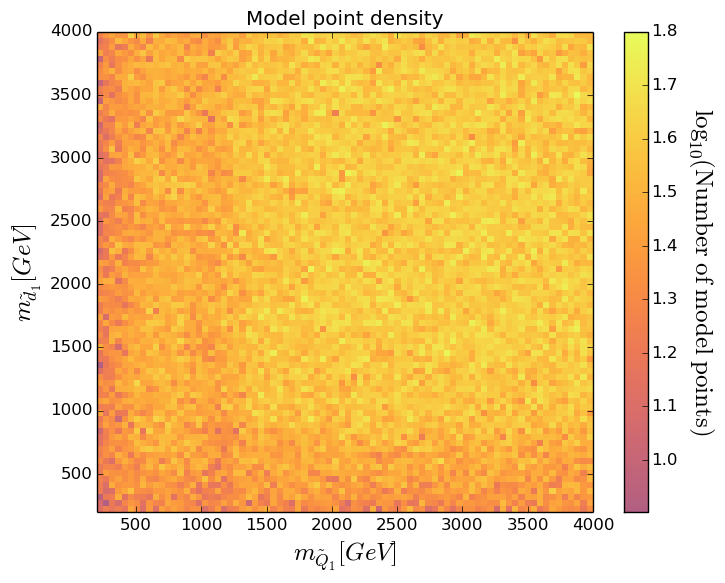}
	\end{subfigure}
	\hfill
	\begin{subfigure}[b]{0.23\textwidth}
	    \includegraphics[width=\textwidth]{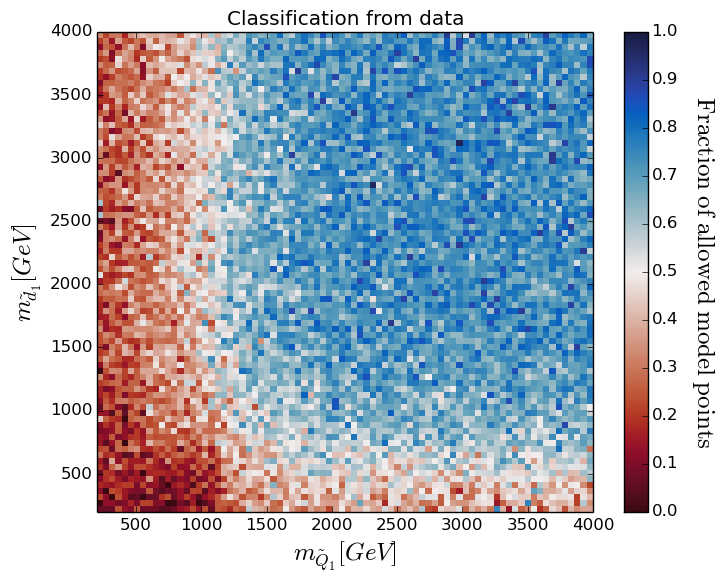}
	\end{subfigure}
	\hfill
	\begin{subfigure}[b]{0.23\textwidth}
	    \includegraphics[width=\textwidth]{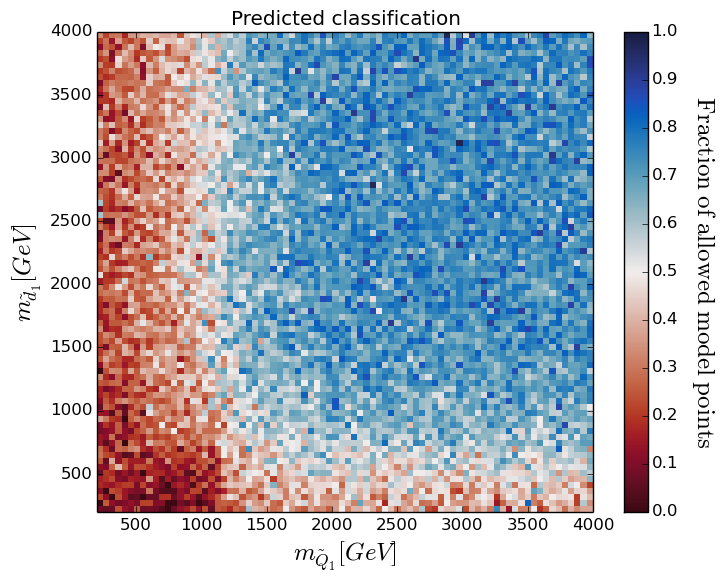}
	\end{subfigure}
	\hfill
	\begin{subfigure}[b]{0.23\textwidth}
	    \includegraphics[width=\textwidth]{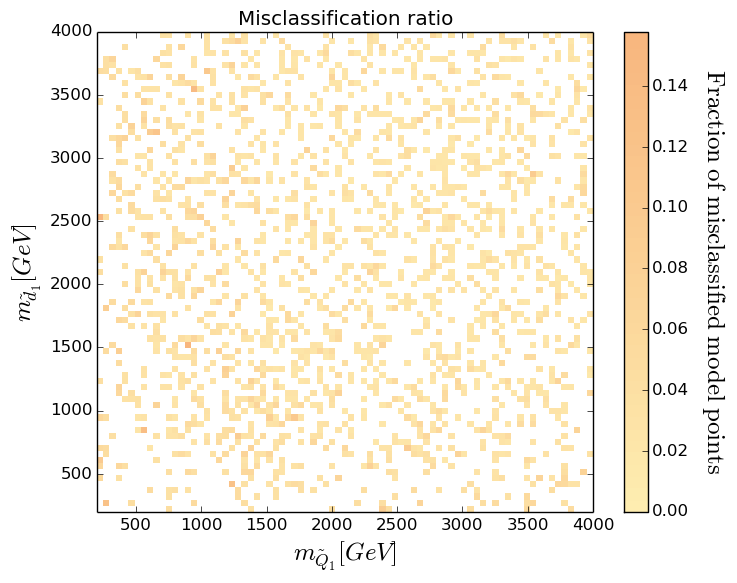}
	\end{subfigure}  
	
	\begin{subfigure}[b]{0.03\textwidth}
	    \rotatebox{90}{\hspace{2.15cm}99\-CL}
	\end{subfigure}
	\hfill
	\begin{subfigure}[b]{0.23\textwidth}
	    \includegraphics[width=\textwidth]{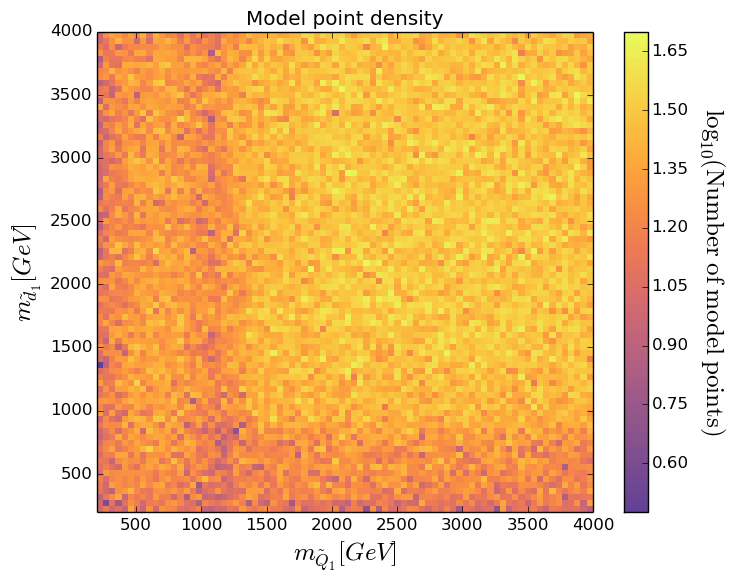}
	\end{subfigure}
	\hfill
	\begin{subfigure}[b]{0.23\textwidth}
	    \includegraphics[width=\textwidth]{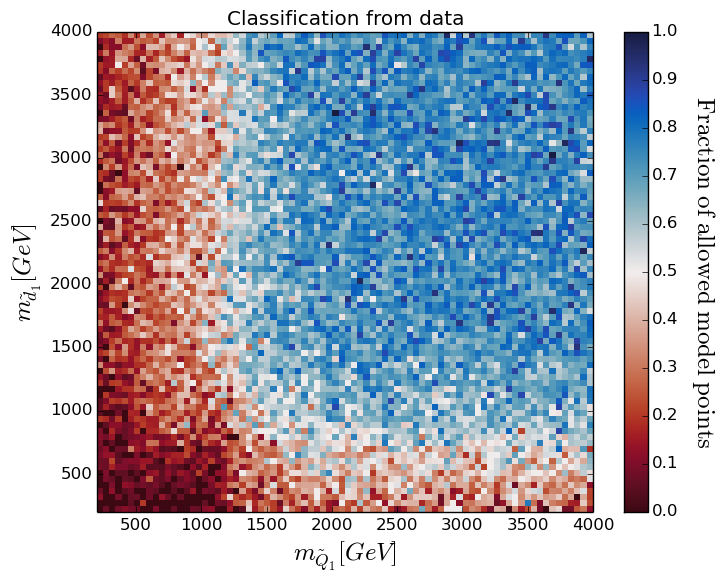}
	\end{subfigure}
	\hfill
	\begin{subfigure}[b]{0.23\textwidth}
	    \includegraphics[width=\textwidth]{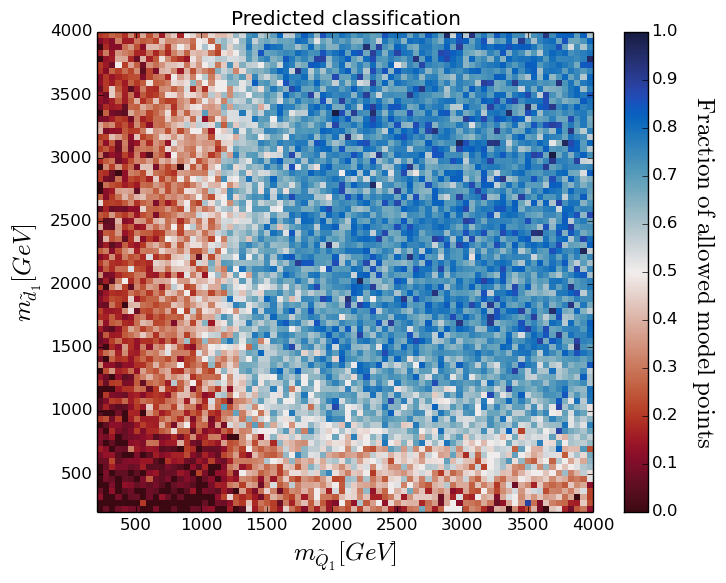}
	\end{subfigure}
	\hfill
	\begin{subfigure}[b]{0.23\textwidth}
	    \includegraphics[width=\textwidth]{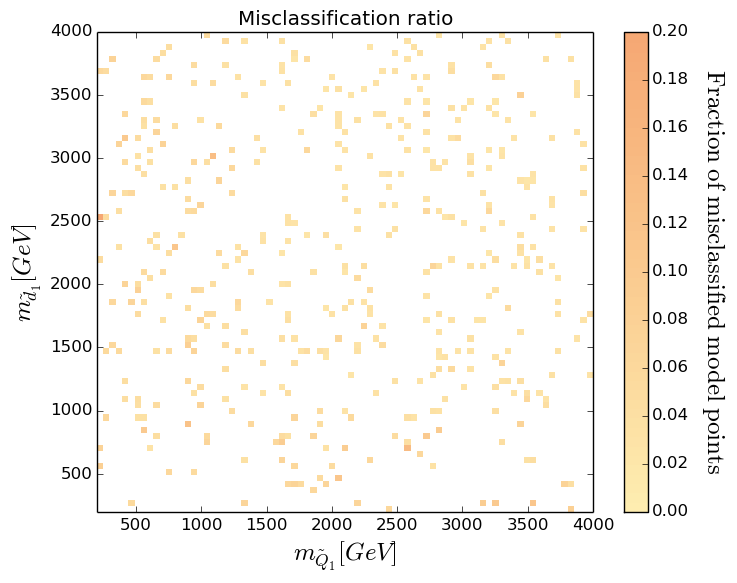}
	\end{subfigure}
	\caption{Color histograms for a projection of the 19-dimensional pMSSM parameter space on the $m_{\widetilde{Q}_1}$--$m_{\widetilde{D}_1}$ plane. The color in the second and third column indicates the fraction of allowed data points for the true classification and the out-of-bag prediction, respectively. The last column shows the fraction of misclassified model points.}
	\label{fig:appplot08}
    \end{sidewaysfigure}
    \clearpage
}
\end{appendices}

\newpage
\bibliography{MSSMml}

\providecommand{\href}[2]{#2}\begingroup\raggedright\begin{thebibliography}{10}

\bibitem{Aad:2015iea}
{\bf ATLAS} Collaboration, G.~Aad et~al., {\it {Summary of the searches for
  squarks and gluinos using $ \sqrt{s}=8 $ TeV pp collisions with the ATLAS
  experiment at the LHC}},  {\em JHEP} {\bf 10} (2015) 054,
  [\href{http://arxiv.org/abs/1507.05525}{{\tt arXiv:1507.05525}}].

\bibitem{Aad:2015eda}
{\bf ATLAS} Collaboration, G.~Aad et~al., {\it {Search for the electroweak
  production of supersymmetric particles in $\sqrt{s}=8$ TeV $pp$ collisions
  with the ATLAS detector}},  {\em Phys. Rev.} {\bf D93} (2016), no.~5 052002,
  [\href{http://arxiv.org/abs/1509.07152}{{\tt arXiv:1509.07152}}].

\bibitem{Chatrchyan:2013sza}
{\bf CMS} Collaboration, S.~Chatrchyan et~al., {\it {Interpretation of Searches
  for Supersymmetry with simplified Models}},  {\em Phys. Rev.} {\bf D88}
  (2013), no.~5 052017, [\href{http://arxiv.org/abs/1301.2175}{{\tt
  arXiv:1301.2175}}].

\bibitem{Aad:2015pfx}
{\bf ATLAS} Collaboration, G.~Aad et~al., {\it {ATLAS Run 1 searches for direct
  pair production of third-generation squarks at the Large Hadron Collider}},
  {\em Eur. Phys. J.} {\bf C75} (2015), no.~10 510,
  [\href{http://arxiv.org/abs/1506.08616}{{\tt arXiv:1506.08616}}]. [Erratum:
  \textit{Eur. Phys. J.} \textbf{C76} (2016) no. 3, 153].

\bibitem{Khachatryan:2015pwa}
{\bf CMS} Collaboration, V.~Khachatryan et~al., {\it {Search for Supersymmetry
  Using Razor Variables in Events with $b$-Tagged Jets in $pp$ Collisions at
  $\sqrt{s} =$ 8 TeV}},  {\em Phys. Rev.} {\bf D91} (2015) 052018,
  [\href{http://arxiv.org/abs/1502.00300}{{\tt arXiv:1502.00300}}].

\bibitem{Aad:2015baa}
{\bf ATLAS} Collaboration, G.~Aad et~al., {\it {Summary of the ATLAS
  experiment's sensitivity to supersymmetry after LHC Run 1 - interpreted in
  the phenomenological MSSM}},  {\em JHEP} {\bf 10} (2015) 134,
  [\href{http://arxiv.org/abs/1508.06608}{{\tt arXiv:1508.06608}}].

\bibitem{Khachatryan:2016nvf}
{\bf CMS} Collaboration, V.~Khachatryan et~al., {\it {Phenomenological MSSM
  interpretation of CMS searches in pp collisions at $\sqrt{s} = 7$ and 8
  TeV}},  \href{http://arxiv.org/abs/1606.03577}{{\tt arXiv:1606.03577}}.

\bibitem{Djouadi:1998di}
{\bf MSSM Working Group} Collaboration, A.~Djouadi et~al., {\it {The Minimal
  supersymmetric standard model: Group summary report}},
  \href{http://arxiv.org/abs/hep-ph/9901246}{{\tt hep-ph/9901246}}.

\bibitem{Papucci:2014rja}
M.~Papucci, K.~Sakurai, A.~Weiler, and L.~Zeune, {\it {Fastlim: a fast LHC
  limit calculator}},  {\em Eur. Phys. J.} {\bf C74} (2014), no.~11 3163,
  [\href{http://arxiv.org/abs/1402.0492}{{\tt arXiv:1402.0492}}].

\bibitem{Kraml:2013mwa}
S.~Kraml, S.~Kulkarni, U.~Laa, A.~Lessa, W.~Magerl, D.~Proschofsky-Spindler,
  and W.~Waltenberger, {\it {SModelS: a tool for interpreting simplified-model
  results from the LHC and its application to supersymmetry}},  {\em Eur. Phys.
  J.} {\bf C74} (2014) 2868, [\href{http://arxiv.org/abs/1312.4175}{{\tt
  arXiv:1312.4175}}].

\bibitem{Beenakker:1996ch}
W.~Beenakker, R.~Hopker, M.~Spira, and P.~M. Zerwas, {\it {Squark and gluino
  production at hadron colliders}},  {\em Nucl. Phys.} {\bf B492} (1997)
  51--103, [\href{http://arxiv.org/abs/hep-ph/9610490}{{\tt hep-ph/9610490}}].

\bibitem{Kulesza:2008jb}
A.~Kulesza and L.~Motyka, {\it {Threshold resummation for squark-antisquark and
  gluino-pair production at the LHC}},  {\em Phys. Rev. Lett.} {\bf 102} (2009)
  111802, [\href{http://arxiv.org/abs/0807.2405}{{\tt arXiv:0807.2405}}].

\bibitem{Kulesza:2009kq}
A.~Kulesza and L.~Motyka, {\it {Soft gluon resummation for the production of
  gluino-gluino and squark-antisquark pairs at the LHC}},  {\em Phys. Rev.}
  {\bf D80} (2009) 095004, [\href{http://arxiv.org/abs/0905.4749}{{\tt
  arXiv:0905.4749}}].

\bibitem{Drees:2013wra}
M.~Drees, H.~Dreiner, D.~Schmeier, J.~Tattersall, and J.~S. Kim, {\it
  {CheckMATE: Confronting your Favourite New Physics Model with LHC Data}},
  {\em Comput. Phys. Commun.} {\bf 187} (2014) 227--265,
  [\href{http://arxiv.org/abs/1312.2591}{{\tt arXiv:1312.2591}}].

\bibitem{Kim:2015wza}
J.~S. Kim, D.~Schmeier, J.~Tattersall, and K.~Rolbiecki, {\it {A framework to
  create customised LHC analyses within CheckMATE}},  {\em Comput. Phys.
  Commun.} {\bf 196} (2015) 535--562,
  [\href{http://arxiv.org/abs/1503.01123}{{\tt arXiv:1503.01123}}].

\bibitem{Conte:2012fm}
E.~Conte, B.~Fuks, and G.~Serret, {\it {MadAnalysis 5, A User-Friendly
  Framework for Collider Phenomenology}},  {\em Comput. Phys. Commun.} {\bf
  184} (2013) 222--256, [\href{http://arxiv.org/abs/1206.1599}{{\tt
  arXiv:1206.1599}}].

\bibitem{deFavereau:2013fsa}
{\bf DELPHES 3} Collaboration, J.~de~Favereau, C.~Delaere, P.~Demin,
  A.~Giammanco, V.~Lemaitre, A.~Mertens, and M.~Selvaggi, {\it {DELPHES 3, A
  modular framework for fast simulation of a generic collider experiment}},
  {\em JHEP} {\bf 02} (2014) 057, [\href{http://arxiv.org/abs/1307.6346}{{\tt
  arXiv:1307.6346}}].

\bibitem{rok2008}
L.~M.~O. Rokach, {\em {Data mining with decision trees: theory and
  applications.}}
\newblock World Scientific, 2008.

\bibitem{Aad:2012tfa}
{\bf ATLAS} Collaboration, G.~Aad et~al., {\it {Observation of a new particle
  in the search for the Standard Model Higgs boson with the ATLAS detector at
  the LHC}},  {\em Phys. Lett.} {\bf B716} (2012) 1--29,
  [\href{http://arxiv.org/abs/1207.7214}{{\tt arXiv:1207.7214}}].

\bibitem{Chatrchyan:2012xdj}
{\bf CMS} Collaboration, S.~Chatrchyan et~al., {\it {Observation of a new boson
  at a mass of 125 GeV with the CMS experiment at the LHC}},  {\em Phys. Lett.}
  {\bf B716} (2012) 30--61, [\href{http://arxiv.org/abs/1207.7235}{{\tt
  arXiv:1207.7235}}].

\bibitem{Bridges:2010de}
M.~Bridges, K.~Cranmer, F.~Feroz, M.~Hobson, R.~R. de~Austri, and R.~Trotta,
  {\it {A Coverage Study of the CMSSM Based on ATLAS Sensitivity Using Fast
  Neural Networks Techniques}},  {\em JHEP} {\bf 03} (2011) 012,
  [\href{http://arxiv.org/abs/1011.4306}{{\tt arXiv:1011.4306}}].

\bibitem{Buckley:2011kc}
A.~Buckley, A.~Shilton, and M.~J. White, {\it {Fast supersymmetry phenomenology
  at the Large Hadron Collider using machine learning techniques}},  {\em
  Comput. Phys. Commun.} {\bf 183} (2012) 960--970,
  [\href{http://arxiv.org/abs/1106.4613}{{\tt arXiv:1106.4613}}].

\bibitem{Bornhauser:2013aya}
N.~Bornhauser and M.~Drees, {\it {Determination of the CMSSM Parameters using
  Neural Networks}},  {\em Phys. Rev.} {\bf D88} (2013) 075016,
  [\href{http://arxiv.org/abs/1307.3383}{{\tt arXiv:1307.3383}}].

\bibitem{deVries:2015hva}
K.~J. de~Vries et~al., {\it {The pMSSM10 after LHC Run 1}},  {\em Eur. Phys.
  J.} {\bf C75} (2015), no.~9 422, [\href{http://arxiv.org/abs/1504.03260}{{\tt
  arXiv:1504.03260}}].

\bibitem{Strege:2014ija}
C.~Strege, G.~Bertone, G.~J. Besjes, S.~Caron, R.~{Ruiz de Austri}, A.~Strubig,
  and R.~Trotta, {\it {Profile likelihood maps of a 15-dimensional MSSM}},
  {\em JHEP} {\bf 09} (2014) 081, [\href{http://arxiv.org/abs/1405.0622}{{\tt
  arXiv:1405.0622}}].

\bibitem{Drees:2004jm}
M.~Drees, R.~Godbole, and P.~Roy, {\em {Theory and phenomenology of sparticles:
  An account of four-dimensional N=1 supersymmetry in high energy physics}}.
\newblock World Scientific, 2004.

\bibitem{Girardello:1981wz}
L.~Girardello and M.~T. Grisaru, {\it {Soft Breaking of Supersymmetry}},  {\em
  Nucl. Phys.} {\bf B194} (1982) 65.

\bibitem{Haber:1997if}
H.~E. Haber, {\it {The Status of the minimal supersymmetric standard model and
  beyond}},  {\em Nucl. Phys. Proc. Suppl.} {\bf 62} (1998) 469--484,
  [\href{http://arxiv.org/abs/hep-ph/9709450}{{\tt hep-ph/9709450}}].

\bibitem{Berger:2008cq}
C.~F. Berger, J.~S. Gainer, J.~L. Hewett, and T.~G. Rizzo, {\it {Supersymmetry
  Without Prejudice}},  {\em JHEP} {\bf 02} (2009) 023,
  [\href{http://arxiv.org/abs/0812.0980}{{\tt arXiv:0812.0980}}].

\bibitem{Aad:2014wea}
{\bf ATLAS} Collaboration, G.~Aad et~al., {\it {Search for squarks and gluinos
  with the ATLAS detector in final states with jets and missing transverse
  momentum using $\sqrt{s}=8$ TeV proton-proton collision data}},  {\em JHEP}
  {\bf 09} (2014) 176, [\href{http://arxiv.org/abs/1405.7875}{{\tt
  arXiv:1405.7875}}].

\bibitem{Aad:2013wta}
{\bf ATLAS} Collaboration, G.~Aad et~al., {\it {Search for new phenomena in
  final states with large jet multiplicities and missing transverse momentum at
  $\sqrt{s}=8$ TeV proton-proton collisions using the ATLAS experiment}},  {\em
  JHEP} {\bf 10} (2013) 130, [\href{http://arxiv.org/abs/1308.1841}{{\tt
  arXiv:1308.1841}}]. [Erratum: \textit{JHEP} \textbf{01} (2014) 109].

\bibitem{Aad:2015mia}
{\bf ATLAS} Collaboration, G.~Aad et~al., {\it {Search for squarks and gluinos
  in events with isolated leptons, jets and missing transverse momentum at
  $\sqrt{s}=8$ TeV with the ATLAS detector}},  {\em JHEP} {\bf 04} (2015) 116,
  [\href{http://arxiv.org/abs/1501.03555}{{\tt arXiv:1501.03555}}].

\bibitem{Aad:2014mra}
{\bf ATLAS} Collaboration, G.~Aad et~al., {\it {Search for supersymmetry in
  events with large missing transverse momentum, jets, and at least one tau
  lepton in 20 fb$^{-1}$ of $\sqrt{s}=$ 8 TeV proton-proton collision data with
  the ATLAS detector}},  {\em JHEP} {\bf 09} (2014) 103,
  [\href{http://arxiv.org/abs/1407.0603}{{\tt arXiv:1407.0603}}].

\bibitem{Aad:2014pda}
{\bf ATLAS} Collaboration, G.~Aad et~al., {\it {Search for supersymmetry at
  $\sqrt{s}=8$ TeV in final states with jets and two same-sign leptons or three
  leptons with the ATLAS detector}},  {\em JHEP} {\bf 06} (2014) 035,
  [\href{http://arxiv.org/abs/1404.2500}{{\tt arXiv:1404.2500}}].

\bibitem{Aad:2014lra}
{\bf ATLAS} Collaboration, G.~Aad et~al., {\it {Search for strong production of
  supersymmetric particles in final states with missing transverse momentum and
  at least three $b$-jets at $\sqrt{s}= 8$ TeV proton-proton collisions with
  the ATLAS detector}},  {\em JHEP} {\bf 10} (2014) 024,
  [\href{http://arxiv.org/abs/1407.0600}{{\tt arXiv:1407.0600}}].

\bibitem{Aad:2015zva}
{\bf ATLAS} Collaboration, G.~Aad et~al., {\it {Search for new phenomena in
  final states with an energetic jet and large missing transverse momentum in
  pp collisions at $\sqrt{s}=$8 TeV with the ATLAS detector}},  {\em Eur. Phys.
  J.} {\bf C75} (2015), no.~7 299, [\href{http://arxiv.org/abs/1502.01518}{{\tt
  arXiv:1502.01518}}]. [Erratum: \textit{Eur. Phys. J.} \textbf{C75} (2015) no.
  9, 408].

\bibitem{Aad:2014bva}
{\bf ATLAS} Collaboration, G.~Aad et~al., {\it {Search for direct pair
  production of the top squark in all-hadronic final states in proton-proton
  collisions at $\sqrt{s}=8$ TeV with the ATLAS detector}},  {\em JHEP} {\bf
  09} (2014) 015, [\href{http://arxiv.org/abs/1406.1122}{{\tt
  arXiv:1406.1122}}].

\bibitem{Aad:2014kra}
{\bf ATLAS} Collaboration, G.~Aad et~al., {\it {Search for top squark pair
  production in final states with one isolated lepton, jets, and missing
  transverse momentum in $\sqrt s =8$ TeV $pp$ collisions with the ATLAS
  detector}},  {\em JHEP} {\bf 11} (2014) 118,
  [\href{http://arxiv.org/abs/1407.0583}{{\tt arXiv:1407.0583}}].

\bibitem{Aad:2014qaa}
{\bf ATLAS} Collaboration, G.~Aad et~al., {\it {Search for direct top-squark
  pair production in final states with two leptons in $pp$ collisions at
  $\sqrt{s}=$ 8 TeV with the ATLAS detector}},  {\em JHEP} {\bf 06} (2014) 124,
  [\href{http://arxiv.org/abs/1403.4853}{{\tt arXiv:1403.4853}}].

\bibitem{Aad:2014nra}
{\bf ATLAS} Collaboration, G.~Aad et~al., {\it {Search for pair-produced
  third-generation squarks decaying via charm quarks or in compressed
  supersymmetric scenarios in $pp$ collisions at $\sqrt{s}=8~$TeV with the
  ATLAS detector}},  {\em Phys. Rev.} {\bf D90} (2014), no.~5 052008,
  [\href{http://arxiv.org/abs/1407.0608}{{\tt arXiv:1407.0608}}].

\bibitem{Aad:2014mha}
{\bf ATLAS} Collaboration, G.~Aad et~al., {\it {Search for direct top squark
  pair production in events with a $Z$ boson, $b$-jets and missing transverse
  momentum in $\sqrt{s}=8$ TeV $pp$ collisions with the ATLAS detector}},  {\em
  Eur. Phys. J.} {\bf C74} (2014), no.~6 2883,
  [\href{http://arxiv.org/abs/1403.5222}{{\tt arXiv:1403.5222}}].

\bibitem{Aad:2013ija}
{\bf ATLAS} Collaboration, G.~Aad et~al., {\it {Search for direct
  third-generation squark pair production in final states with missing
  transverse momentum and two $b$-jets in $\sqrt{s} =$ 8 TeV $pp$ collisions
  with the ATLAS detector}},  {\em JHEP} {\bf 10} (2013) 189,
  [\href{http://arxiv.org/abs/1308.2631}{{\tt arXiv:1308.2631}}].

\bibitem{Aad:2015jqa}
{\bf ATLAS} Collaboration, G.~Aad et~al., {\it {Search for direct pair
  production of a chargino and a neutralino decaying to the 125 GeV Higgs boson
  in $\sqrt{s} = 8$ TeV ${pp}$ collisions with the ATLAS detector}},  {\em Eur.
  Phys. J.} {\bf C75} (2015), no.~5 208,
  [\href{http://arxiv.org/abs/1501.07110}{{\tt arXiv:1501.07110}}].

\bibitem{Aad:2014vma}
{\bf ATLAS} Collaboration, G.~Aad et~al., {\it {Search for direct production of
  charginos, neutralinos and sleptons in final states with two leptons and
  missing transverse momentum in $pp$ collisions at $\sqrt{s} =$ 8 TeV with the
  ATLAS detector}},  {\em JHEP} {\bf 05} (2014) 071,
  [\href{http://arxiv.org/abs/1403.5294}{{\tt arXiv:1403.5294}}].

\bibitem{Aad:2014yka}
{\bf ATLAS} Collaboration, G.~Aad et~al., {\it {Search for the direct
  production of charginos, neutralinos and staus in final states with at least
  two hadronically decaying taus and missing transverse momentum in $pp$
  collisions at $\sqrt{s}$ = 8 TeV with the ATLAS detector}},  {\em JHEP} {\bf
  10} (2014) 096, [\href{http://arxiv.org/abs/1407.0350}{{\tt
  arXiv:1407.0350}}].

\bibitem{Aad:2014nua}
{\bf ATLAS} Collaboration, G.~Aad et~al., {\it {Search for direct production of
  charginos and neutralinos in events with three leptons and missing transverse
  momentum in $\sqrt{s} = 8$ TeV $pp$ collisions with the ATLAS detector}},
  {\em JHEP} {\bf 04} (2014) 169, [\href{http://arxiv.org/abs/1402.7029}{{\tt
  arXiv:1402.7029}}].

\bibitem{Aad:2014iza}
{\bf ATLAS} Collaboration, G.~Aad et~al., {\it {Search for supersymmetry in
  events with four or more leptons in $\sqrt{s}$ = 8 TeV pp collisions with the
  ATLAS detector}},  {\em Phys. Rev.} {\bf D90} (2014), no.~5 052001,
  [\href{http://arxiv.org/abs/1405.5086}{{\tt arXiv:1405.5086}}].

\bibitem{Aad:2013yna}
{\bf ATLAS} Collaboration, G.~Aad et~al., {\it {Search for charginos nearly
  mass degenerate with the lightest neutralino based on a disappearing-track
  signature in pp collisions at $\sqrt{s}=8$ TeV with the ATLAS detector}},
  {\em Phys. Rev.} {\bf D88} (2013), no.~11 112006,
  [\href{http://arxiv.org/abs/1310.3675}{{\tt arXiv:1310.3675}}].

\bibitem{Aad:2012pra}
{\bf ATLAS} Collaboration, G.~Aad et~al., {\it {Searches for heavy long-lived
  sleptons and R-Hadrons with the ATLAS detector in $pp$ collisions at
  $\sqrt{s}=7$ TeV}},  {\em Phys. Lett.} {\bf B720} (2013) 277--308,
  [\href{http://arxiv.org/abs/1211.1597}{{\tt arXiv:1211.1597}}].

\bibitem{ATLAS:2014fka}
{\bf ATLAS} Collaboration, G.~Aad et~al., {\it {Searches for heavy long-lived
  charged particles with the ATLAS detector in proton-proton collisions at $
  \sqrt{s}=8 $ TeV}},  {\em JHEP} {\bf 01} (2015) 068,
  [\href{http://arxiv.org/abs/1411.6795}{{\tt arXiv:1411.6795}}].

\bibitem{Aad:2014vgg}
{\bf ATLAS} Collaboration, G.~Aad et~al., {\it {Search for neutral Higgs bosons
  of the minimal supersymmetric standard model in pp collisions at $\sqrt{s}$ =
  8 TeV with the ATLAS detector}},  {\em JHEP} {\bf 11} (2014) 056,
  [\href{http://arxiv.org/abs/1409.6064}{{\tt arXiv:1409.6064}}].

\bibitem{CahillRowley:2012cb}
M.~W. Cahill-Rowley, J.~L. Hewett, S.~Hoeche, A.~Ismail, and T.~G. Rizzo, {\it
  {The New Look pMSSM with Neutralino and Gravitino LSPs}},  {\em Eur. Phys.
  J.} {\bf C72} (2012) 2156, [\href{http://arxiv.org/abs/1206.4321}{{\tt
  arXiv:1206.4321}}].

\bibitem{Baak:2012kk}
M.~Baak, M.~Goebel, J.~Haller, A.~Hoecker, D.~Kennedy, R.~Kogler, K.~Moenig,
  M.~Schott, and J.~Stelzer, {\it {The Electroweak Fit of the Standard Model
  after the Discovery of a New Boson at the LHC}},  {\em Eur. Phys. J.} {\bf
  C72} (2012) 2205, [\href{http://arxiv.org/abs/1209.2716}{{\tt
  arXiv:1209.2716}}].

\bibitem{Amhis:2012bh}
{\bf Heavy Flavor Averaging Group} Collaboration, Y.~Amhis et~al., {\it
  {Averages of B-Hadron, C-Hadron, and tau-lepton properties as of early
  2012}},  \href{http://arxiv.org/abs/1207.1158}{{\tt arXiv:1207.1158}}.

\bibitem{DeBruyn:2012wk}
K.~{De Bruyn}, R.~Fleischer, R.~Knegjens, P.~Koppenburg, M.~Merk,
  A.~Pellegrino, and N.~Tuning, {\it {Probing New Physics via the $B^0_s\to
  \mu^+\mu^-$ Effective Lifetime}},  {\em Phys. Rev. Lett.} {\bf 109} (2012)
  041801, [\href{http://arxiv.org/abs/1204.1737}{{\tt arXiv:1204.1737}}].

\bibitem{CMS:2014xfa}
{\bf LHCb, CMS} Collaboration, V.~Khachatryan et~al., {\it {Observation of the
  rare $B^0_s\to\mu^+\mu^-$ decay from the combined analysis of CMS and LHCb
  data}},  {\em Nature} {\bf 522} (2015) 68--72,
  [\href{http://arxiv.org/abs/1411.4413}{{\tt arXiv:1411.4413}}].

\bibitem{Mahmoudi:2008tp}
F.~Mahmoudi, {\it {SuperIso v2.3: A Program for calculating flavor physics
  observables in Supersymmetry}},  {\em Comput. Phys. Commun.} {\bf 180} (2009)
  1579--1613, [\href{http://arxiv.org/abs/0808.3144}{{\tt arXiv:0808.3144}}].

\bibitem{Aoyama:2012wk}
T.~Aoyama, M.~Hayakawa, T.~Kinoshita, and M.~Nio, {\it {Complete Tenth-Order
  QED Contribution to the Muon g-2}},  {\em Phys. Rev. Lett.} {\bf 109} (2012)
  111808, [\href{http://arxiv.org/abs/1205.5370}{{\tt arXiv:1205.5370}}].

\bibitem{Bennett:2006fi}
{\bf Muon g-2} Collaboration, G.~W. Bennett et~al., {\it {Final Report of the
  Muon E821 Anomalous Magnetic Moment Measurement at BNL}},  {\em Phys. Rev.}
  {\bf D73} (2006) 072003, [\href{http://arxiv.org/abs/hep-ex/0602035}{{\tt
  hep-ex/0602035}}].

\bibitem{ALEPH:2005ab}
{\bf SLD Electroweak Group, DELPHI, ALEPH, SLD, SLD Heavy Flavour Group, OPAL,
  LEP Electroweak Working Group, L3} Collaboration, S.~Schael et~al., {\it
  {Precision electroweak measurements on the $Z$ resonance}},  {\em Phys.
  Rept.} {\bf 427} (2006) 257--454,
  [\href{http://arxiv.org/abs/hep-ex/0509008}{{\tt hep-ex/0509008}}].

\bibitem{LEPconst}
``{The LEP SUSY Working Group and the ALEPH, DELPHI, L3 and OPAL experiments,
  note LEPSUSYWG/01-03.1}.'' \url{http://lepsusy.web.cern.ch/lepsusy}.

\bibitem{Ade:2015xua}
{\bf Planck} Collaboration, P.~A.~R. Ade et~al., {\it {Planck 2015 results.
  XIII. Cosmological parameters}},  \href{http://arxiv.org/abs/1502.01589}{{\tt
  arXiv:1502.01589}}.

\bibitem{Aad:2015zhl}
{\bf ATLAS, CMS} Collaboration, G.~Aad et~al., {\it {Combined Measurement of
  the Higgs Boson Mass in $pp$ Collisions at $\sqrt{s}=7$ and 8 TeV with the
  ATLAS and CMS Experiments}},  {\em Phys. Rev. Lett.} {\bf 114} (2015) 191803,
  [\href{http://arxiv.org/abs/1503.07589}{{\tt arXiv:1503.07589}}].

\bibitem{Hahn:2013ria}
T.~Hahn, S.~Heinemeyer, W.~Hollik, H.~Rzehak, and G.~Weiglein, {\it
  {High-Precision Predictions for the Light CP -Even Higgs Boson Mass of the
  Minimal Supersymmetric Standard Model}},  {\em Phys. Rev. Lett.} {\bf 112}
  (2014), no.~14 141801, [\href{http://arxiv.org/abs/1312.4937}{{\tt
  arXiv:1312.4937}}].

\bibitem{Breiman:RF}
``{Random Forests - classification description}.''
  \url{https://www.stat.berkeley.edu/~breiman/RandomForests/cc\_home.htm}.
\newblock Accessed: 2015-12-07.

\bibitem{Freund:1997}
Y.~Freund and R.~E. Schapire, {\it {A Decision-Theoretic Generalization of
  On-Line Learning and an Application to Boosting}},  {\em J. Comput. Syst.
  Sci.} {\bf 55} (Aug., 1997) 119--139.

\bibitem{Cover:1967}
T.~Cover and P.~Hart, {\it {Nearest neightbor pattern classification}},  {\em
  IEEE Transactions on Information Theory} {\bf IT-11} (1967) 21--27.

\bibitem{Boser:1992}
B.~E. Boser, I.~M. Guyon, and V.~N. Vapnik, {\it {A Training Algorithm for
  Optimal Margin Classifiers}},  {\em {Proceedings of the Fifth Annual Workshop
  on Computational Learning Theory}} (1992) 144--152.

\bibitem{Witten:2005}
I.~H. Witten and E.~Frank, {\em {Data mining: practical machine learning tools
  and techniques}}.
\newblock Elsevier/Morgan Kaufmann, 2nd~ed., 2005.

\bibitem{Bishop:2006}
C.~M. Bishop, {\em {Pattern recognition and machine learning}}.
\newblock Springer, Oct., 2006.

\bibitem{breimano}
L.~Breiman, ``{Out-of-bag estimation}.''
  \url{https://www.stat.berkeley.edu/~breiman/OOBestimation.pdf}.

\bibitem{scikit-learn}
F.~Pedregosa, G.~Varoquaux, A.~Gramfort, V.~Michel, B.~Thirion, O.~Grisel,
  M.~Blondel, P.~Prettenhofer, R.~Weiss, V.~Dubourg, J.~Vanderplas, A.~Passos,
  D.~Cournapeau, M.~Brucher, M.~Perrot, and E.~Duchesnay, {\it {Scikit-learn:
  Machine Learning in {P}ython}},  {\em Journal of Machine Learning Research}
  {\bf 12} (2011) 2825--2830.

\bibitem{fawcett:2006roc}
T.~Fawcett, {\it {An introduction to ROC analysis}},  {\em Pattern recognition
  letters} {\bf 27} (2006), no.~8 861--874.

\bibitem{Skands:2003cj}
P.~Z. Skands et~al., {\it {SUSY Les Houches accord: Interfacing SUSY spectrum
  calculators, decay packages, and event generators}},  {\em JHEP} {\bf 07}
  (2004) 036, [\href{http://arxiv.org/abs/hep-ph/0311123}{{\tt
  hep-ph/0311123}}].

\bibitem{Gansner00anopen}
E.~R. Gansner and S.~C. North, {\it {An open graph visualization system and its
  applications to software engineering}},  {\em {Software -- practice and
  experience}} {\bf 30} (2000), no.~11 1203--1233.

\bibitem{Drees:2015aeo}
M.~Drees and J.~S. Kim, {\it {Minimal natural supersymmetry after the LHC8}},
  {\em Phys. Rev.} {\bf D93} (2016), no.~9 095005,
  [\href{http://arxiv.org/abs/1511.04461}{{\tt arXiv:1511.04461}}].

\bibitem{Drees:1990dx}
M.~Drees and K.~Hagiwara, {\it {Supersymmetric Contribution to the Electroweak
  $\rho$ Parameter}},  {\em Phys. Rev.} {\bf D42} (1990) 1709--1725.

\bibitem{Abbiendi:2002vz}
{\bf OPAL} Collaboration, G.~Abbiendi et~al., {\it {Search for nearly mass
  degenerate charginos and neutralinos at LEP}},  {\em Eur. Phys. J.} {\bf C29}
  (2003) 479--489, [\href{http://arxiv.org/abs/hep-ex/0210043}{{\tt
  hep-ex/0210043}}].

\bibitem{Abdallah:2003xe}
{\bf DELPHI} Collaboration, J.~Abdallah et~al., {\it {Searches for
  supersymmetric particles in e+ e- collisions up to 208-GeV and interpretation
  of the results within the MSSM}},  {\em Eur. Phys. J.} {\bf C31} (2003)
  421--479, [\href{http://arxiv.org/abs/hep-ex/0311019}{{\tt hep-ex/0311019}}].

\bibitem{Abbiendi:2003sc}
{\bf OPAL} Collaboration, G.~Abbiendi et~al., {\it {Search for chargino and
  neutralino production at $\sqrt{s} = 192$ GeV to 209 GeV at LEP}},  {\em Eur.
  Phys. J.} {\bf C35} (2004) 1--20,
  [\href{http://arxiv.org/abs/hep-ex/0401026}{{\tt hep-ex/0401026}}].

\bibitem{Sjostrand:2014zea}
T.~Sjostrand, S.~Ask, J.~R. Christiansen, R.~Corke, N.~Desai, P.~Ilten,
  S.~Mrenna, S.~Prestel, C.~O. Rasmussen, and P.~Z. Skands, {\it {An
  Introduction to PYTHIA 8.2}},  {\em Comput. Phys. Commun.} {\bf 191} (2015)
  159--177, [\href{http://arxiv.org/abs/1410.3012}{{\tt arXiv:1410.3012}}].

\bibitem{Alwall:2014hca}
J.~Alwall, R.~Frederix, S.~Frixione, V.~Hirschi, F.~Maltoni, O.~Mattelaer,
  H.~S. Shao, T.~Stelzer, P.~Torrielli, and M.~Zaro, {\it {The automated
  computation of tree-level and next-to-leading order differential cross
  sections, and their matching to parton shower simulations}},  {\em JHEP} {\bf
  07} (2014) 079, [\href{http://arxiv.org/abs/1405.0301}{{\tt
  arXiv:1405.0301}}].

\bibitem{Sjostrand:2006za}
T.~Sjostrand, S.~Mrenna, and P.~Z. Skands, {\it {PYTHIA 6.4 Physics and
  Manual}},  {\em JHEP} {\bf 05} (2006) 026,
  [\href{http://arxiv.org/abs/hep-ph/0603175}{{\tt hep-ph/0603175}}].

\bibitem{Chatrchyan:2013mys}
{\bf CMS} Collaboration, S.~Chatrchyan et~al., {\it {Search for supersymmetry
  in hadronic final states with missing transverse energy using the variables
  $\alpha_T$ and b-quark multiplicity in pp collisions at $\sqrt s=8$ TeV}},
  {\em Eur. Phys. J.} {\bf C73} (2013), no.~9 2568,
  [\href{http://arxiv.org/abs/1303.2985}{{\tt arXiv:1303.2985}}].

\bibitem{ATLAS-CONF-2012-104}
{\it {Search for supersymmetry at $\sqrt{s} = 8$ TeV in final states with jets,
  missing transverse momentum and one isolated lepton}},  Tech. Rep.
  ATLAS-CONF-2012-104, CERN, Geneva, Aug, 2012.

\bibitem{ATLAS-CONF-2013-024}
{\it {Search for direct production of the top squark in the all-hadronic
  $t\bar{t}$ + $E_T^\text{miss}$ final state in 21 fb$^{-1}$ of $pp$ collisions
  at $\sqrt{s}=8$ TeV with the ATLAS detector}},  Tech. Rep.
  ATLAS-CONF-2013-024, CERN, Geneva, Mar, 2013.

\bibitem{ATLAS-CONF-2013-047}
{\it {Search for squarks and gluinos with the ATLAS detector in final states
  with jets and missing transverse momentum and 20.3 fb$^{-1}$ of $\sqrt{s}=8$
  TeV proton-proton collision data}},  Tech. Rep. ATLAS-CONF-2013-047, CERN,
  Geneva, May, 2013.

\bibitem{ATLAS-CONF-2013-061}
{\it {Search for strong production of supersymmetric particles in final states
  with missing transverse momentum and at least three b-jets using 20.1
  fb$^{-1}$ of pp collisions at $\sqrt{s} = 8$ TeV with the ATLAS Detector.}},
  Tech. Rep. ATLAS-CONF-2013-061, CERN, Geneva, Jun, 2013.

\bibitem{ATLAS-CONF-2013-062}
{\it {Search for squarks and gluinos in events with isolated leptons, jets and
  missing transverse momentum at $\sqrt{s}=8$ TeV with the ATLAS detector}},
  Tech. Rep. ATLAS-CONF-2013-062, CERN, Geneva, Jun, 2013.

\bibitem{CMS-PAS-SUS13-016}
{\it {Search for supersymmetry in pp collisions at a center-of-mass energy of 8
  TeV in events with two opposite sign leptons, large number of jets, b-tagged
  jets, and large missing transverse energy}},  Tech. Rep. CMS-SUS-13-016,
  CERN, Geneva, Nov, 2013.

\bibitem{Read:2002hq}
A.~L. Read, {\it {Presentation of search results: The CL(s) technique}},  {\em
  J. Phys.} {\bf G28} (2002) 2693--2704.

\bibitem{Chamseddine:1982jx}
A.~H. Chamseddine, R.~L. Arnowitt, and P.~Nath, {\it {Locally Supersymmetric
  Grand Unification}},  {\em Phys. Rev. Lett.} {\bf 49} (1982) 970.

\bibitem{Barbieri:1982eh}
R.~Barbieri, S.~Ferrara, and C.~A. Savoy, {\it {Gauge Models with Spontaneously
  Broken Local Supersymmetry}},  {\em Phys. Lett.} {\bf B119} (1982) 343.

\bibitem{Hall:1983iz}
L.~J. Hall, J.~D. Lykken, and S.~Weinberg, {\it {Supergravity as the Messenger
  of Supersymmetry Breaking}},  {\em Phys. Rev.} {\bf D27} (1983) 2359--2378.

\bibitem{Djouadi:2002ze}
A.~Djouadi, J.-L. Kneur, and G.~Moultaka, {\it {SuSpect: A Fortran code for the
  supersymmetric and Higgs particle spectrum in the MSSM}},  {\em Comput. Phys.
  Commun.} {\bf 176} (2007) 426--455,
  [\href{http://arxiv.org/abs/hep-ph/0211331}{{\tt hep-ph/0211331}}].

\bibitem{Barr:2016inz}
A.~Barr and J.~Liu, {\it {First interpretation of 13 TeV supersymmetry searches
  in the pMSSM}},  \href{http://arxiv.org/abs/1605.09502}{{\tt
  arXiv:1605.09502}}.

\bibitem{Barr:2016sho}
A.~Barr and J.~Liu, {\it {Analysing parameter space correlations of recent 13
  TeV gluino and squark searches in the pMSSM}},  {\em Eur. Phys. J.} {\bf C77}
  (2017), no.~3 202, [\href{http://arxiv.org/abs/1608.05379}{{\tt
  arXiv:1608.05379}}].

\end{thebibliography}\endgroup
\bibliographystyle{JHEP}

\end{document}